\documentclass[11pt,twoside,a4paper,cmspaper,final,collab]{cms-tdr}
\def\svnVersion{1a10a72-D}\def\svnDate{2026/10/30}\def\cmsCernNoTag{CERN-EP-2026-263}\def\cmsCernDate{\today}\def\cmsMessage{Submitted to the Journal of High Energy Physics}
\begin{document}\cmsNoteHeader{SUS-23-002}

\newlength\cmsFigWidth
\providecommand{\cmsTable}[1]{\resizebox{\textwidth}{!}{#1}}
\newlength\cmsTabSkip\setlength{\cmsTabSkip}{1ex}

\newcommand{\FastSim}{{\textsc{FastSim}}\xspace}
\newcommand{\invM}[1]{\ensuremath{m_{#1}}\xspace}
\newcommand{\mll}{\invM{\Pell\Pell}}
\newcommand{\MT}[1]{\ensuremath{m_\mathrm{#1}}\xspace}
\newcommand{\mtll}{\ensuremath{\MT{T2}(\Pell\Pell)}\xspace}
\newcommand{\tvec}[1]{\ensuremath{{\vec p}^{\,\text{#1}}_{\mathrm{T}}}}
\newcommand{\mumu}{\ensuremath{\PGm\PGm}\xspace}
\newcommand{\ee}{\ensuremath{\Pe\Pe}\xspace}
\newcommand{\emu}{\ensuremath{\Pe\PGm}\xspace}
\newcommand{\tautau}{\ensuremath{\PGt\PGt}\xspace}
\newcommand{\pp}{\ensuremath{\Pp\Pp}\xspace}
\newcommand{\tW}{\ensuremath{\PQt\PW}\xspace}
\newcommand{\WW}{\ensuremath{\PW\PW}\xspace}
\newcommand{\WZ}{\ensuremath{\PW\PZ}\xspace}
\newcommand{\ZZ}{\ensuremath{\PZ\PZ}\xspace}
\newcommand{\ttW}{\ensuremath{\ttbar\PW}\xspace}
\newcommand{\ttZ}{\ensuremath{\ttbar\PZ}\xspace}
\newcommand{\DY}{\text{Drell--Yan}\xspace}
\newcommand{\Dm}{\ensuremath{\Delta m}\xspace}
\newcommand{\Dphi}{\ensuremath{\Delta\phi}\xspace}
\newcommand{\DZmass}{\ensuremath{\abs{\mll - \invM{\PZ}}}}
\newcommand{\N}[1]{\ensuremath{N_\text{#1}}\xspace}
\newcommand{\Reg}[3]{\ensuremath{\text{#1}_\text{#2}^\text{#3}}\xspace}
\newcommand{\TChipmSlep}{\ensuremath{\PSGcpmDo\to\PGn\PSell\to\PGn\Pell\PSGczDo}\xspace}
\newcommand{\TChipmSneu}{\ensuremath{\PSGcpmDo\to\Pell\sNu\to\Pell\PGn\PSGczDo}\xspace}
\newcommand{\TChipmDecay}{\ensuremath{\PSGcpmDo\to\PSell\PGn/\Pell\sNu\to\Pell\PGn\PSGczDo}}
\newcommand{\BF}{\ensuremath{\mathcal{B}}\xspace}
\newcommand{\dphimin}{\ensuremath{\Delta\phi^{\text{min}}(\Pell,\ptmiss)}\xspace}
\newcommand{\ptvecmissz}{\ensuremath{{\vec p}_{\mathrm{T}}^{\kern1pt\text{miss,\PZ}}}\xspace}

\cmsNoteHeader{SUS-23-002}
\title{Search for supersymmetric particle pair production in final states with two oppositely charged leptons and large missing transverse momentum in proton-proton collisions at \texorpdfstring{$\sqrt{s}=13\TeV$}{sqrt(s)=13 TeV}}

\date{\today}

\abstract{
A search is presented for direct production of pairs of supersymmetric particles in final states with a pair of oppositely charged leptons (electrons or muons) and missing transverse momentum. The analysis uses proton-proton collision data at $\sqrt{s}={13}\TeV$, corresponding to an integrated luminosity of 138\fbinv, recorded with the CMS detector at the CERN LHC. The observed data do not exhibit any excess over the standard model predictions. Constraints are derived on chargino, top squark, and slepton pair production in the context of simplified models with $R$-parity conservation. In these models, the neutralino is taken to be the lightest supersymmetric particle. For chargino production, an exclusion region in the chargino-neutralino mass plane is determined, which extends to chargino masses of 1120\GeV and neutralino masses of 490\GeV. For mass splittings between the lighter top squark and the neutralino ranging from the \PW boson mass to the top quark mass, top squark masses up to 500--540\GeV are excluded. Slepton masses are excluded up to 700\GeV for a massless neutralino.
}

\hypersetup{
pdfauthor={CMS Collaboration},
pdftitle={Search for supersymmetric particle pair production in final states with two oppositely charged leptons and large missing transverse momentum in proton-proton collisions at sqrt(s)=13 TeV},
pdfsubject={CMS},
pdfkeywords={CMS, supersymmetry}} 

\maketitle

\section{Introduction}\label{sec:introduction}

The standard model (SM) of particle physics provides a highly precise description of a wide range of high-energy phenomena. Despite this success, it possesses fundamental limitations, such as the lack of a viable dark-matter candidate capable of explaining cosmological observations~\cite{DarkMatter1,DarkMatter2,DarkMatter3} and its inability to account for the hierarchy between the Higgs boson mass and the Planck scale~\cite{PhysRevD.13.3333,Hierarchy1}.
Supersymmetry~\cite{Ramond:1971gb,Golfand:1971iw,Neveu:1971rx,Volkov:1972jx,Wess:1973kz,Wess:1974tw,Fayet:1974pd,Nilles:1983ge,SUSYPrimer} is an extension of the SM that addresses these limitations by embedding each SM field into a supermultiplet containing a partner whose spin differs by 1/2, providing fermionic (bosonic) superpartners for bosonic (fermionic) SM fields.
The fermionic superpartners of the SM electroweak gauge and Higgs fields (gauginos and higgsinos) mix to form mass eigenstates known as charginos and neutralinos. In a broad class of models with supersymmetry, the lightest neutralino (\PSGczDo) is the lightest supersymmetric particle and, when $R$-parity~\cite{Farrar:1978xj} is preserved, remains stable, and thus serves as a suitable dark matter candidate. 
The scalar superpartners of the left- and right-handed top quarks combine into two mass eigenstates. For values of the lightest top squark (\stone) mass below or around the TeV scale, the radiative contributions from top squark loops to the Higgs boson mass become sufficiently large to cancel the dominant top quark loop corrections, stabilizing the Higgs boson mass at the electroweak scale~\cite{Hierarchy2,Hierarchy3,Hierarchy4}.

In this paper, a search for supersymmetric particle production in final states with two leptons ($\Pell=\Pe$ or \PGm) with opposite sign of charge (OS) and missing transverse momentum is presented. 
Proton-proton (\pp) collision data at $\sqrt{s}=13\TeV$ collected between 2016 and 2018 by the CMS experiment~\cite{Chatrchyan:2008zzk} at the CERN LHC are used.  The data correspond to an integrated luminosity of 138\fbinv~\cite{CMS-LUM-17-003,CMS-LUM-20-001}. The analysis is an extension of a previously published CMS search~\cite{CMSTChipm13TeV} for pair production of the lightest chargino (\PSGcpmDo) and of the lightest top squark, which was based on data collected in 2016. An optimization of the event selection is performed to capture the specific event topologies of the different signal scenarios, and additional control samples for the estimation of the main SM processes in data are introduced. 
An additional signal scenario with the pair production of sleptons (\PSell) is considered.

\begin{figure}[hbtp]
  \centering
    \includegraphics[width=0.4\textwidth]{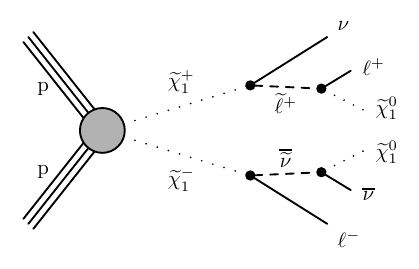}    
    \includegraphics[width=0.4\textwidth]{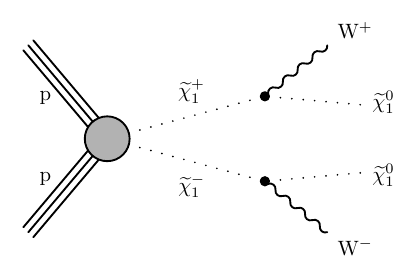}                                                                                                                
    \caption{Simplified-model diagrams of chargino pair production with two benchmark decay modes: the left plot shows decays through intermediate charged sleptons or sneutrinos, while the right one displays prompt decays into a {\PW} boson and the lightest neutralino.}
    \label{Fig:TChipm}
\end{figure}

Results are interpreted in terms of various simplified supersymmetric model spectra (SMS)~\cite{SMS1,SMS2,SMS3} scenarios.
Three classes of signal processes are considered: chargino pair production, top squark pair production, and slepton pair production. Two decay scenarios are considered for charginos and top squarks, while one is considered for sleptons.

The search for chargino pair production considers, as a reference, a model (Fig.~\ref{Fig:TChipm}, left)
where the charginos decay into a lepton, a neutrino (\PGn), and the lightest neutralino, assumed to be the lightest supersymmetric particle, via an intermediate charged slepton (\TChipmSlep) or sneutrino (\TChipmSneu).
Each of these decay chains is assigned a 50\% branching fraction ({\BF}).
The three generations of sleptons are assumed to be degenerate, each with a mass equal to the average of the chargino and neutralino masses. 
This slepton mass assignment generally leads to an intermediate scenario in terms of sensitivity, since a larger mass splitting between the intermediate slepton and the lightest neutralino results in a higher-momentum \PSGczDo, and so in a higher missing transverse momentum.
Results are also interpreted in terms of a second model involving charginos, shown in Fig.~\ref{Fig:TChipm} (right). In this model, each chargino decays only into the lightest neutralino and a {\PW} boson. 
This paper complements recent results for hadronic final states~\cite{CMSTChiWWHadrWX} and for compressed mass spectra in inclusive final states~\cite{CMSCompressedSUS}. 
The ATLAS Collaboration has also published several searches for these signal models~\cite{ref:ATLASTChiWWSlep2lepCompressedRun2,ref:ATLASTChiSlep2lepRun2,ref:ATLASTChiSlepTauhRun2,ATLASTChiWW1lep,ATLASTChipmWWCombi}.

\begin{figure}[hbtp]
  \centering
    \includegraphics[width=0.4\textwidth]{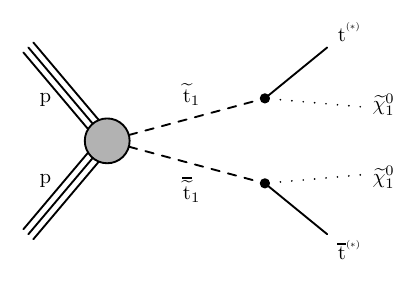}
    \includegraphics[width=0.4\textwidth]{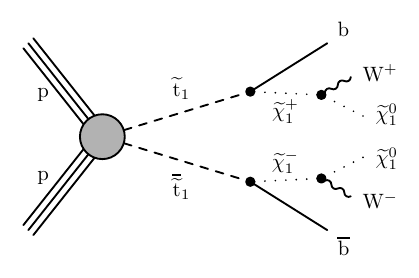}
    \caption{Simplified-model diagrams of top squark pair production with two benchmark decay modes of the top squark: the left plot shows decays into a top quark and the lightest neutralino, while the right one displays prompt decays into a bottom quark, and a chargino that subsequently decays into a neutralino and a {\PW} boson.}
    \label{Fig:T2}
\end{figure}

The main target of the top squark pair production search is an SMS model where the top squark decays via a top quark and the lightest neutralino, as presented in Fig.~\ref{Fig:T2} (left).
The analysis is tailored for a specific mass spectrum where the mass difference (\Dm) between the top squark and the lightest neutralino lies between the {\PW} boson and top quark masses $(\invM{\PW}<\Dm\lesssim\invM{\PQt})$. In this ``compressed'' scenario, top quarks are produced off-shell, leading to final states characterized by low-momentum bottom quarks that frequently elude detection.
Further interpretations of the results are given in terms of an additional model involving top squarks, shown in Fig.~\ref{Fig:T2} (right). In this model, each of the pair-produced top squarks decays into a bottom quark and a chargino, which in turn decays into a {\PW} boson and the lightest neutralino. In this scenario, the chargino's mass is taken to be the mean of the top squark and neutralino masses, similarly to what was done for the model with pair production of charginos decaying via intermediate sleptons.
This work is complementary to other OS dilepton searches published by CMS~\cite{ref:t2tt2lep,ref:t2tt2lepRun2}, aimed at testing models where $\Dm>\invM{\PQt}$, which result in signatures with on-shell top quarks and higher-momentum particles.
With respect to that analysis, additional sensitivity is obtained in the compressed-mass region by loosening the requirements on the jets from bottom quark hadronization and optimizing the signal event selection for the lower momentum carried by the neutralinos.
Other searches targeting the same signal models in different final states have been published by CMS~\cite{ref:t2tt0lep,ref:t2tt1lep,ref:t2tt1lepRun2,CMSCompressedSUS}, and a combination of the results has been performed~\cite{CMSStopComb}.
Several searches addressing these signal models have also been published by ATLAS using various final states~\cite{ref:ATLAS0lep,ref:ATLAS1lep,ATLASStop,ref:ATLAS0lepRun2,ref:ATLAS1lepRun2,ATLASStopRun2,ATLASStop02LepMET}.

\begin{figure}[hbtp]
  \centering
    \includegraphics[width=0.4\textwidth]{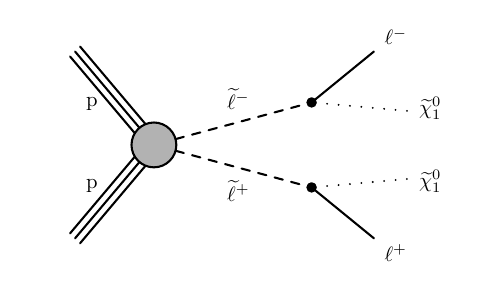}
  \caption{Simplified-model diagram of charged slepton pair production. We focus on a scenario where each slepton decays into a charged lepton and a neutralino.}
  \label{Fig:TSlepSlep}
\end{figure}

Finally, results are interpreted in terms of a model with pair production of charged sleptons, each of them subsequently decaying into a charged lepton and a neutralino, as shown in Fig.~\ref{Fig:TSlepSlep}. This search extends the phase space explored by other studies previously published by CMS~\cite{edgerun2,ref:TwoThreeLeptoSoft,CMSCompressedSUS} and their combination~\cite{CMSEWKCombination}. Several results on this model are also available from ATLAS~\cite{ref:ATLASTChiSlep2lepRun2,ref:ATLASTChiWWSlep2lepCompressedRun2,ref:ATLASTChiSlep2lepRun2,ref:ATLASCompSlep}.

The paper is organized as follows: Section~\ref{sec:detector} introduces the experimental apparatus;
Sections~\ref{sec:datasets} and~\ref{sec:reconstruction} describe the data and simulated event samples used in this search and the details on the event reconstruction, respectively;
Section~\ref{sec:searchstrategy} presents the general strategy of the analysis;
Section~\ref{sec:bkgestimation} discusses the estimates of the contributions from SM processes to the selected events; Section~\ref{sec:systematics} details the sources of systematic uncertainties for signal and background processes;
Section~\ref{sec:results} reports the results and their interpretation in terms of the considered SMS;
and finally Section~\ref{sec:summary} summarizes the results of the search.
Tabulated results are accessible in the HEPData record~\cite{hepdata} for this analysis.

\section{The CMS detector}\label{sec:detector}

The CMS apparatus is a multipurpose, nearly hermetic detector, designed to trigger on and identify electrons, muons, photons, and (charged and neutral) hadrons. Its central feature is a superconducting solenoid of 6\unit{m} internal diameter, providing a magnetic field of 3.8\unit{T}.
A silicon pixel and strip tracker, a lead tungstate crystal electromagnetic calorimeter (ECAL), and a brass and scintillator hadron calorimeter (HCAL) are within the solenoid volume, each composed of a barrel and two endcap sections. Forward calorimeters extend the pseudorapidity $\eta$ coverage provided by the barrel and endcap detectors. Muons are reconstructed in gas-ionization detectors embedded in the steel flux-return yoke outside the solenoid. More detailed descriptions of the CMS detector, together with a definition of the coordinate system used and the relevant kinematic variables, can be found in Ref.~\cite{Chatrchyan:2008zzk}.

Events of interest are selected using a two-tiered trigger system. The first level (L1), composed of custom hardware processors, uses information from the calorimeters and muon detectors to select events at a rate of around 100\unit{kHz} within a fixed latency of approximately 4\mus~\cite{CMS:2020cmk}. The second level, known as the high-level trigger, consists of a farm of processors running a version of the full event reconstruction software optimized for fast processing, and reduces the event rate to around 1\unit{kHz} before data storage~\cite{CMS:2016ngn,CMS:2024aqx}.

\section{Data and simulated samples}\label{sec:datasets}

Events of interest are selected using high-level trigger algorithms that require the presence of two isolated leptons (\ee, \mumu, \emu).
The threshold on the transverse momentum (\pt) of the leading lepton is 23\GeV for the \ee and \emu triggers, and 17\GeV for the \mumu triggers. 
The threshold for the trailing lepton is 12\GeV for electrons in all triggers and for muons in the \emu triggers during the 2017 and 2018 data-taking periods, while it is 8\GeV for muons in the \emu triggers during 2016 data-taking period and in the \mumu triggers.
To increase the efficiency of the trigger selection, events are also accepted by triggers requiring at least one electron or muon with higher transverse momentum and passing tighter identification criteria than required in the double-lepton triggers.
The thresholds on the lepton \pt are 27 (24), 35 (27), and 32 (24)\GeV for the single-electron (-muon) triggers during the 2016, 2017 and 2018 data-taking periods, respectively. For the 2016 data set, an additional single-electron trigger with a \pt threshold at 25\GeV is also used to recover efficiencies for electrons with $\abs{\eta}<2.1$.
The performance of each trigger is measured in events collected through large missing transverse momentum requirements.
The combined efficiency of the dilepton and single-lepton triggers for signal events is found to range 75--99\%, depending on the \pt of the leptons.

Samples of Monte Carlo (MC) simulated events are used to study the contribution of SM processes to the selected data set and the expected acceptance for the different signal models. Events from top quark-antiquark pair (\ttbar) production are generated with \POWHEG v2~\cite{bib:powhegv2a,bib:powhegv2b,bib:powhegv2c} and normalized to the expected cross section calculated at next-to-next-to-leading order (NNLO) in perturbative quantum chromodynamics (QCD), including resummation of next-to-next-to-leading logarithmic (NNLL) soft-gluon terms~\cite{Czakon:2011xx}.
Events containing a single top quark produced in association with a {\PW} boson (\tW) are generated using \POWHEG v2~\cite{bib:powheg3} and normalized to an approximate NNLO cross section calculation~\cite{bib:tWxsec}.
The \WW and \ZZ production via quark-antiquark annihilation is simulated at next-to-leading order (NLO) using \POWHEG v2~\cite{bib:powhegVVa,bib:powhegVVb}.
The yields of events from \WW production are scaled to the NNLO cross section~\cite{bib:WWxsec}.
Events from $\Pq\Paq\to\PZ\PZ$ production are reweighted via NNLO/NLO \emph{K}-factors~\cite{HIG16041}.
Diboson production via gluon fusion is simulated using \MCFM v7~\cite{bib:MCFM}, and leading order (LO) cross sections obtained from the generator are corrected with the NNLO/LO \emph{K}-factors~\cite{bib:ggWWkfactors,HIG16041}.
\DY events are generated with \MGvATNLO v2.6.5~\cite{Artoisenet:2012st} at LO, and event yields are scaled to the NNLO cross section~\cite{bib:DYxsec}.
Events from \WZ, \ttW, \ttZ, triboson, and $\PH\to\WW$ production are generated at NLO with the \MGvATNLO v2.6.5 generator~\cite{bib:ttVxsec,bib:HWWxsec}.

Signal events for the chargino, top squark, and slepton pair production models are generated using \MGvATNLO v2.6.5 at LO with up to two extra partons in the matrix element calculations, and are normalized to the respective cross sections computed assuming all the other sparticles to be heavy and decoupled. Cross sections for chargino pair production and slepton pair production are computed at NLO plus next-to-leading logarithmic (NLL) precision~\cite{Beenakker:1999xh,Fuks:2012qx,Fuks:2013vua,Fuks:2013lya}. In the case of chargino pair production, calculations are performed in a limit of mass-degenerate wino \PSGczDt and \PSGcpmDo, and light bino \PSGczDo. For the slepton pair production, cross sections are computed for any single generation of left- and right-handed sleptons. The top squark pair production cross sections are determined at approximate NNLO plus NNLL precision~\cite{Beenakker:1996ch,Kulesza:2008jb,Kulesza:2009kq,Beenakker:2009ha,Beenakker:2011fu,Borschensky:2014cia,Beenakker:2011sf,Beenakker:2013mva,Beenakker:2014sma,Beenakker:2016lwe,Beenakker:1997ut,Beenakker:2010nq,Beenakker:2016gmf,Beenakker:2024}. 

All processes are generated using the NNPDF3.1~\cite{NNPDF31} parton distribution function (PDF) set.
Parton showering and hadronization are simulated using \PYTHIA 8.240~\cite{Sjostrand:2014zea} for all the processes but for \ttbar events, for which the first emission is done at the matrix element level with \POWHEG v2. The underlying event is modeled with the CP5~\cite{CMS:2019csb} tune.
Weights for the estimation of theoretical systematic uncertainties, including those related to the choice of PDFs, and renormalization and factorization scales, are included in the simulated events.

The detector response to the generated events is simulated using a realistic model of the CMS detector based on \GEANTfour~\cite{Agostinelli:2002hh} for SM processes, while for signal events a fast simulation (\FastSim)~\cite{Abdullin:2011zz} of the detector based on a parametrization of the average response to particles is used. Simulated events are subsequently reconstructed using the same algorithms applied to data.

An average of 32 collisions per bunch crossing were produced at the LHC during the 2016--2018 data-taking period.
In order to model the effect of these multiple interactions (pileup), signal and background simulated events are mixed with minimum-bias events simulated with \PYTHIA, and reweighted so as to match the observed rate of multiple interactions.

A decrease in trigger efficiency was observed in 2016 and 2017 with respect to preceding data-taking periods because of a gradual shift in the timing of the inputs of the ECAL L1 trigger, resulting in a significant fraction of high-$\eta$ trigger objects being mistakenly associated to the previous bunch crossing (prefiring). To model this effect in simulated events, weights are applied as a function of the probability of prefiring measured in data.

The modeling and normalization of the main background processes are studied in data, as discussed in Section~\ref{sec:bkgestimation}. The modeling of \ttbar, \tW, and \WW production is studied in data control regions (CRs), and their normalization is determined via a maximum likelihood (ML) fit to data. The normalization of the yields of events from \WZ, \ZZ, and \ttZ production is also constrained in the fit to the event rates observed in dedicated CRs.

Simulated signal events are reweighted to improve the modeling of the jets from  initial-state radiation (ISR), based on the observed agreement between data and simulated events in dedicated CRs. In particular, \ttbar events in dilepton final states are used for the ISR jet modeling of supersymmetric signals produced via strong interactions such as top squark events, while \DY events are used for signals produced via electroweak interactions such as chargino and slepton pair production. 
The reweighting factors range 1.12--0.65 (0.96--1.47) for multiplicity of ISR jets between 1 and 5 in signal events produced via strong (electroweak) interactions.

\section{Event reconstruction}\label{sec:reconstruction}

The global event reconstruction (also called particle-flow event reconstruction~\cite{CMS:2017yfk}) aims to reconstruct and identify each individual particle in an event, with an optimized combination of all subdetector information. In this process, the identification of the particle type (photon, electron, muon, charged hadron, neutral hadron) plays an important role in the determination of the particle direction and energy. 
Photons (\eg coming from \PGpz decays or from electron bremsstrahlung) are identified as ECAL energy clusters not linked to the extrapolation of any charged-particle trajectory to the ECAL. Electrons are identified as a primary charged particle track and potentially many ECAL energy clusters corresponding to this track extrapolation to the ECAL and to possible bremsstrahlung photons emitted along the way through the tracker material. Muons are identified as tracks in the central tracker consistent with either a track or several hits in the muon system, and associated with calorimeter deposits compatible with the muon hypothesis. Charged hadrons are identified as charged particle tracks identified neither as electrons, nor as muons. Finally, neutral hadrons are identified as HCAL energy clusters not linked to any charged hadron trajectory, or as a combined ECAL and HCAL energy excess with respect to the expected charged hadron energy deposit.

The energy of photons is obtained from the ECAL measurement. The energy of electrons is determined from a combination of the track momentum at the main interaction vertex, the corresponding ECAL cluster energy, and the energy sum of all bremsstrahlung photons attached to the track. The energy of muons is obtained from the corresponding track momentum. The energy of charged hadrons is determined from a combination of the track momentum and the corresponding ECAL and HCAL energies, corrected for the response function of the calorimeters to hadronic showers. 
Finally, the energy of neutral hadrons is computed from the residual energy in the ECAL and HCAL that is not assigned to other particles.

The primary interaction vertex is taken to be the vertex corresponding to the hardest scattering in the event, evaluated using tracking information alone, as described in Section 9.4.1 of Ref.~\cite{CMS-TDR-15-02}.

Electron identification relies on quality criteria of the electron track, matching between the electron trajectory and the associated cluster in the calorimeter, and shape observables of the electromagnetic shower observed in the ECAL~\cite{CMS:2020uim,CMS-DP-2020-021}. 
The muons utilized in the analysis are identified based on the number of reconstructed energy deposits in the tracker and in the muon system, and on the fit quality of the muon track~\cite{CMS:2018rym}.
The average efficiency for lepton reconstruction and selection is found to be around 80\% for electrons and 95\% for muons.

The lepton selection is further optimized to select prompt leptons. The leptons are required to be isolated from other particles in the event using their relative isolation ($I_\text{rel}$), defined as the ratio of the scalar \pt sum of the photons and of the neutral and charged hadrons within a cone of radius $R = \sqrt{\smash[b]{(\Delta\phi)^2 + (\Delta\eta)^2}}=0.4$ around the candidate lepton to the \pt of the lepton itself, where $\phi$ is azimuthal angle in radians. 
To mitigate the impact of particles stemming from pileup interactions, only charged hadrons consistent with originating from the event primary vertex are taken into account. Additionally, adjustments are made to correct for the expected contribution of neutral particles from pileup~\cite{CMS:2018rym,CMS:2020uim}. 
The trajectories of the candidate leptons are required to be compatible with the primary interaction vertex. This is achieved by imposing constraints on their transverse ($d_{0}$) and longitudinal ($d_{z}$) impact parameters, as well as the three-dimensional (3D) impact parameter significance ($S_\mathrm{3D}^\mathrm{d}$), determined by the ratio of the 3D impact parameter to its uncertainty. Both electrons and muons are required to satisfy the conditions $\abs{d_{0}}<0.05\cm$, $\abs{d_{z}}<0.10\cm$, and $S_\mathrm{3D}^\mathrm{d}<4$.
Finally, in order to reject electrons that arise from photon conversions, two specific criteria are applied. First, the electron track must not have missing hits in the innermost layers of the tracker. Second, it must not form a conversion vertex with another electron candidate within the same event~\cite{CMS:2020uim}.
A set of looser requirements is also used to define a veto lepton selection~\cite{CMS:2020uim,CMS-DP-2020-021,CMS:2018rym}.

For each event, hadronic jets are clustered from the reconstructed particles using the infrared- and collinear-safe anti-\kt algorithm~\cite{Cacciari:2008gp, Cacciari:2011ma} with a distance parameter of 0.4. Jet momentum is determined as the vectorial sum of all particle momenta in the jet, and is found from simulation to be, on average, within 5 to 10\% of the true momentum over the entire \pt spectrum and detector acceptance. Pileup can contribute additional tracks and calorimetric energy depositions to the jet momentum. To mitigate this effect, charged particles identified to be originating from pileup vertices are discarded and an offset correction~\cite{CMS:2020ebo} is applied to correct for remaining contributions. Jet energy corrections are derived from simulation to bring the measured response of jets to that of particle-level jets on average. In situ measurements of the momentum balance in dijet, photon$+$jet, $\PZ+$jet, and multijet events are used to account for any residual differences in the jet energy scale between data and simulation~\cite{CMS:2016lmd}. The jet energy resolution amounts typically to 15--20\% at 30\GeV, and 10\% at 100\GeV~\cite{CMS:2016lmd}. Additional selection criteria are applied to each jet to remove jets potentially dominated by anomalous contributions from various subdetector components or reconstruction failures~\cite{CMS:2020ebo}.

Jets originating from the hadronization of bottom quarks (\PQb jets) are identified using the medium operating point of a version of the combined secondary vertex \PQb tagging algorithm based on deep learning techniques~\cite{BTV-16-002}. This criterion yields an efficiency in identifying \PQb jets that ranges 55--70\% for jets with \pt of 20--100\GeV. The misidentification rate for jets originating from light quarks and gluons is approximately 1\% in the same \pt range.

The momentum imbalance of the event in the transverse plane is referred to as missing transverse momentum (\ptvecmiss) and it is defined as the negative vectorial \pt sum of all particle-flow candidates in the event, taking into account the energy corrections applied to the jets~\cite{CMS:2019ctu}. The magnitude of \ptvecmiss is denoted as \ptmiss.

Simulated events are reweighted to account for differences with respect to data in the efficiencies of the lepton reconstruction, identification, and isolation requirements, and in the performance of \PQb jet identification.
The values of the data-to-simulation scale factors differ from unity by less than 10\% with typical efficiency corrections of 2--3 (5)\% for the identification of leptons (\PQb\ jets) with $\pt>20\GeV$ and $\abs{\eta}<2.4$.

\section{Search strategy}\label{sec:searchstrategy}

Two signal production mechanisms are considered in this paper: the pair production of supersymmetric particles via the electroweak interaction (charginos and sleptons), and the pair production of top squarks via the strong interaction. The former is studied for the whole accessible mass spectrum, while for the latter the analysis is focused on the compressed scenario, where the mass difference between the top squark and the lightest neutralino falls between the top quark and {\PW} boson masses. The search strategies developed for both hypotheses involve the same techniques for the background estimation and the signal extraction, while they select slightly different search regions (SRs) in order to improve their respective sensitivity.

The signal events for all models are characterized by a common final state with two OS leptons and two lightest neutralinos, which provide additional contributions to \ptvecmiss. Given this scenario, a general high-acceptance baseline selection is defined, requiring two OS isolated leptons (\Pe, \PGm) with $\abs{\eta}<2.4$ and  $\pt>25$ (20)\GeV for the leading (trailing) lepton.
Events with electrons or muons originating from cascade decays through \PGt leptons are included in the selection and considered part of the signal.
To reduce the contributions from low-mass resonances, $\PZ\to\PGt\PGt$ production, and nonprompt leptons from hadronic jets, the invariant mass \mll of the lepton pair is required to be greater than 20\GeV. If both leptons have the same flavor (SF), \mll is further required to satisfy $\DZmass>15\GeV$, where $\invM{\PZ}=91.1879\GeV$ is the \PZ boson mass~\cite{PDG2026}.
High \ptmiss (${>}160\GeV$) is required. Events are further rejected if they contain a third lepton with $\pt>10\GeV$, $\abs{\eta}<2.4$, and satisfying the looser veto lepton selection introduced in Section~\ref{sec:reconstruction}. A summary of the baseline selection is found in Table~\ref{Tab:BaseSel}.

\begin{table}[htb]
\centering
\topcaption[center]{Baseline selection used to define the SRs for all considered models of chargino, slepton, and top squark pair production.}
\begin{tabular}{lcc}
\hline
Variable & \multicolumn{1}{c}{Selection} \\
\hline
 Lepton flavor         & \EE, \MM, \Pepm\PGmmp\\
 Leading lepton        & $\pt> 25\GeV$, $\abs{\eta}<2.4$ \\
 Trailing lepton       & $\pt> 20\GeV$, $\abs{\eta}<2.4$ \\
 Third lepton veto    &  $\pt> 10\GeV$, $\abs{\eta}<2.4$    \\
 \mll                  & ${>}20\GeV$ \\
 \DZmass               & ${>}15\GeV$ only for \ee and \mumu events\\
 \ptmiss               & ${>}160\GeV$\\
\hline
\end{tabular}
\label{Tab:BaseSel}
\end{table}

The largest contributions from SM processes to the events passing the baseline selection come from \ttbar, \tW, and \WW production. 
For all of these backgrounds, the lepton pair and \ptvecmiss come from the leptonic decays of a {\PW} boson pair.
Consequently, we consider the variable \MT{T2}~\cite{Lester:1999tx}, defined as a generalization of the transverse mass (\MT{T}) for a system with two invisible particles, using the two leptons as the two visible systems:
\begin{equation}
\mtll = \min_{\tvec{miss1} + \tvec{miss2} = \ptvecmiss} \left( \max \left[ \MT{T}(\tvec{lep1},\tvec{miss1}), \MT{T}(\tvec{lep2},\tvec{miss2}) \right] \right),
\end{equation}
where the minimum is taken over all the possible partitions of \ptvecmiss into two components \tvec{miss1} and \tvec{miss2} associated with the two invisible particles. This observable reaches a kinematic endpoint at the mass of the \PW boson  (\invM{\PW}) for the considered backgrounds. Signal events, instead, present \mtll spectra without such an endpoint because of the additional contribution to \ptvecmiss provided by the neutralinos. The sensitivity of the analysis is further enhanced by dividing the search regions (SR) into four bins of \ptmiss (numbered from SR1 to SR4 in order of increasing \ptmiss with the same boundaries used for all the signal models). This allows the analysis not only to exploit the larger tails in the \ptmiss distribution of the signal events, but also to optimize the sensitivity to signal hypotheses with different mass separation between the produced supersymmetric particle and the neutralino. Every \ptmiss bin is further split into events containing SF and different-flavor (DF) leptons to take advantage of the fact that the contribution of \WZ, \ZZ, and \DY production is significantly smaller in the DF channel.

The SRs are further subdivided based on the specific characteristics of each signal model.
A veto on \PQb-tagged jets is applied to reject \ttbar, \tW, \ttZ, and \ttW events in the chargino and slepton searches.
Driven by the different behavior in terms of jet multiplicity for signal and \ttbar production, selected events in the \ptmiss bins below 280\GeV are additionally split into two different subregions, depending on the presence or absence of a jet with $\pt>20\GeV$ and $\abs{\eta}<2.5$.
Events with \PQb-tagged jets are kept as a CR for the normalization of the background from \ttbar and \tW production, discussed in Section~\ref{sec:bkgestimation}.

The final states produced in the top squark decays differ from the ones in chargino and slepton pair production by the presence of two bottom quarks. The difference in the mass of the top squark and the neutralino across the compressed region affects significantly the reconstruction of the arising \PQb\ jets: for \Dm just above \invM{\PW}, they have a relatively low momentum and therefore a lower probability to be tagged, while for larger \Dm, their momentum increases and the final states resemble those of \ttbar events. Consequently, we can enhance the sensitivity to various \Dm regions of top squark production by defining SRs with no \PQb-tagged jets (\Reg{SR}{0tag}{}), suited to signal hypotheses with a small mass splitting between the top squark and the neutralino, and SRs with at least one \PQb-tagged jet (\Reg{SR}{tags}{}), tuned to scenarios with larger \Dm. The \Reg{SR}{0tag}{} are the same as for chargino searches, with the only difference that the \Reg{SR1}{0tag}{} and \Reg{SR2}{0tag}{} for the chargino analysis are split into events with (\Reg{SR1}{0tag}{jets} and \Reg{SR2}{0tag}{jets}) and without (\Reg{SR1}{0tag}{0jet} and \Reg{SR2}{0tag}{0jet}) jets, as discussed above. The two SRs with $\ptmiss<280\GeV$ and \PQb-tagged jets (\Reg{SR1}{tags}{} and \Reg{SR2}{tags}{}) are the same as the CRs used to constrain the \ttbar and \tW backgrounds at low \ptmiss in the chargino analysis (\Reg{CR1}{tags}{} and \Reg{CR2}{tags}{}, respectively).

Another useful means to discriminate top squark production from SM processes in the compressed scenario is given by the presence of high-\pt jets from ISR in the events. 
In the absence of hard ISR, the top squark pair has relatively little transverse recoil, and the small mass splitting causes the neutralinos and neutrinos from the decay chain to have relatively low transverse momenta.
However, events with harder neutralinos can occur when the top squark pair system recoils against a high-\pt ISR jet. In this hard-ISR regime, background is still constrained by the kinematic \invM{\PW} endpoint in \mtll, and can be effectively separated from the signal. Hard ISR events are selected by requiring that the leading jet satisfies $\pt>150\GeV$ and fails to pass the \PQb tagging algorithm requirement. In order to favor the topology in which the jet recoils against the rest of the system, the $\Dphi$ between the jet and the \ptvecmiss is required to be larger than 2.5. This requirement is found to be effective in discriminating top squark production from background events at high \ptmiss, and is therefore applied only for events with $\ptmiss>280\GeV$. In this way, the \Reg{SR3}{0tag}{ISR} used in the top squark analysis is defined by the same selection requirements as the \Reg{SR3}{0tag}{} of the chargino analysis with the addition of the ISR jet requirement, and analogously for \Reg{SR3}{tags}{ISR}, \Reg{SR4}{0tag}{ISR}, and \Reg{SR4}{tags}{ISR}.

A summary of the SRs in the searches for charginos and sleptons and for top squarks is given in Table~\ref{Tab:AllSR}, indicating the \ptmiss range, the selection on the multiplicity of jets (\N{jets}) and \PQb-tagged jets (\N{\PQb\ jets}) in the event, and the ISR jet requirement in the case of the top squark search. 

\begin{table}[htb]
      \centering
      \topcaption{Definition of the SRs as a function of the \ptmiss value, the \PQb-tagged jet multiplicity and jet multiplicity, and the lepton pair channels, as well as the ISR requirement used for the top squark search. Also shown are the CRs with \PQb-tagged jets used for the normalization of the \ttbar and \tW backgrounds for the chargino and slepton searches. The last row for each search type shows the \mtll binning used in the ML fit performed to extract the signal.}\label{Tab:AllSR}
      \cmsTable{
      \begin{tabular}{lcccccccccc}
      \hline
      \multicolumn{1}{c}{ }&\multicolumn{3}{c}{SR1} & \multicolumn{3}{c}{SR2} &\multicolumn{2}{c}{SR3} &\multicolumn{2}{c}{SR4} \\ 
      \ptmiss [{\GeVns}]&\multicolumn{3}{c}{160--220} & \multicolumn{3}{c}{220--280} &\multicolumn{2}{c}{280--380} &\multicolumn{2}{c}{${>}380$} \\
      Channels&\multicolumn{3}{c}{SF, DF} & \multicolumn{3}{c}{SF, DF}  &\multicolumn{2}{c}{SF, DF}  &\multicolumn{2}{c}{SF, DF} \\
      \hline
      \multicolumn{1}{c}{ }&\multicolumn{10}{c}{Chargino and slepton searches} \\
      
      \multicolumn{1}{c}{ }& \Reg{SR1}{0tag}{0jet} & \Reg{SR1}{0tag}{jets} & \Reg{CR1}{tags}{}  & \Reg{SR2}{0tag}{0jet} & \Reg{SR2}{0tag}{jets} & \Reg{CR2}{tags}{}  & \Reg{SR3}{0tag}{} & \Reg{CR3}{tags}{} & \Reg{SR4}{0tag}{} & \Reg{CR4}{tags}{} \\
      \N{\PQb\ jets}  &  0  &  0  & ${\geq}1$ & 0 & 0 & ${\geq}1$  &  0 & ${\geq}1$ &  0 & ${\geq}1$\\
      \N{jets}    &  0  & ${\geq}1$  & ${\geq}1$ & 0 & ${\geq}1$ & ${\geq}1$ & ${\geq}0$ & ${\geq}1$ & ${\geq}0$ & ${\geq}1$ \\
      \mtll [{\GeVns}] & \multicolumn{10}{c}{0--20, 20--40, 40--60, 60--80, 80--100, 100--160, 160--240, 240--370, ${>}370$} \\[\cmsTabSkip]
      \multicolumn{1}{c}{ }&\multicolumn{10}{c}{Top squark search}\\
      \multicolumn{1}{c}{ }& \multicolumn{2}{c}{\Reg{SR1}{0tag}{}} & \Reg{SR1}{tags}{} & \multicolumn{2}{c}{\Reg{SR2}{0tag}{}} & \Reg{SR2}{tags}{} & \Reg{SR3}{0tag}{ISR} & \Reg{SR3}{tags}{ISR} & \Reg{SR4}{0tag}{ISR} & \Reg{SR4}{tags}{ISR}\\
      \N{\PQb\ jets}  & \multicolumn{2}{c}{0} & ${\geq}1$ &\multicolumn{2}{c}{0} & ${\geq}1$ & 0 & ${\geq}1$ & 0 & ${\geq}1$ \\
      \N{jets}    &  \multicolumn{2}{c}{${\geq}0$}  & ${\geq}1$ & \multicolumn{2}{c}{${\geq}0$} & ${\geq}1$ & ${\geq}1$ & ${\geq}2$ & ${\geq}1$ & ${\geq}2$ \\
      ISR jets  & \multicolumn{2}{c}{${\geq}0$} & ${\geq}0$ & \multicolumn{2}{c}{${\geq}0$} & ${\geq}0$ & ${\geq}1$ & ${\geq}1$ & ${\geq}1$ & ${\geq}1$ \\
      \mtll [{\GeVns}] & \multicolumn{10}{c}{0--20, 20--40, 40--60, 60--80, 80--100, 100--160, ${>}160$} \\
      \hline
      \end{tabular}
      }
\end{table}
      
A simultaneous ML fit to the \mtll distribution in all the SRs and CRs for the corresponding signal process is performed to extract the signal (as described in Section~\ref{sec:results}). 
The \mtll binning used in the fit is optimized independently in the searches for charginos and sleptons and for top squarks and is also shown in Table~\ref{Tab:AllSR}. Since the first \mtll bins have a low signal contribution, we exploit them to constrain the contributions of the dominant backgrounds in the SRs with one \PQb-tagged jet (dominated by \ttbar and \tW production) and without \PQb-tagged jets (where \WW production becomes relevant) through the fit.

\section{Background estimation}\label{sec:bkgestimation}

The main contributions from SM processes to the SRs come from \ttbar, \tW, and \WW production. The normalization of these backgrounds is determined by the ML fit, as mentioned in Section~\ref{sec:searchstrategy}. Their \mtll shape has a natural endpoint at the \PW mass, and events enter into the relevant region for signal extraction, $\mtll>80\GeV$, mainly because of detector resolution effects, whose contributions may be not accurately modeled by the simulation.
For this reason, we study the modeling of the \mtll distribution for these processes in dedicated CRs in data described in Section~\ref{sec:mt2llShape}. 
The contributions of subleading backgrounds from $\PZ+\text{X}$ production (such as \ttZ, \WZ, \ZZ, and \DY events) are also tested in CRs, which are included in the ML fit to constrain their normalizations, as discussed in Section~\ref{Sec:bckNorma}.
Remaining minor backgrounds from \ttW, $\PH\to\WW$, $\PH\to\tautau$, and triboson production give small contributions in the SRs, and the estimates for these processes are taken directly from simulation.
Background contributions from other SM processes are found to be negligible.
The contribution of signal to any of the CRs used is below the percent level.

\subsection{Modeling of \texorpdfstring{\mtll}{mT2(ll)} in \texorpdfstring{\ttbar}{ttbar}, \texorpdfstring{\tW}{tW}, and \texorpdfstring{\WW}{WW} events}\label{sec:mt2llShape}

The reconstruction of the \mtll observable for \ttbar, \tW, and \WW events in the SRs is sensitive to various detector resolution effects that can drive its value beyond the \invM{\PW} kinematic endpoint: \ptmiss mismeasurements and nonprompt leptons. 

\begin{figure}
\centering
  \includegraphics[width=0.49\textwidth]{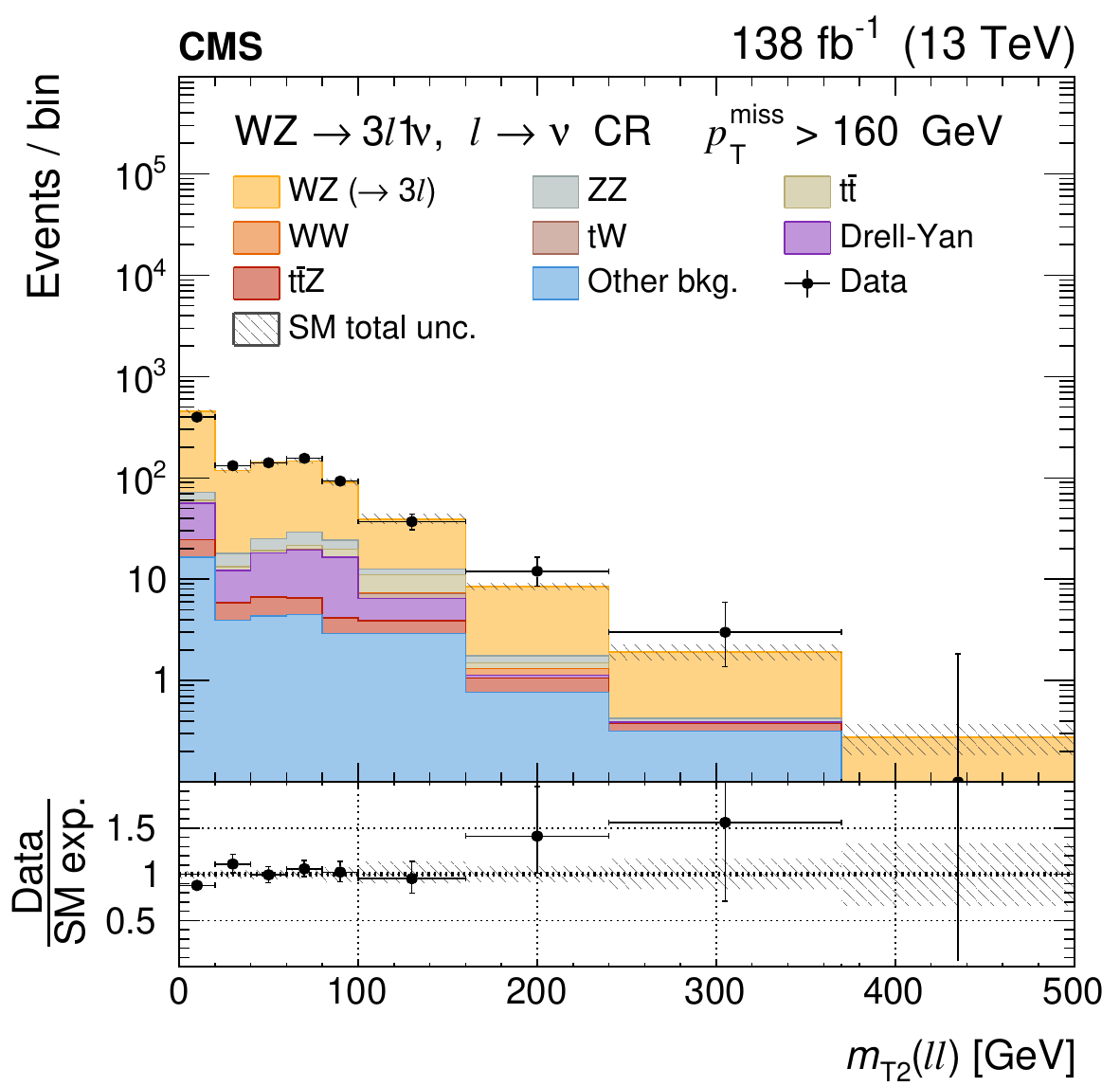}
  \includegraphics[width=0.49\textwidth]{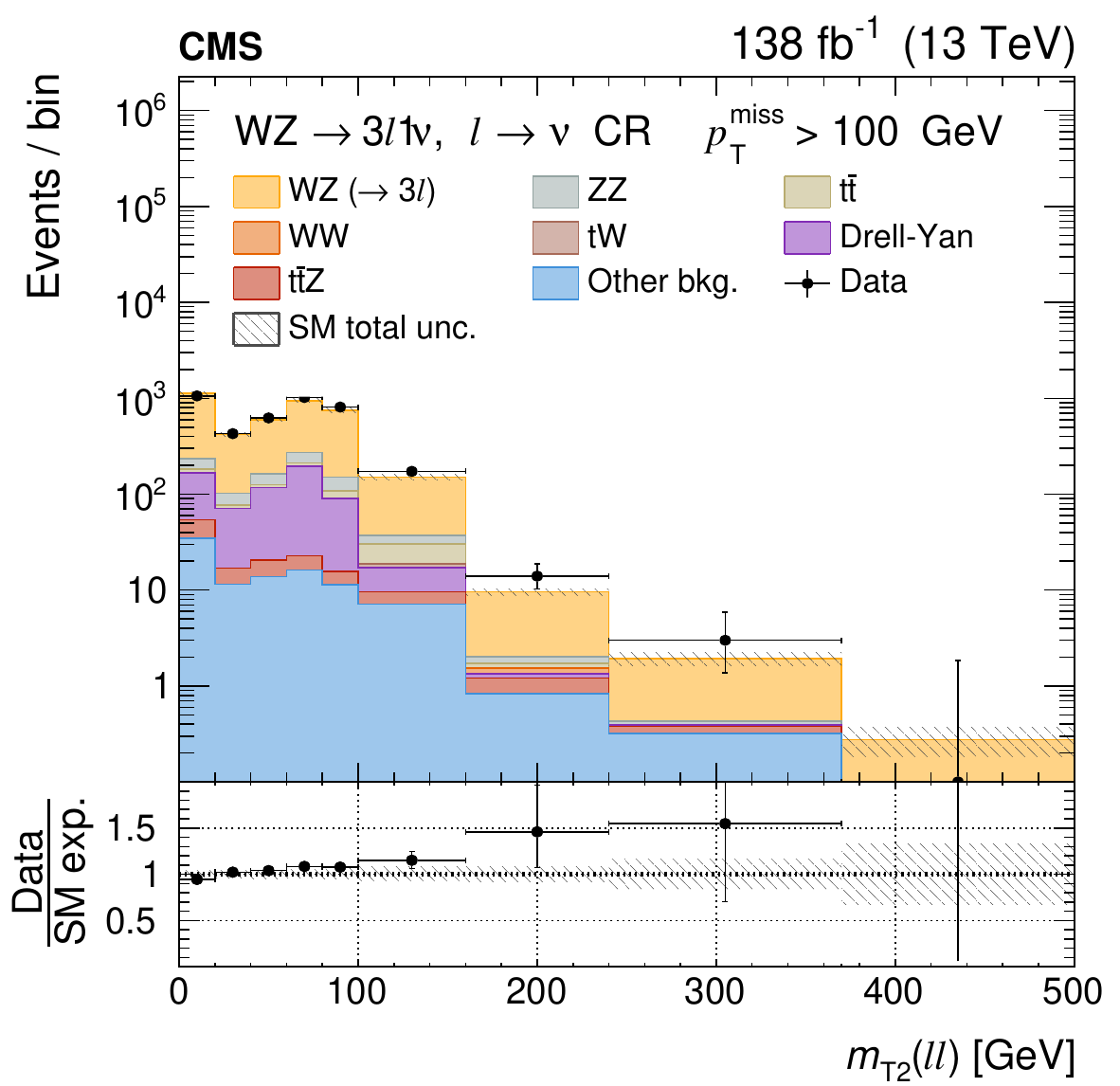}
	\caption{distributions of \mtll in $\WZ\to 3\Pell 1\PGn$ events with $\ptmiss>160\GeV$ (left) and $\ptmiss>100\GeV$ (right) after the \WW emulation. The last bin includes the overflow entries. The ratio of the data to the SM expectation is shown in the lower panel. The hatched band represents the uncertainty in the expected SM yields, including the contributions described in Section~\ref{sec:systematics} except for the ones in the background normalization described in Section~\ref{sec:mt2llShape}.}
\label{Fig:WZtoWW}
\end{figure}

At large \ptmiss, the contribution from mismeasurements to the momentum imbalance can become significant. To build a CR where we can test the modeling of the \mtll distributions in this regime, we select events from $\WZ\to 3\Pell 1\PGn$ production and emulate the \mtll shape of \WW and top quark events by taking the lepton from the \PZ boson with the same charge as the lepton from the {\PW} boson, and adding its \ptvec vectorially to \ptvecmiss, effectively treating it like a neutrino. 
These events are selected by requiring exactly three leptons, with no further leptons passing the looser veto requirements.
A veto is applied to events with \PQb-tagged jets to remove residual \ttbar events. Among the three leptons, a pair of OS SF leptons with an invariant mass within 15\GeV of the \PZ boson mass is required to identify the \PZ boson.
The simulation is found to slightly underestimate the data in the tail of the \mtll distribution in events with $\ptmiss>160\GeV$, as shown in Fig.~\ref{Fig:WZtoWW} (left). 
We lower the threshold on the \ptmiss to 100\GeV to enlarge the statistical power of the test while retaining a high purity of the sample in \WZ events (Fig.~\ref{Fig:WZtoWW}, right), and derive a correction by a linear fit to the ratio of the (background-subtracted) observed events to the expected \WZ events.
The correction is only applied for an \mtll value of 50\GeV and above, and is defined to be 1.0 at 50\GeV. The slope of the correction is $0.0019 \pm 0.0015\GeV^{-1}$. An imperfect agreement is also observed in the bulk of the \mtll distribution for events with $\ptmiss>380\GeV$ after the aforementioned correction is applied, which was found to be related to some mismodeling in the minimum azimuthal distance between the \ptmiss and any lepton in the event, \dphimin. Because of the limited number of events in this region, an uncertainty is set to cover the observed mismodeling, ranging up to 40\% across the \mtll bins.

The second potential source of mismodeling in the tails of the \mtll distributions arises from nonprompt leptons originating, for instance, from semileptonic decays of \PQb hadrons in \PQb jets or from hadronic jets accidentally passing the lepton selection. 
In \ttbar, \tW, and \WW events where one prompt lepton from the decay of a \PW boson does not satisfy the required selection, and a nonprompt lepton is instead selected, the value of \mtll will not be bound by the \invM{\PW} endpoint.
The fraction of simulated events with a nonprompt lepton is observed to be less than 1\% of the expected background across the different SRs. It becomes more relevant only at large values of \mtll and \ptmiss, where it amounts up to 20\% of the \ttbar background. We study the modeling of the rate of nonprompt leptons in simulation by selecting events with two leptons with same charge but otherwise passing all SR selections.
The largest contribution to this sample comes from \ttbar events with a nonprompt lepton.
Based on the observed agreement with data, correction factors are derived for the nonprompt-lepton rate in simulation, whose values range 1.1--1.4 across the data-taking periods.

\subsection{Normalization and modeling of backgrounds from \texorpdfstring{$\PZ+\text{X}$}{Z+X} production}\label{Sec:bckNorma}

The contributions of the subleading \WZ, \ZZ, \ttZ, and \DY backgrounds are also tested in CRs, which are included in the ML fit to constrain their normalizations.
Events from \WZ production enter the signal event selection when both bosons decay leptonically and one of the three charged leptons fails the veto lepton requirements. We test the modeling of this source of background in a CR with exactly three leptons, $\ptmiss>160\GeV$, and no \PQb-tagged jets.
Good agreement between data and SM expectations is observed for both the \ptmiss and \mtll distributions. 
In order to constrain the normalization of the \WZ background in the SRs, CRs are introduced in the ML fit used to extract the signal, based on the event selection presented in this section and divided into the same \ptmiss and jet multiplicity bins used in the SR definition.

The \ZZ background is dominated by events with one boson decaying into charged leptons and the other one decaying into neutrinos. This contribution is studied by mimicking the $\ZZ\to 2\Pell2\PGn$ production via $\ZZ\to 4\Pell$ events, where the \ptvec of one of the reconstructed \PZ bosons (randomly chosen between the ones satisfying the $\DZmass<15$\GeV condition) is added vectorially to the \ptvecmiss. The resulting observable is referred to as \ptvecmissz in the following. Events are selected by requiring four leptons, with one lepton allowed to pass the looser veto lepton requirement in order to increase the acceptance for \ZZ production. The events are retained if the four leptons can be arranged into two pairs of OS SF leptons, both with an invariant mass greater than 30\GeV, and at least one within 15\GeV of the \PZ boson mass. 
Just as we do for the \WZ production, CRs are introduced in the ML fit to constrain the normalization of the \ZZ background, based on the $\ZZ\to 2\Pell2\PGn$ emulation presented in this section and divided in the same \ptmiss and jet multiplicity bins used in the SR definition.

The production of \ttZ events where the two \PW bosons decay leptonically and the \PZ boson decays into neutrinos leads to final states with the same experimental signature as the signal events. No natural endpoint for the reconstructed \mtll distribution is present for this background, due to the additional contribution of the neutrinos from the \PZ boson decay to the \ptvecmiss. 
We study this background in events with three or four leptons and at least two jets with $\pt>20\GeV$, including at least one \PQb-tagged jet. At least one pair of OS SF leptons with an invariant mass not further than 15\GeV from the \PZ boson mass is also required, and the \ptvecmissz is computed. 
The CRs for constraining the normalization of the \ttZ production in the ML fit are obtained by further requiring two \PQb-tagged jets if exactly three leptons are present in the event.
The modeling of the \mtll distribution for the \ttZ background is validated in events with a fourth lepton. 

\DY events can pass the baseline selection when large \ptmiss is produced in $\PZ/\gamma^{*}\to\tautau$ events or because of mismeasurements in the recoil system.
The value of the \mtll observable in the former events, however, is constrained by a kinematic endpoint at the mass of the \PGt lepton.
We study the modeling of the \mtll tails for the latter events in a CR with two OS SF leptons with $\DZmass<15$\GeV, no additional leptons, and no \PQb-tagged jets.
The \mtll distribution observed in data is well modeled by the simulations. Since the \DY contribution in the \mtll tails is found to be at most 25\% compared to the diboson contribution, no dedicated CR is added in the ML fit to constrain the normalization of this background, to which we just assign a 50\% uncertainty. A summary of the CRs used to constrain the normalization of the $\PZ+\text{X}$ production processes in the ML fit is given in Table~\ref{Tab:AllCR}.
To preserve statistical power, these CRs are not subdivided into \mtll bins.

\begin{table}[htb]
  \centering
	\topcaption{Definition of the CRs for the normalization of the \WZ, \ZZ, and \ttZ backgrounds in the ML fit, as a function of the \ptmiss value and of the lepton, the \PQb-tagged jet, and jet multiplicities. For the \ZZ and \ttZ processes, the \ptvecmissz is used to define the CRs.}\label{Tab:AllCR}
      \begin{tabular}{llcccccc}    
      \hline
      \multicolumn{2}{c}{ }&\multicolumn{2}{c}{CR1} & \multicolumn{2}{c}{CR2} &\multicolumn{1}{c}{CR3} &\multicolumn{1}{c}{CR4} \\
      \multicolumn{2}{l}{\ptmiss [{\GeVns}]}&\multicolumn{2}{c}{160--220} & \multicolumn{2}{c}{220--280} &\multicolumn{1}{c}{280--380} &\multicolumn{1}{c}{${>}380$} \\
      \hline
      \multicolumn{2}{c}{ }&\multicolumn{6}{c}{\WZ CRs}\\    
	      \multicolumn{2}{c}{ }& \Reg{CR1}{\WZ}{0jet} & \Reg{CR1}{\WZ}{jets} & \Reg{CR2}{\WZ}{0jet} & \Reg{CR2}{\WZ}{jets} & \Reg{CR3}{\WZ}{} & \Reg{CR4}{\WZ}{} \\
	      \multicolumn{2}{l}{\N{lep.}} & 3 & 3 & 3 & 3 & 3 & 3 \\ 
	      \multicolumn{2}{l}{\N{\PQb\ jets}}  &  0  &  0  & 0 & 0 & 0 & 0 \\
	      \multicolumn{2}{l}{\N{jets}}  &  0  & ${\geq}1$ & 0 & ${\geq}1$ & ${\geq}0$ & ${\geq}0$ \\[\cmsTabSkip]
      \multicolumn{2}{c}{ }&\multicolumn{6}{c}{\ZZ CRs}\\
	      \multicolumn{2}{c}{ }& \Reg{CR1}{\ZZ}{0jet} & \Reg{CR1}{\ZZ}{jets} & \Reg{CR2}{\ZZ}{0jet} & \Reg{CR2}{\ZZ}{jets} & \Reg{CR3}{\ZZ}{} & \Reg{CR4}{\ZZ}{} \\
	      \multicolumn{2}{l}{\N{lep.}} & 4 & 4 & 4 & 4 & 4 & 4 \\
	      \multicolumn{2}{l}{\N{\PQb\ jets}}  &  0  &  0  & 0 & 0 & 0 & 0 \\
	      \multicolumn{2}{l}{\N{jets}}  &  0  & ${\geq}1$ & 0 & ${\geq}1$ & ${\geq}0$ & ${\geq}0$ \\[\cmsTabSkip]
      \multicolumn{2}{c}{ }&\multicolumn{6}{c}{\ttZ CRs}\\
      \multicolumn{2}{c}{ }& \multicolumn{2}{c}{\Reg{CR1}{\ttZ}{}} & \multicolumn{2}{c}{\Reg{CR2}{\ttZ}{}} & \Reg{CR3}{\ttZ}{} & \Reg{CR4}{\ttZ}{} \\
	      \multicolumn{2}{l}{\N{lep.}} & \multicolumn{2}{c}{${\geq}3$} & \multicolumn{2}{c}{${\geq}3$} & ${\geq}3$ & ${\geq}3$ \\
	      \multirow{2}{*}{\N{\PQb\ jets}}  & ($\N{lep.}\ge4$) & \multicolumn{2}{c}{${\geq}1$} & \multicolumn{2}{c}{${\geq}1$} & ${\geq}1$ & ${\geq}1$ \\
	       & ($\N{lep.}=3$) & \multicolumn{2}{c}{${\geq}2$} & \multicolumn{2}{c}{${\geq}2$} & ${\geq}2$ & ${\geq}2$ \\
	      \multicolumn{2}{l}{\N{jets}}  &  \multicolumn{2}{c}{${\geq}2$} & \multicolumn{2}{c}{${\geq}2$} & ${\geq}2$ & ${\geq}2$ \\
      \hline
      \end{tabular}
\end{table}

\section{Systematic uncertainties}\label{sec:systematics}

Several sources of systematic uncertainty that affect both the normalizations and the \mtll shapes of the background and signal events are evaluated in the analysis.
\begin{itemize}
\item The integrated luminosities for the 2016, 2017, and 2018 data-taking years have 0.82--1.2\% individual uncertainties~\cite{CMS-LUM-17-003,CMS-LUM-20-001}, while the overall uncertainty for the 2016--2018 period is 0.73\%.
\item The uncertainty in the measured trigger efficiency is 2\%.
\item The recommended \pp inelastic cross section is varied by ${\pm} 4.6\%$~\cite{inppxsec13TeV}, and new pileup weights are computed and propagated through the analysis.
\item The jet energy scale and jet energy resolution are varied within their uncertainties~\cite{CMS:2016lmd}, and the effect propagated to all the observables in the event.
\item The energy scale of the low-\pt particles that are not clustered in jets is varied within its uncertainty~\cite{CMS:2019ctu}, and the effect propagated to \ptvecmiss.
\item The uncertainties in the L1 prefiring weights are propagated in the 2016 and 2017 samples.
\item Lepton identification and isolation efficiencies are corrected by data-to-simulation scale factors measured in $\PZ\to\Pell\Pell$ events. The corresponding uncertainties are typically smaller than 3\% per lepton~\cite{CMS:2020uim,CMS-DP-2020-021,CMS:2018rym}.
\item The efficiencies and misidentification rates of the \PQb jet identification algorithms are also corrected by data-to-simulation scale factors measured in inclusive jet and \ttbar events~\cite{BTV-16-002}. Their uncertainties are in the 1--6\% range, depending on the \pt and $\eta$ of the jets.
\item The effect of the simulated data sample sizes on the modeling of the \mtll distributions is taken into account by treating the statistical uncertainty in each bin for each process as an additional uncorrelated uncertainty.
\item Uncertainties in the renormalization and factorization scales, as well as in PDFs, are propagated by taking the largest changes in the acceptance when individually or simultaneously doubling and halving the renormalization and factorization scales~\cite{RFScale1,RFScale2}, and by using the NNPDF3.1 replicas of the PDFs according to PDF4LHC recommendations~\cite{PDF4LHC}. The PDF uncertainties are not applied for signal models, as they are found to be redundant, once the uncertainty in the ISR modeling is included.
\end{itemize}
{\tolerance=10 The estimates of the SM backgrounds are also affected by specific uncertainties in the modeling of the different processes\par}.
\begin{itemize}
\item The normalizations of the \ttbar, \tW, and \WW production processes are determined by the ML fit in the low \mtll bins of the SRs, as described in Section~\ref{sec:results}. Common normalization parameters are used for \ttbar and \tW events, as their contributions are not easily separable. An additional 10\% uncertainty is therefore set for the \tW process to take into account possible mismodeling on its relative normalization with respect to the \ttbar production as well as any interference effect between them. This value corresponds to the maximum possible variation in the ratio of the theoretical cross sections for the two processes under their current uncertainties.
\item In the SR1 and SR2 of the chargino analysis, additional free parameters are introduced to control top quark and \WW event migration between \Reg{SR}{0tag}{0jet} and \Reg{SR}{0tag}{jets}, as no suitable CR is available to study the modeling of the jet multiplicity at high \ptmiss for these backgrounds. These parameters allow the ML fit to vary the relative contributions of top quark and \WW events with and without jets while keeping the total number of events in \Reg{SR}{0tag}{} constant.
\item The normalizations of the \WZ, \ZZ, and \ttZ background processes are described by free parameters constrained in the dedicated CRs discussed in Section~\ref{Sec:bckNorma}.
\item Backgrounds from \DY, \ttW, $\PH\to\WW$, $\PH\to\tautau$, and triboson production for which no CR is built are assigned an uncertainty of 50\%. This choice has been found to have negligible effect on the results.
\item An uncertainty in the \mtll tail correction for the backgrounds with a kinematic endpoint at the \PW boson mass is derived by the uncertainty in the fitted value of the slope, $0.0019\pm0.0015\GeV^{-1}$, described in Section~\ref{sec:mt2llShape}.
\item An uncertainty to account for the observed mismodeling in \dphimin for events with $\ptmiss>380\GeV$ is also set, ranging up to 40\% across the \mtll bins, as discussed in Section~\ref{sec:mt2llShape}.
\item The weight of events with nonprompt leptons in simulated samples is varied within the 30\% uncertainty in the correction factors derived in events with two same-charge leptons, as described in Section~\ref{sec:mt2llShape}.
\item The spectrum of top quark \pt in \ttbar events has been observed to be softer in data than in simulated events~\cite{bib:toppt1,bib:toppt3,bib:toppt4}. An uncertainty is derived from the observed variations when reweighting the \ttbar events to the \pt distribution observed in data.
\end{itemize}

\begin{table}[ht]
  \centering                                                                                                              
  \topcaption{Sizes of systematic uncertainties in the predicted yields and \mtll shapes for SM processes. The column labeled ``Change in yields'' shows the range of the uncertainties in the global background normalization across the different SRs. The column labeled ``Change in \mtll shape'' quantifies the effect on the \mtll shape. This is computed by taking the maximum variation across the \mtll bins (after adjusting for the global normalization change) in each SR. The range of this variation across the SRs is given.}
\cmsTable{
  \begin{tabular}{lcc}
  \hline
   \multirow{2}{*}{Source of uncertainty}   & \multicolumn{2}{c}{SM processes} \\
    & Change in yields & Change in \mtll shape \\
  \hline
      Integrated luminosity & 1\% & \NA \\
      Trigger efficiency & 2\% & ${<}1\%$\\
      Pileup & ${\le}2\%$ & 2--20\%\\
      Jet energy scale & 3--8\% & 2--10\%\\
      Jet energy resolution & 1--2\% & 2--8\%\\
      Unclustered energy & 1--2\% & 2--13\%\\
      L1 prefiring & 1--2\% & ${\le}3\%$\\
      Lepton ident./isolation & 2--4\% & ${\le}15\%$\\
      b tagging uncertainties & ${\le}5\%$ & ${\le}6\%$\\
      b tagging uncertainties (light-quark/gluon jets) & ${<}1\%$ & ${\le}3\%$\\
      Simulated samples' size & ${\le}3\%$ & 4--37\%\\
      Renorm./fact. scales & 2--23\% & 1--15\%\\
      PDFs & ${\le}2\%$ & ${\le}9\%$\\
      \DY normalization & ${\le}7\%$ & ${\le}26\%$\\
      \tW normalization & 1--2\% & ${\le}3\%$\\
      Minor bkg. normalization & ${\le}3\%$ & 1--8\%\\
      \mtll tails & 1--2\% & 5--20\%\\
      \dphimin mismodeling & ${\le}2\%$ & 33--41\%\\
      Nonprompt leptons & ${<}1\%$ & ${\le}8\%$\\
      \ttbar \pt reweighting & 1--6\% & 1--6\%\\
   \hline
  \end{tabular}
  }
  \label{tab:BackSyst}
\end{table}

{Finally, additional uncertainties in the modeling of signal events are taken into account, mostly related to the performance of the event reconstruction in \FastSim.\par}
\begin{itemize}
\item The uncertainty in the lepton identification efficiency in events simulated with \FastSim, relative to the full detector simulation, is estimated to be 2\% per lepton.
\item The analogous uncertainty in the \PQb tagging efficiency in \FastSim samples ranges between 0.2--0.5\%.
\item Uncertainties in the ISR reweighting procedure described in Section~\ref{sec:datasets} are derived to be 50\% of the deviation in the reweighting factors from unity, based on closure tests of the agreement between the observed and expected distributions of other ISR related observables such as the scalar sum of the ISR jet \pt's.
\end{itemize}
In the ML fit, the correlations of each systematic uncertainty source across SRs and data-taking periods are taken into account as follows: uncertainties of a theoretical nature (\pp inelastic cross section, renormalization and factorization scales, PDFs) are treated as correlated across all data-taking periods and SRs; uncertainties related to the detector performance (trigger efficiencies, object reconstruction efficiencies, jet energy scale and resolutions, etc.) are treated as uncorrelated across data-taking periods (but for specific cases such as luminosity and \PQb tagging efficiencies where the uncertainties are broken down into correlated and uncorrelated contributions); uncertainties in the modeling of the \mtll observable at high \ptmiss are treated as uncorrelated across data-taking periods and \ptmiss bins.

Tables~\ref{tab:BackSyst} and~\ref{tab:SignSyst} summarize the systematic uncertainties in the predicted yields and \mtll shapes for SM processes and for two reference signals, respectively.
The sources of systematic uncertainty with the largest impact in the ML fit are found to be the ones associated with background normalization and \mtll tail modeling.

\begin{table}[htb]
\centering                                                                                                             
\topcaption{Sizes of systematic uncertainties in the predicted yields and \mtll shapes for two representative signal points, one for chargino pair production and one for top squark pair production. For each signal point, the column labeled ``Yields'' shows the range of the uncertainties in the global background normalization across the different SRs. The column labeled ``\mtll shape'' quantifies the effect on the \mtll shape. This is computed by taking the maximum variation across the \mtll bins (after adjusting for the global normalization change) in each SR. The range of this variation across the SRs is given.}
\cmsTable{
\begin{tabular}{lcccc}
\hline
 \multirow{3}{*}{Source of uncertainty}   & \multicolumn{2}{c}{\TChipmDecay} & \multicolumn{2}{c}{$\stone\to \cPqt\PSGczDo$} \\
    & \multicolumn{2}{c}{($\invM{\PSGcpmDo}=1050\GeV$, $\invM{\PSGczDo}=1\GeV$)} & \multicolumn{2}{c}{ ($\invM{\stone}=475\GeV$, $\invM{\PSGczDo}=388\GeV$)} \\
  & \,\,\,\,\,\,Yields\,\,\,\,\,\,\, & \mtll shape & \,\,\,\,\,\,Yields\,\,\,\,\,\,\, & \mtll shape \\
\hline
    Integrated luminosity & 1\% & \NA & 1\% & \NA \\
    Trigger efficiency & 2\% & ${<}1\%$ & $2\%$ & ${<}1\%$\\
    Pileup & ${\le}4\%$ & ${\le}1\%$ & ${\le}8\%$ & ${\le}5\%$\\
    Jet energy scale & ${\le}6\%$ & ${\le}2\%$ & ${\le}11\%$ & ${\le}6\%$\\
    Jet energy resolution & ${\le}4\%$ & ${\le}2\%$ & ${\le}6\%$ & ${\le}6\%$\\
    Unclustered energy & ${\le}4\%$ & ${\le}2\%$ & ${\le}4\%$ & ${\le}4\%$\\
    Lepton ident./isolation & 2--7\% & ${\le}2\%$ & 1--3\% & ${\le}3\%$\\
    b tagging uncertainties & ${<}1\%$ & ${\le}1\%$  & ${\le}4\%$ & ${\le}1\%$\\
    b tagging uncertainties (light-quark/gluon jets) & ${<}1\%$ & ${\le}1\%$ & ${\le}4\%$ & ${\le}2\%$\\
    Simulated samples' size & 2--12\% & 3--7\% & 2--15\% & 6--16\%\\
    Renorm./fact. scales & ${\le}2\%$ & ${<}1\%$ & ${\le}3\%$ & ${<}1\%$\\
    Nonprompt leptons & ${<}1\%$ & ${\le}1\%$  & ${<}1\%$ & ${\le}1\%$\\
    Lepton ident./isolation (\FastSim) & 4.0\% & ${<}1\%$ & 4.0\% & ${<}1\%$\\
    b tagging uncertainties (\FastSim) & ${<}1\%$ & ${\le}2\%$ & ${\le}5\%$ & ${\le}3\%$\\
    ISR reweighting & ${\le}2\%$ & ${<}1\%$ & 2--11\% & ${\le}3\%$\\
\hline
\end{tabular}
}
\label{tab:SignSyst}
\end{table}

\section{Results and interpretation}\label{sec:results}

A simultaneous binned ML fit to the \mtll distribution in all the SRs and CRs is performed for each signal model and mass hypothesis considered.
Uncertainties due to signal and background normalizations are included through nuisance parameters with log-normal prior distributions, while uncertainties in the shape of the \mtll distributions are included using a template morphing technique~\cite{ConwayNuis} with Gaussian constrained morphing parameters.
Uncertainties due to the limited size of the simulated samples are introduced into the ML fit following the Barlow--Beeston approach~\cite{BarlowBeeston,ConwayNuis}. 
As explained in Section~\ref{sec:bkgestimation}, the normalizations of the main backgrounds from top quark and \WW production are left to be determined in the fit via the constraint provided by the low-\mtll region with and without \PQb-tagged jets, while the normalizations of the subleading backgrounds from \WZ, \ZZ, and \ttZ events are constrained in dedicated CRs included in the fit.

\begin{figure}
\centering
\includegraphics[width=0.49\textwidth]{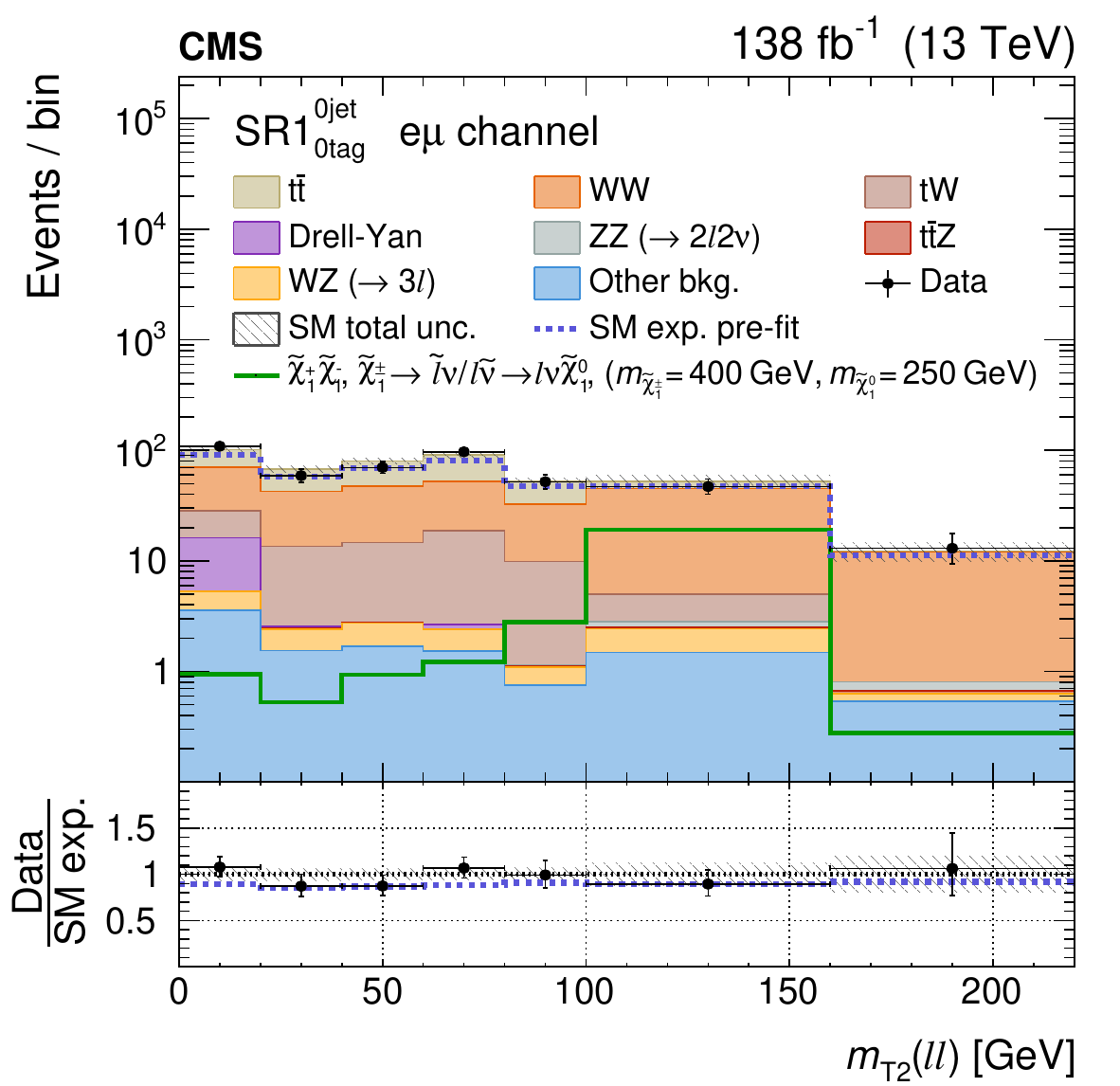}
\includegraphics[width=0.49\textwidth]{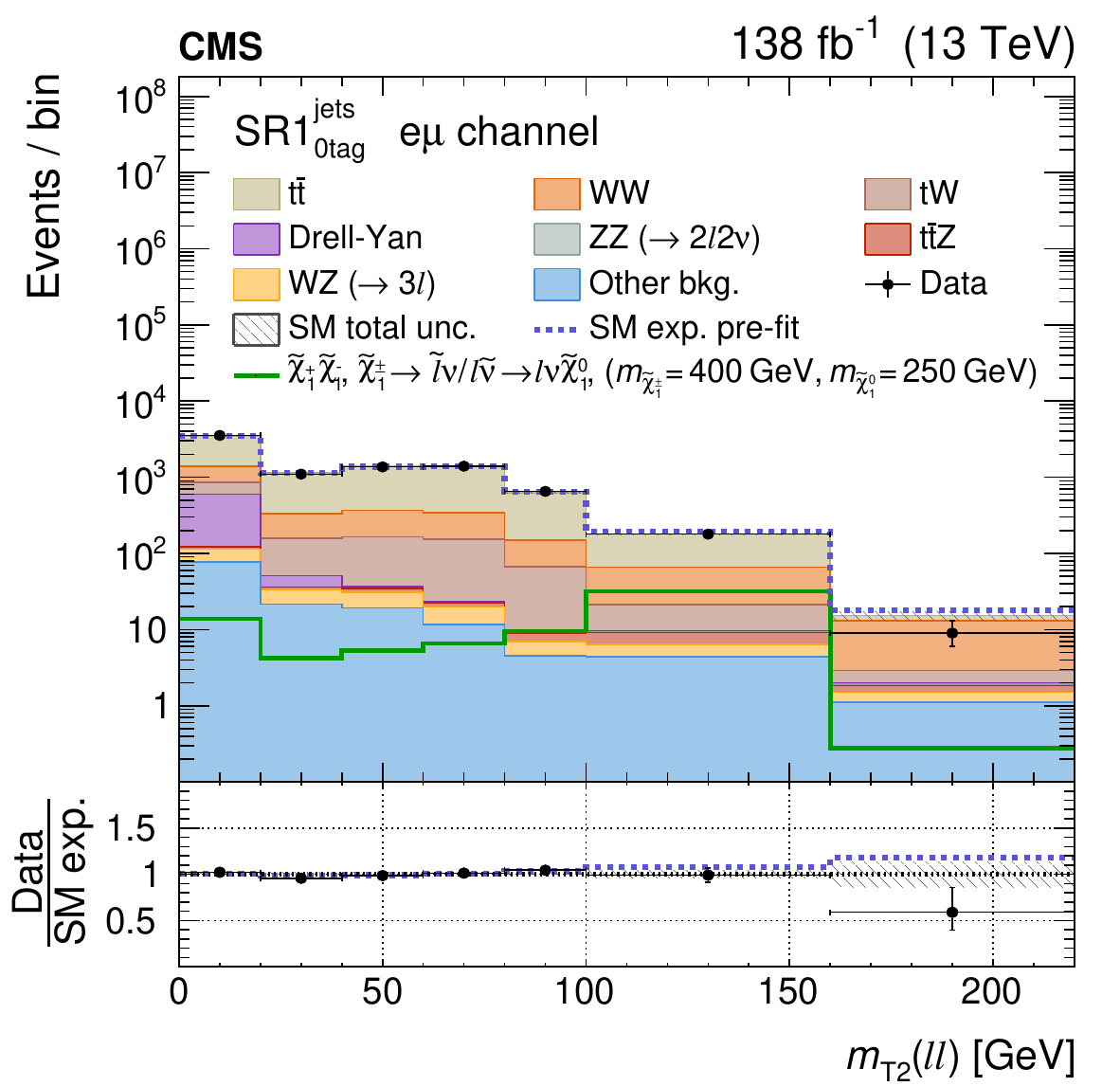}
\includegraphics[width=0.49\textwidth]{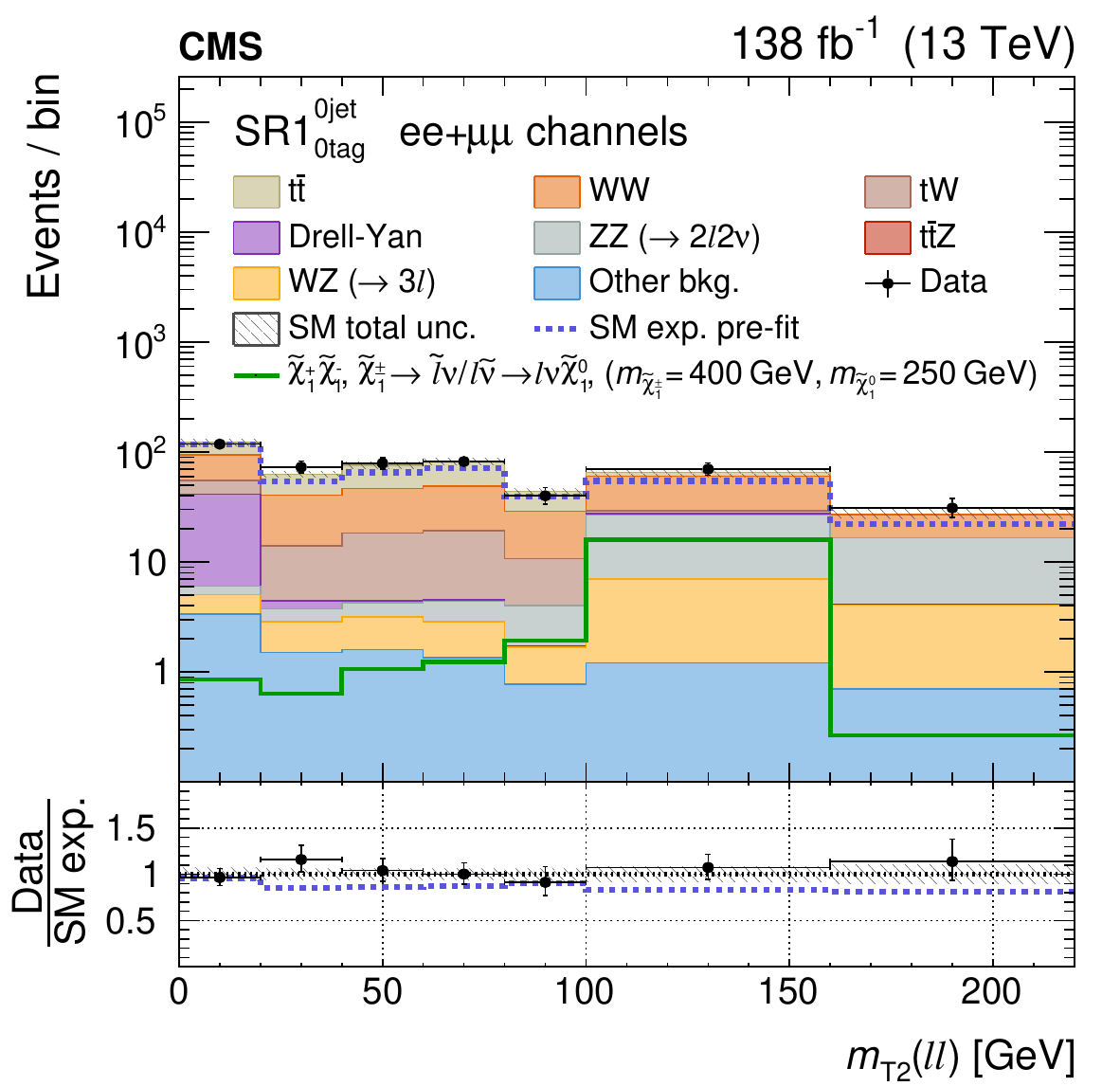}
\includegraphics[width=0.49\textwidth]{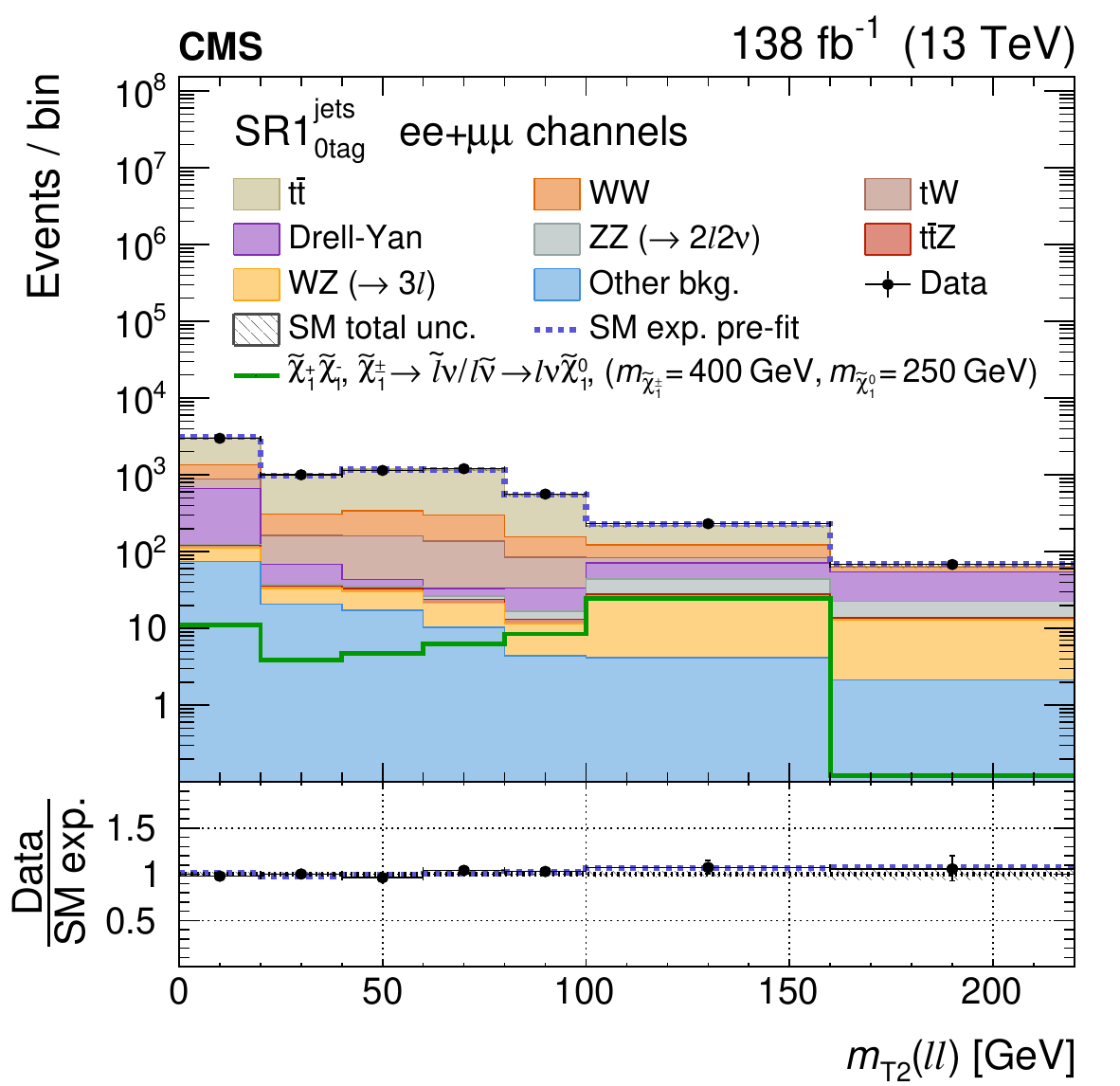}
\caption{Distributions of \mtll after the fit to data in the chargino and slepton SRs. The top row shows DF events in SR1, divided into \Reg{SR1}{0tag}{0jet} (left) and \Reg{SR1}{0tag}{jets} (right). The bottom row shows the same for SF events. The green histogram shows the expected \mtll distribution for chargino pair production with $\invM{\PSGcpmDo}=400\GeV$ and $\invM{\PSGczDo}=250\GeV$. The expected total SM contributions before the fit (blue dashed line) is also shown. The last bin includes the overflow entries. In the bottom panels, the ratio of data and SM expectations is shown for the expected total SM contribution after the fit using the background-only hypothesis (black markers with error bars). Also shown is the ratio of the SM expectations before and after the fit (blue dashed line). The hatched band represents the total uncertainty after the fit.} 
	\label{Fig:MT2ll_TChi_SR1}
\end{figure}

\begin{figure}
\centering
\includegraphics[width=0.49\textwidth]{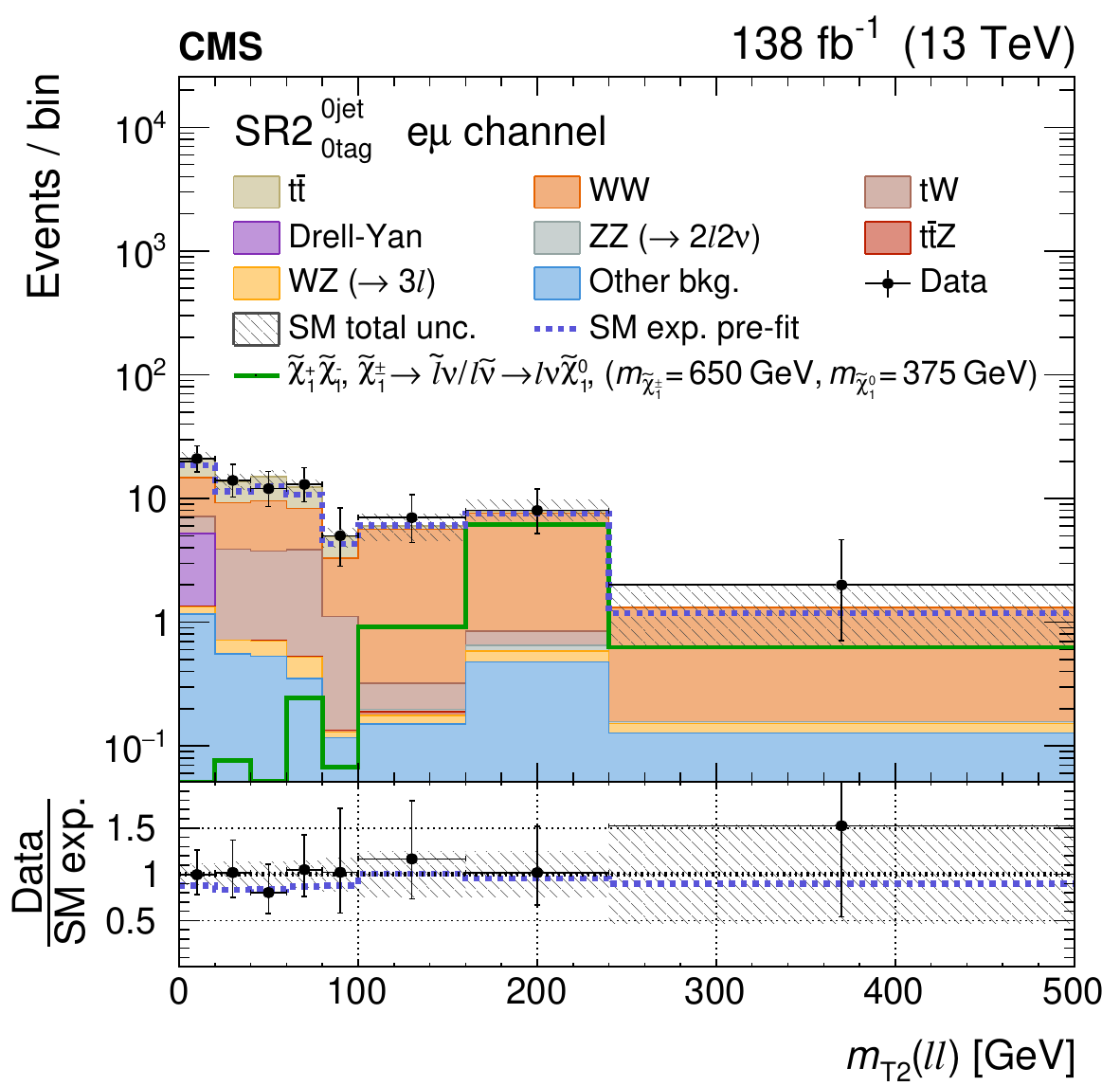}
\includegraphics[width=0.49\textwidth]{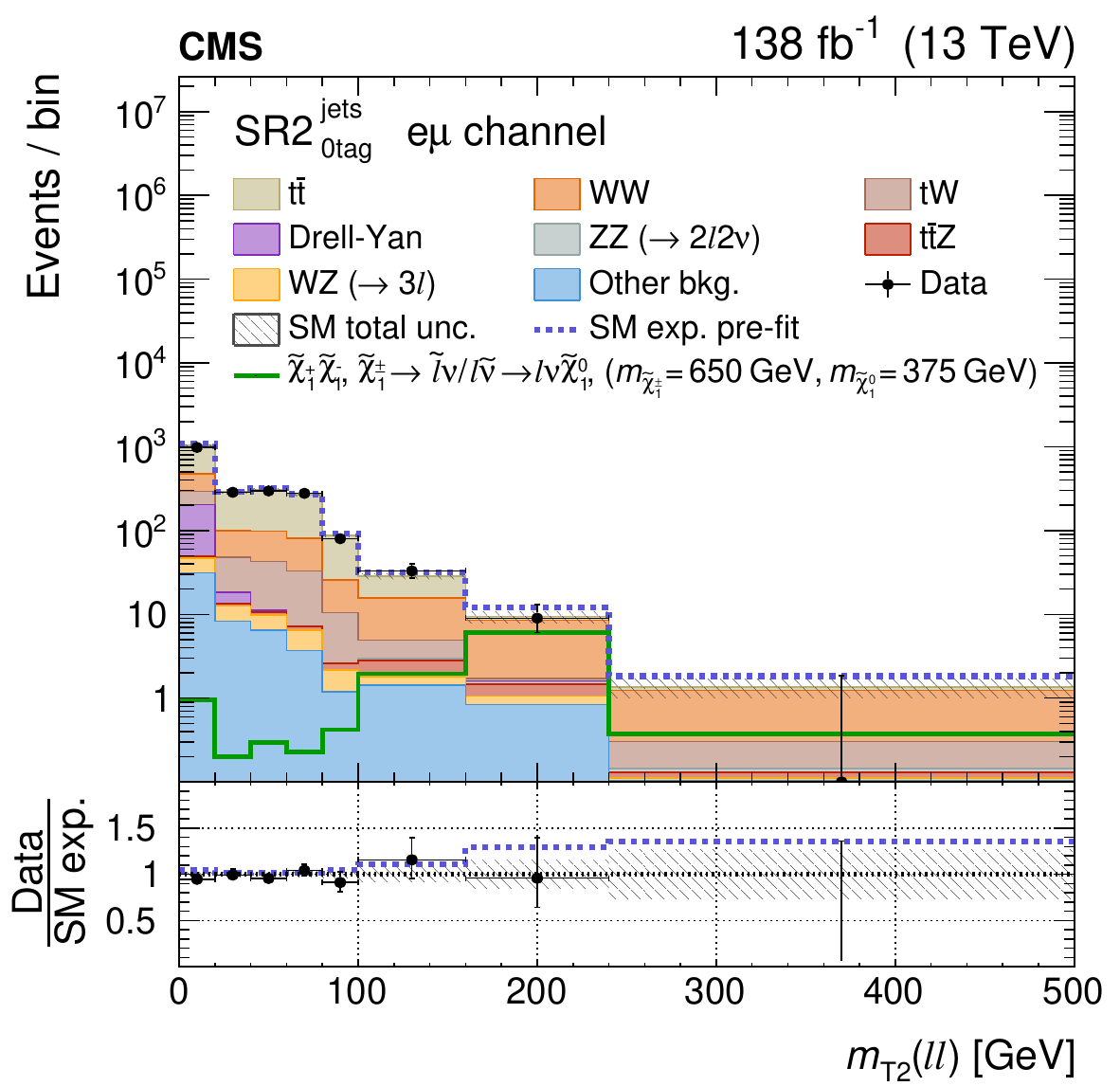}
\includegraphics[width=0.49\textwidth]{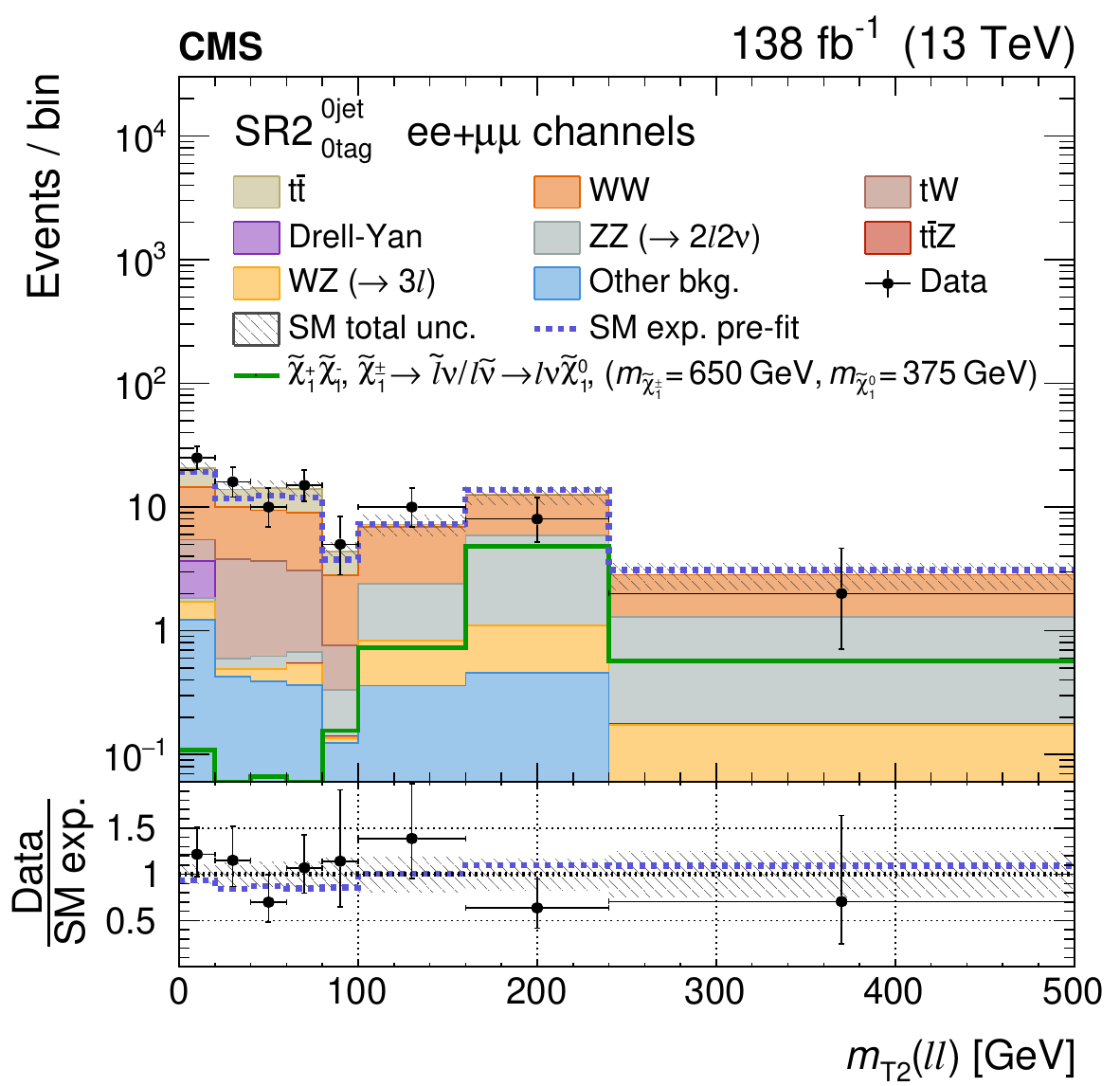}
\includegraphics[width=0.49\textwidth]{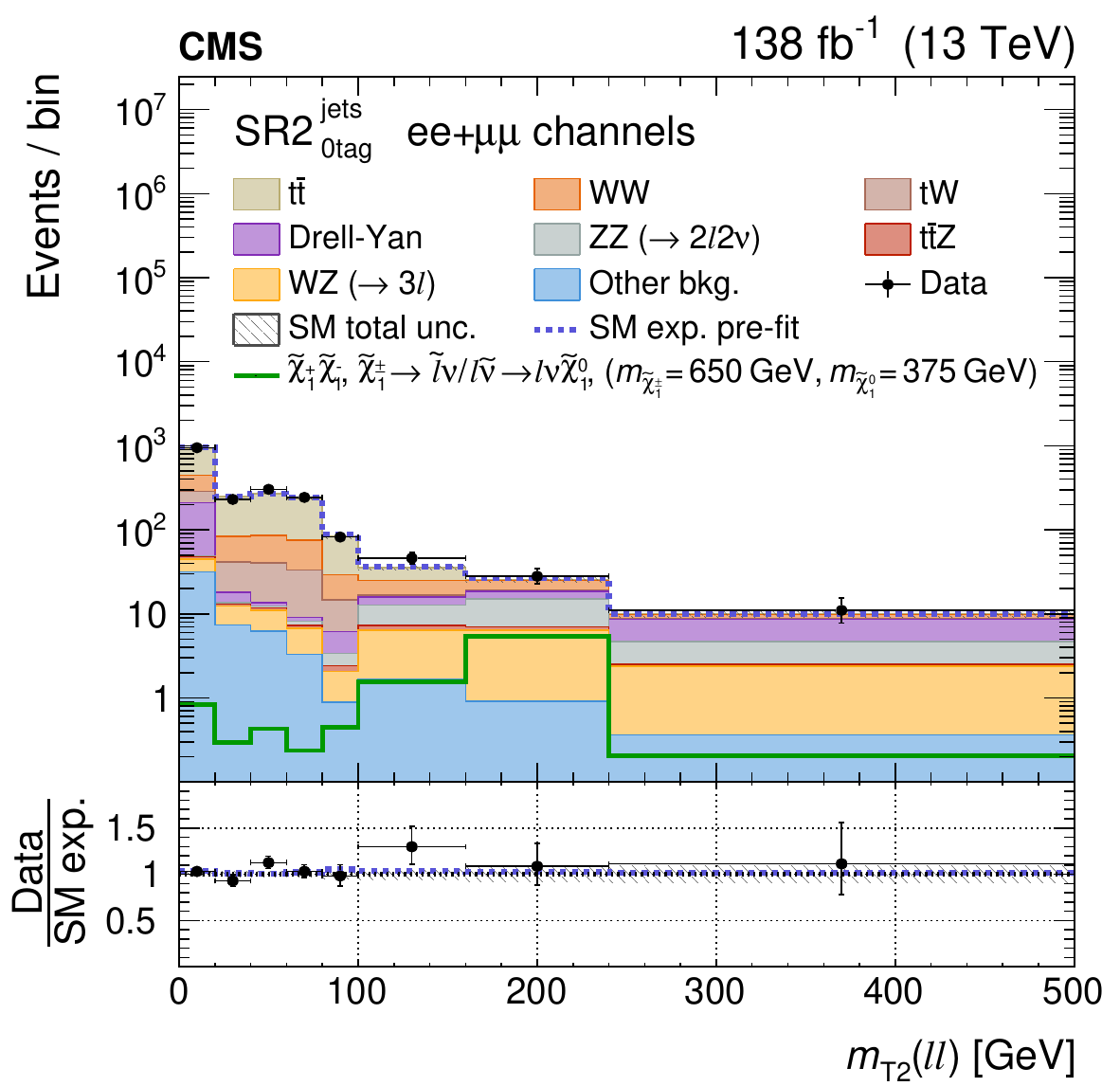}
\caption{Distributions of \mtll after the fit to data in the chargino and slepton SRs. The top row shows DF events in SR2, divided into \Reg{SR2}{0tag}{0jet} (left) and \Reg{SR2}{0tag}{jets} (right). The bottom row shows the same for SF events. The green histogram shows the expected \mtll distribution for chargino pair production with $\invM{\PSGcpmDo}=650\GeV$ and $\invM{\PSGczDo}=375\GeV$. The expected total SM contributions before the fit (blue dashed line) is also shown. The last bin includes the overflow entries. In the bottom panels, the ratio of data and SM expectations is shown for the expected total SM contribution after the fit using the background-only hypothesis (black markers with error bars). Also shown is the ratio of the SM expectations before and after the fit (blue dashed line). The hatched band represents the total uncertainty after the fit.}
\label{Fig:MT2ll_TChi_SR2}
\end{figure}

\begin{figure}
\centering
\includegraphics[width=0.49\textwidth]{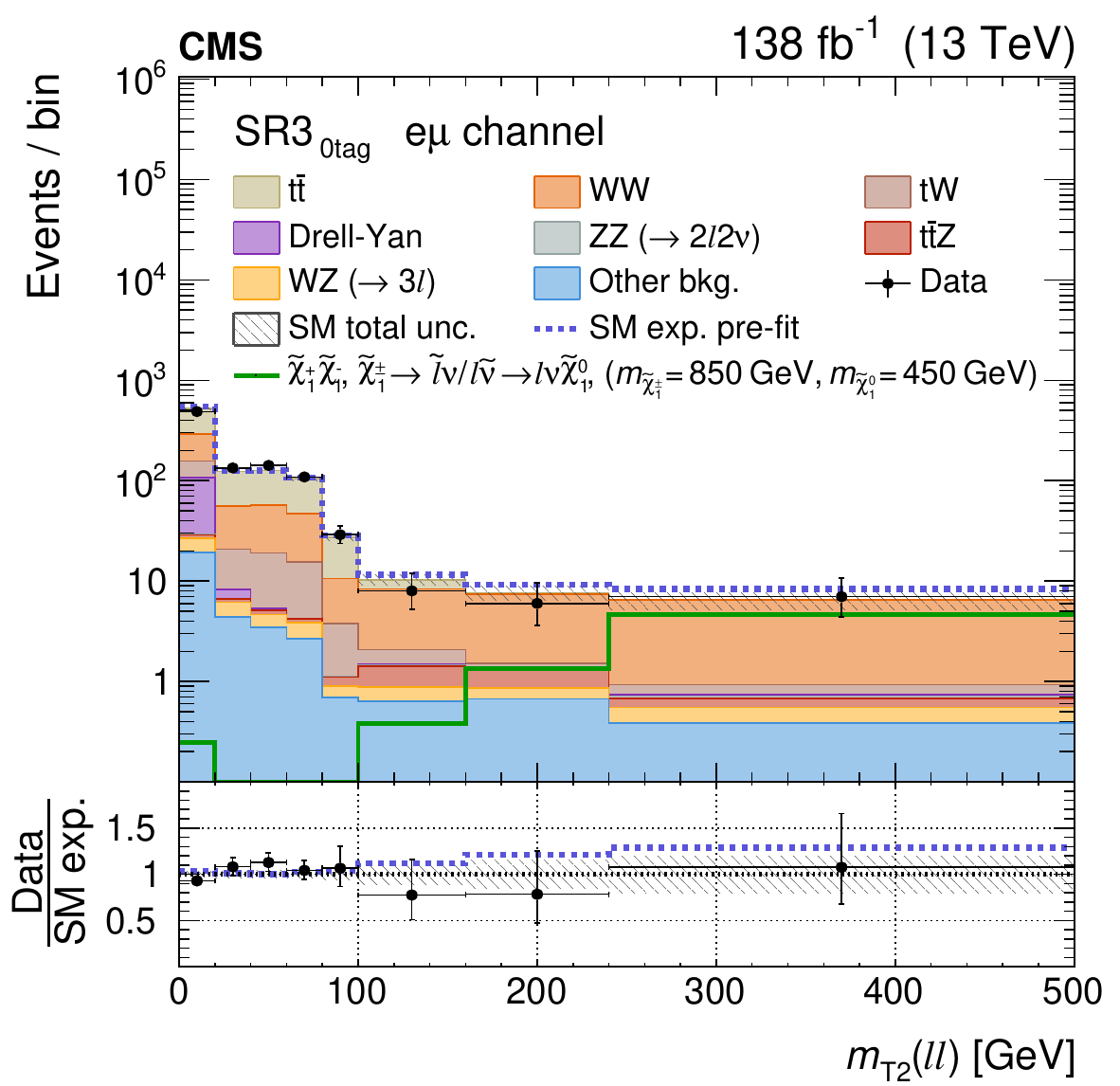}
\includegraphics[width=0.49\textwidth]{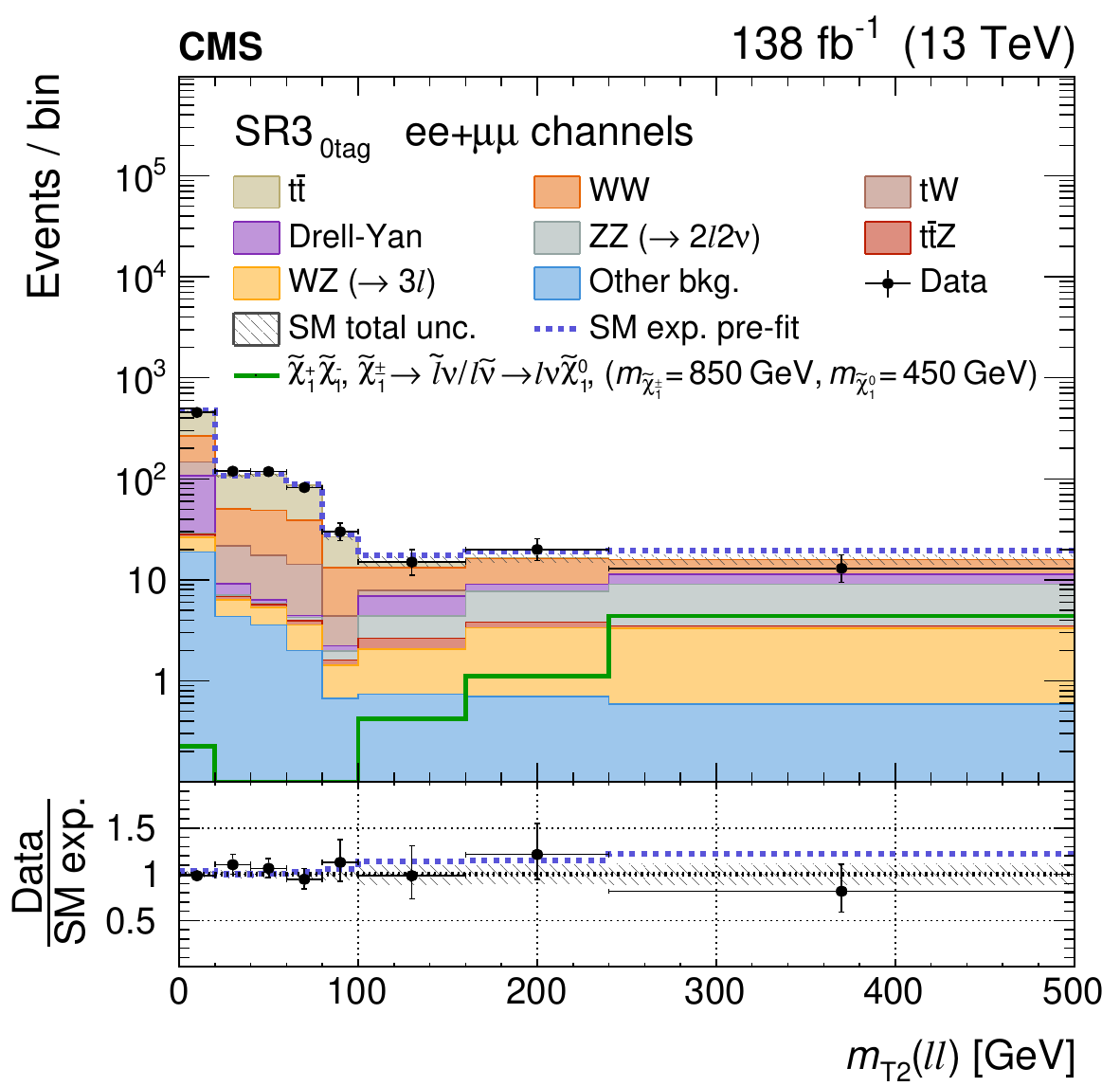}
\includegraphics[width=0.49\textwidth]{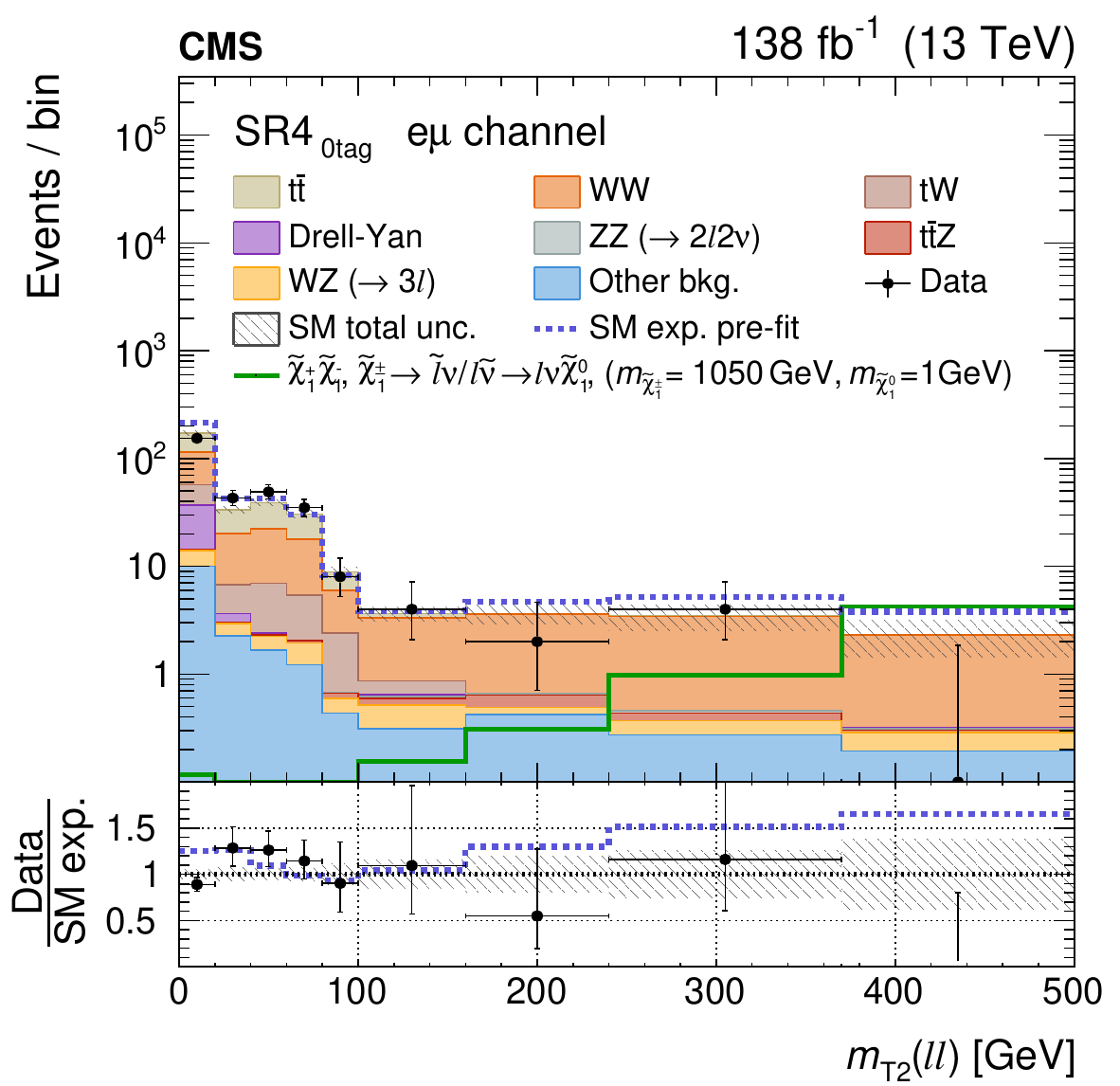}
\includegraphics[width=0.49\textwidth]{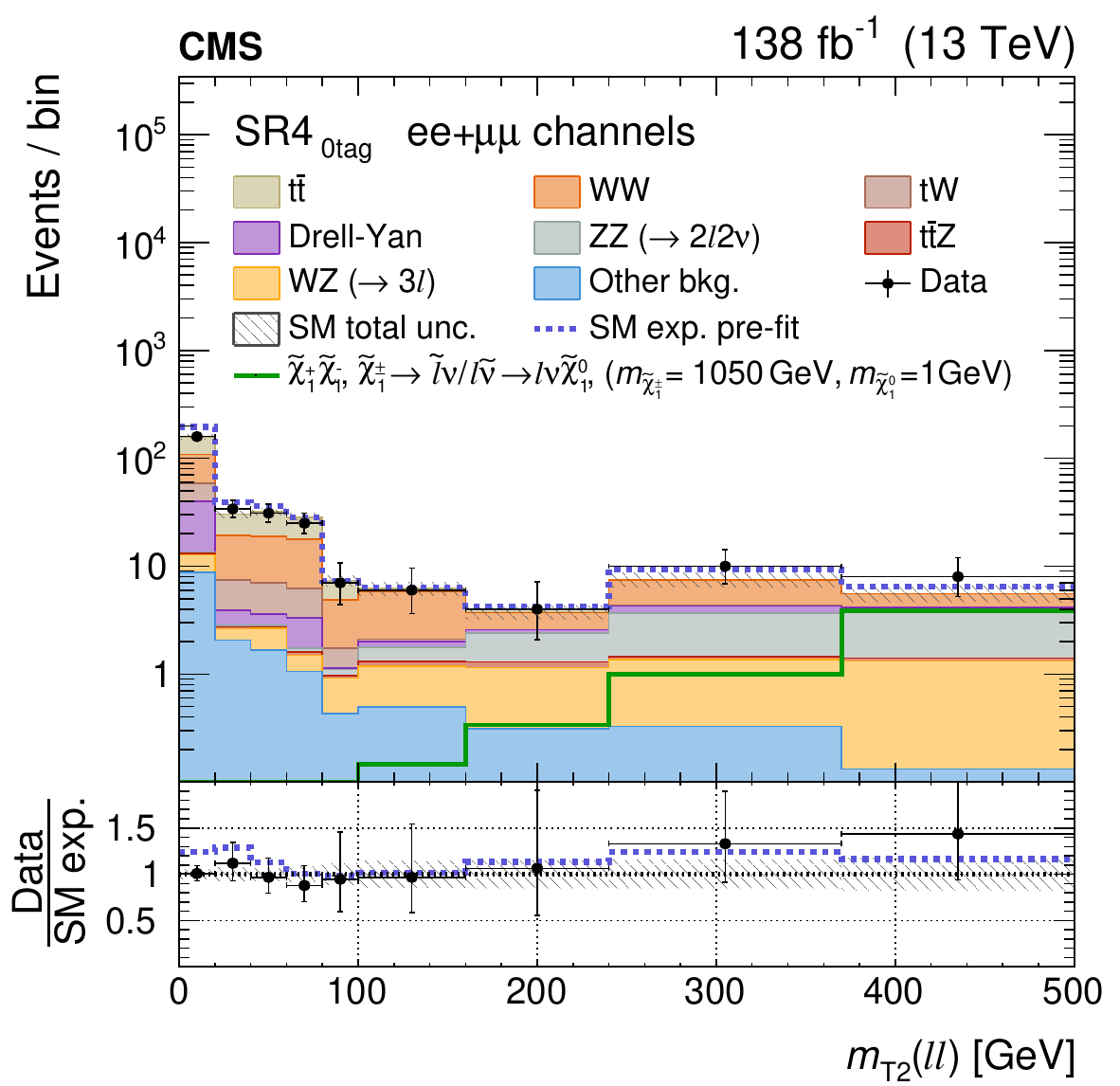}
	\caption{Distributions of \mtll after the fit to data in the chargino and slepton SRs. The top row shows DF (left) and SF (right) events in \Reg{SR3}{0tag}{}. The bottom row shows the same for \Reg{SR4}{0tag}{}. The green histogram shows the expected \mtll distribution for chargino pair production with $\invM{\PSGcpmDo}=850\GeV$ and $\invM{\PSGczDo}=450\GeV$ ($\invM{\PSGcpmDo}=1050\GeV$ and $\invM{\PSGczDo}=1\GeV$) in \Reg{SR3}{0tag}{} (\Reg{SR4}{0tag}{}). The expected total SM contributions before the fit (blue dashed line) is also shown. The last bin includes the overflow entries. In the bottom panels, the ratio of data and SM expectations is shown for the expected total SM contribution after the fit using the background-only hypothesis (black markers with error bars). Also shown is the ratio of the SM expectations before and after the fit (blue dashed line). The hatched band represents the total uncertainty after the fit.}
\label{Fig:MT2ll_TChi_SR34}
\end{figure}

\begin{figure}
\centering
\includegraphics[width=0.49\textwidth]{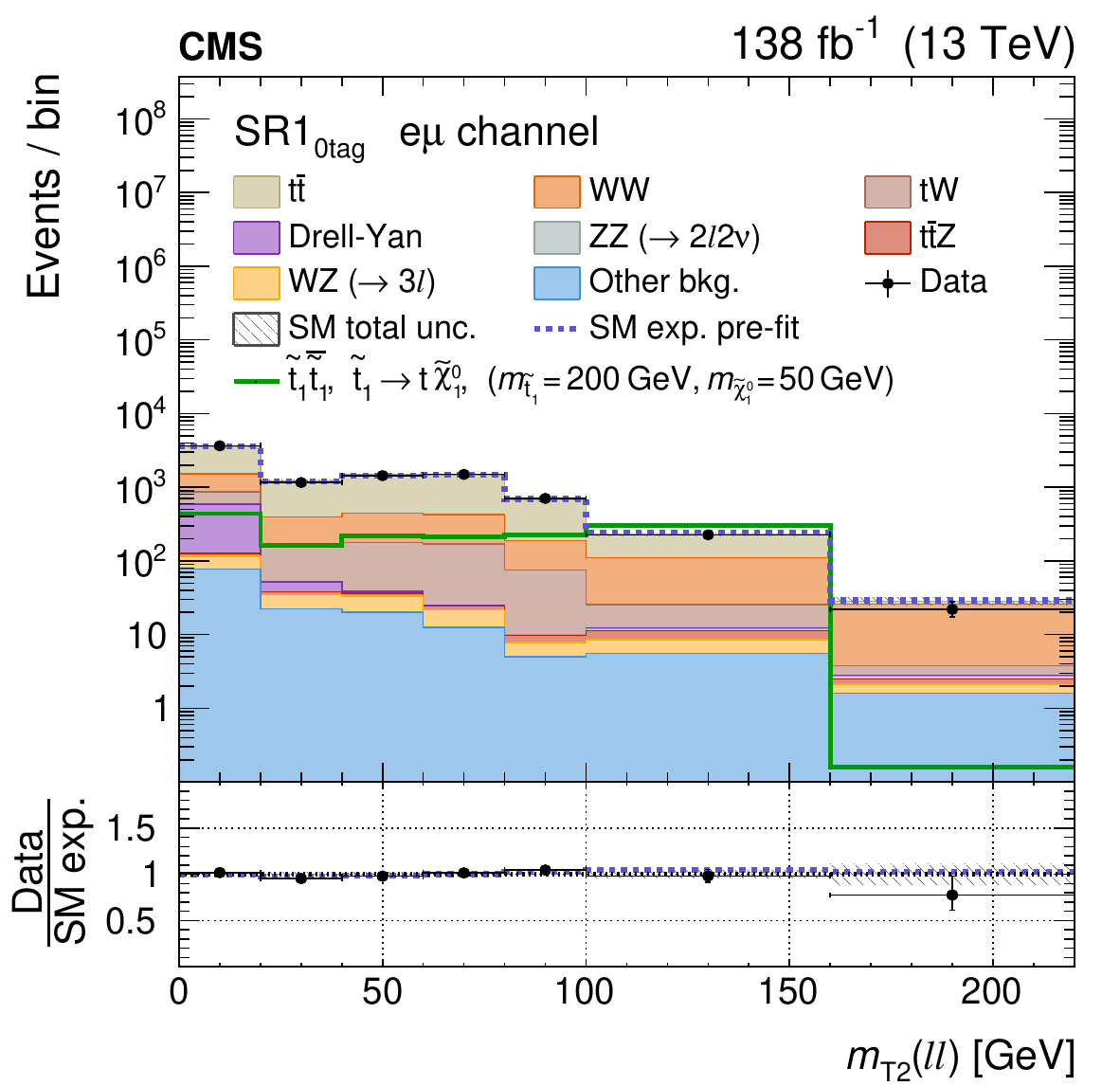}
\includegraphics[width=0.49\textwidth]{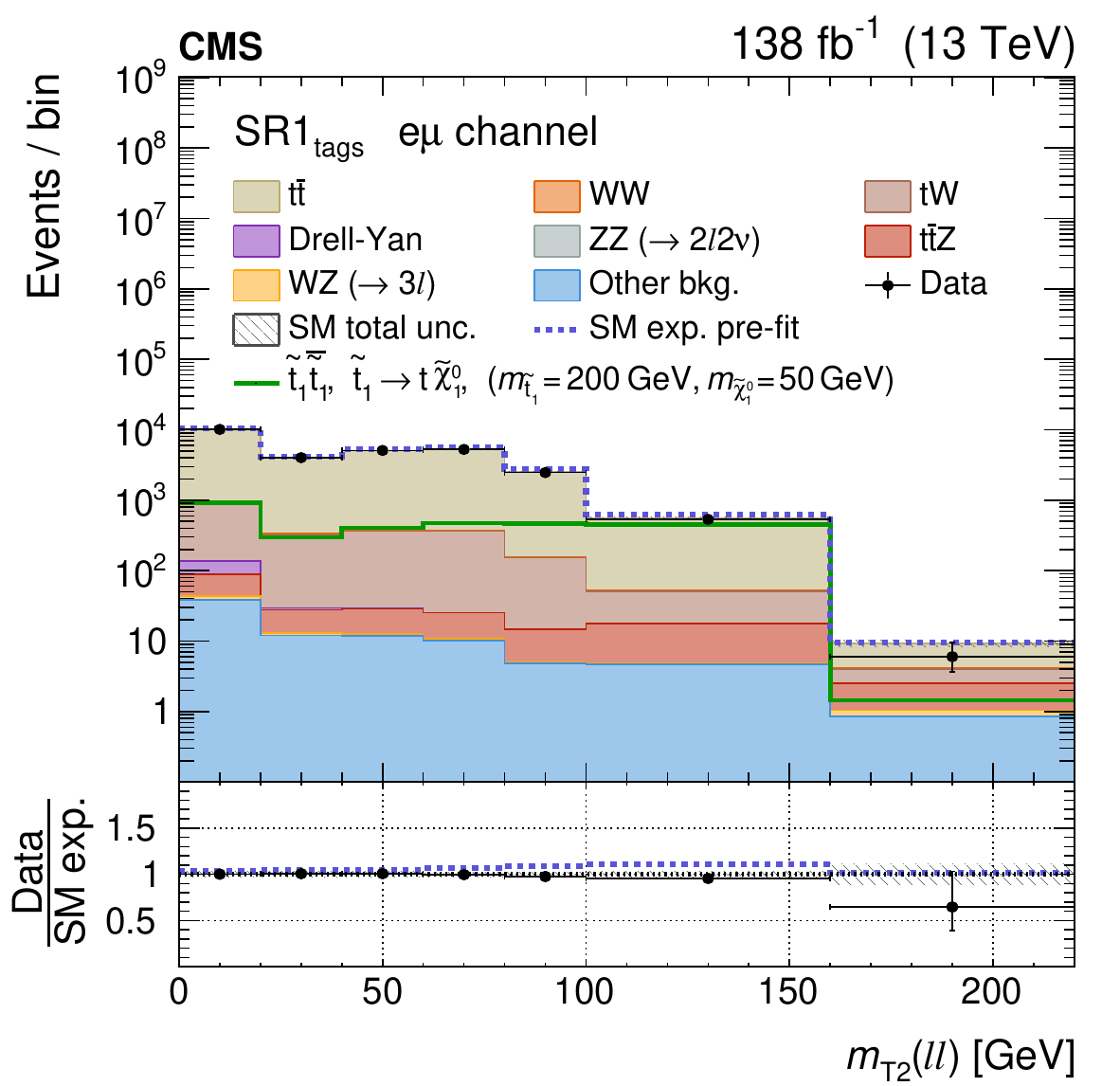}

\includegraphics[width=0.49\textwidth]{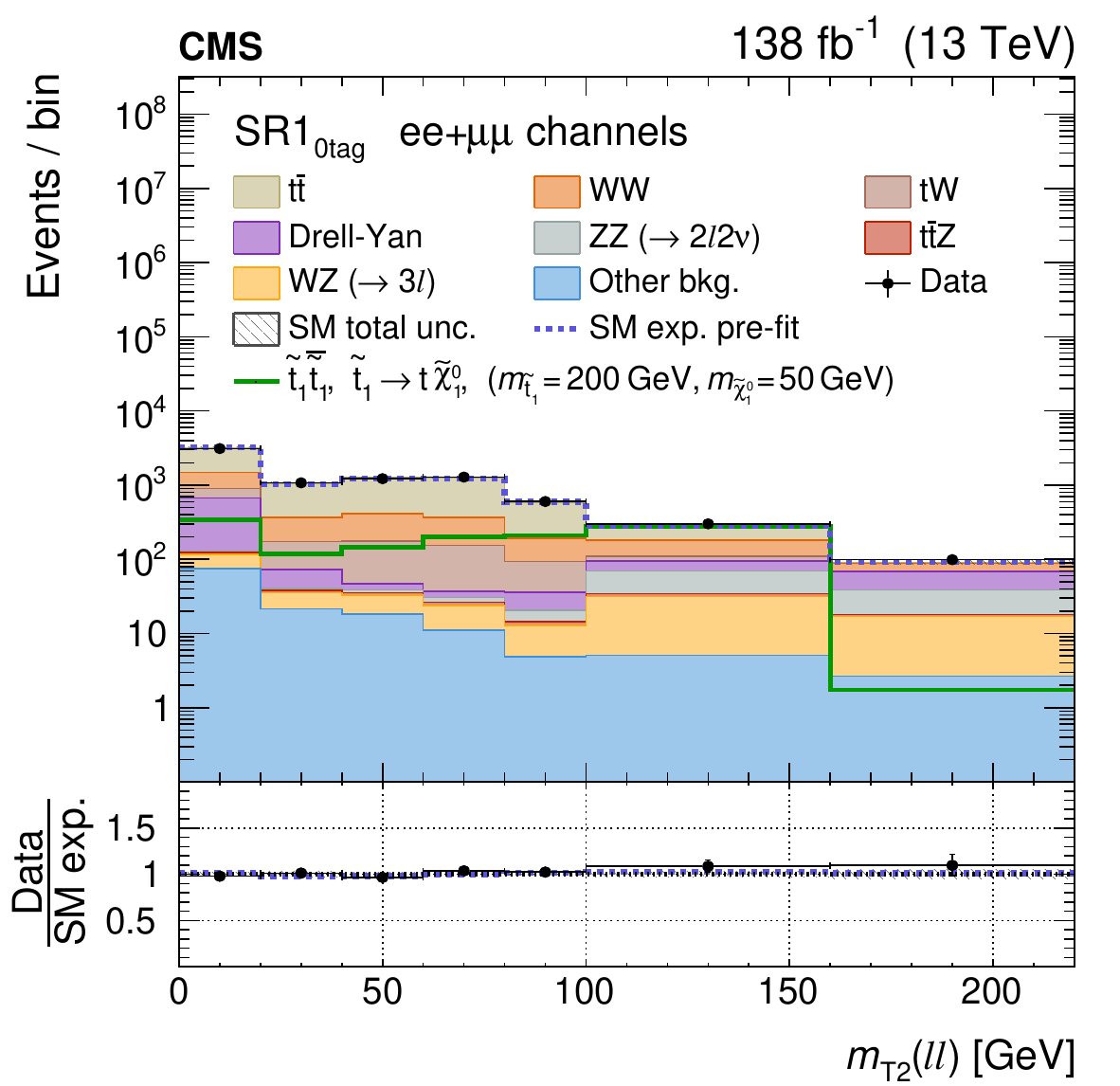}
\includegraphics[width=0.49\textwidth]{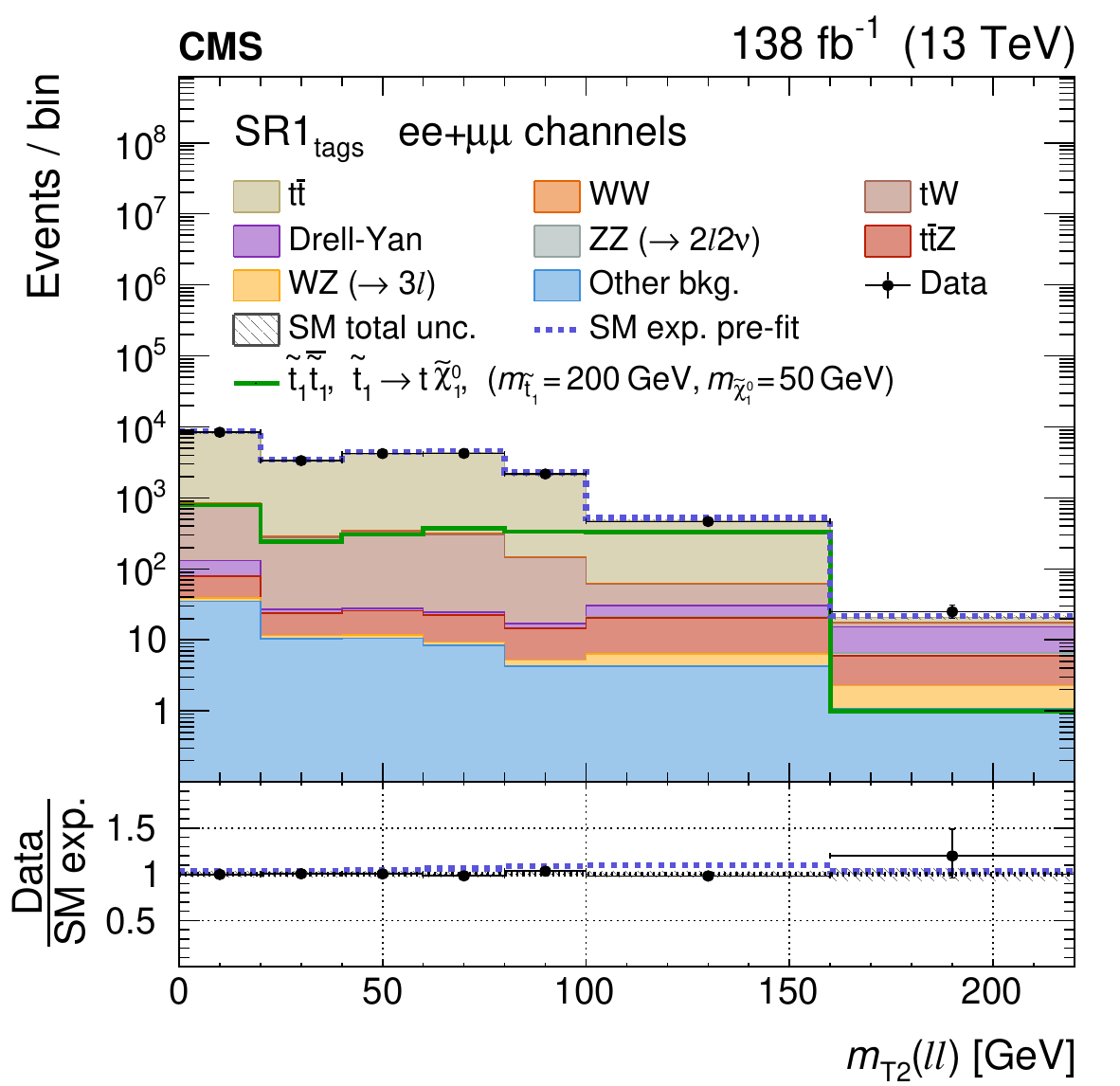}
      \caption{Distributions of \mtll after the fit to data in the top squark SRs. The first row shows DF events in SR1, divided into \Reg{SR1}{0tag}{} (left) and \Reg{SR1}{tags}{} (right). The second row shows the same for SF events. The magenta histogram shows the expected \mtll distribution for top squark pair production with $\invM{\stone}=200\GeV$ and $\invM{\PSGczDo}=50\GeV$. The expected total SM contributions before the fit (blue dashed line) is also shown. The last bin includes the overflow entries. In the bottom panels, the ratio of data and SM expectations is shown for the expected total SM contribution after the fit using the background-only hypothesis (black markers with error bars). Also shown is the ratio of the SM expectations before and after the fit (blue dashed line). The hatched band represents the total uncertainty after the fit.} 
\label{Fig:MT2ll_T2_SR1}
\end{figure}

\begin{figure}
\centering
\includegraphics[width=0.49\textwidth]{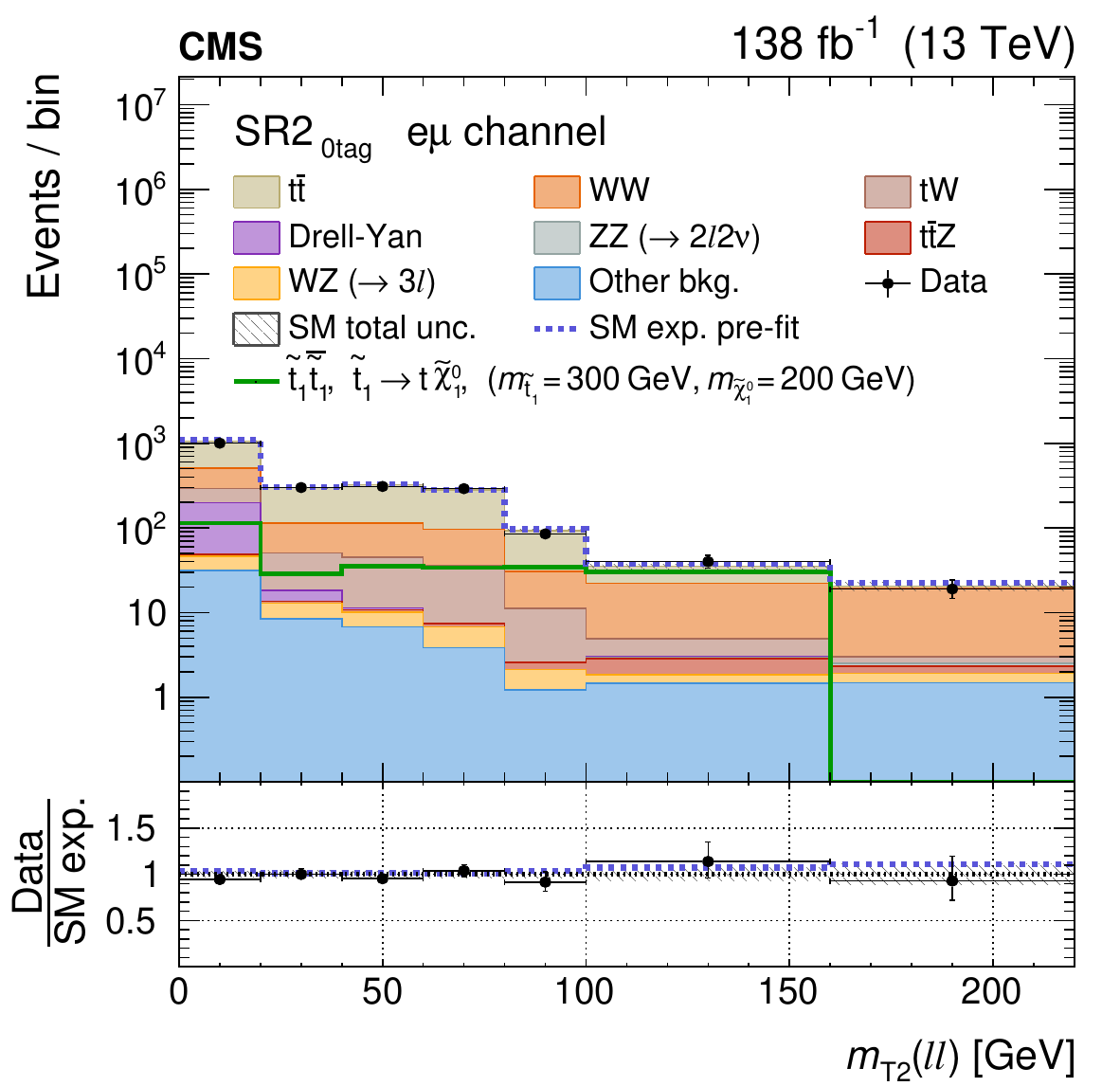}
\includegraphics[width=0.49\textwidth]{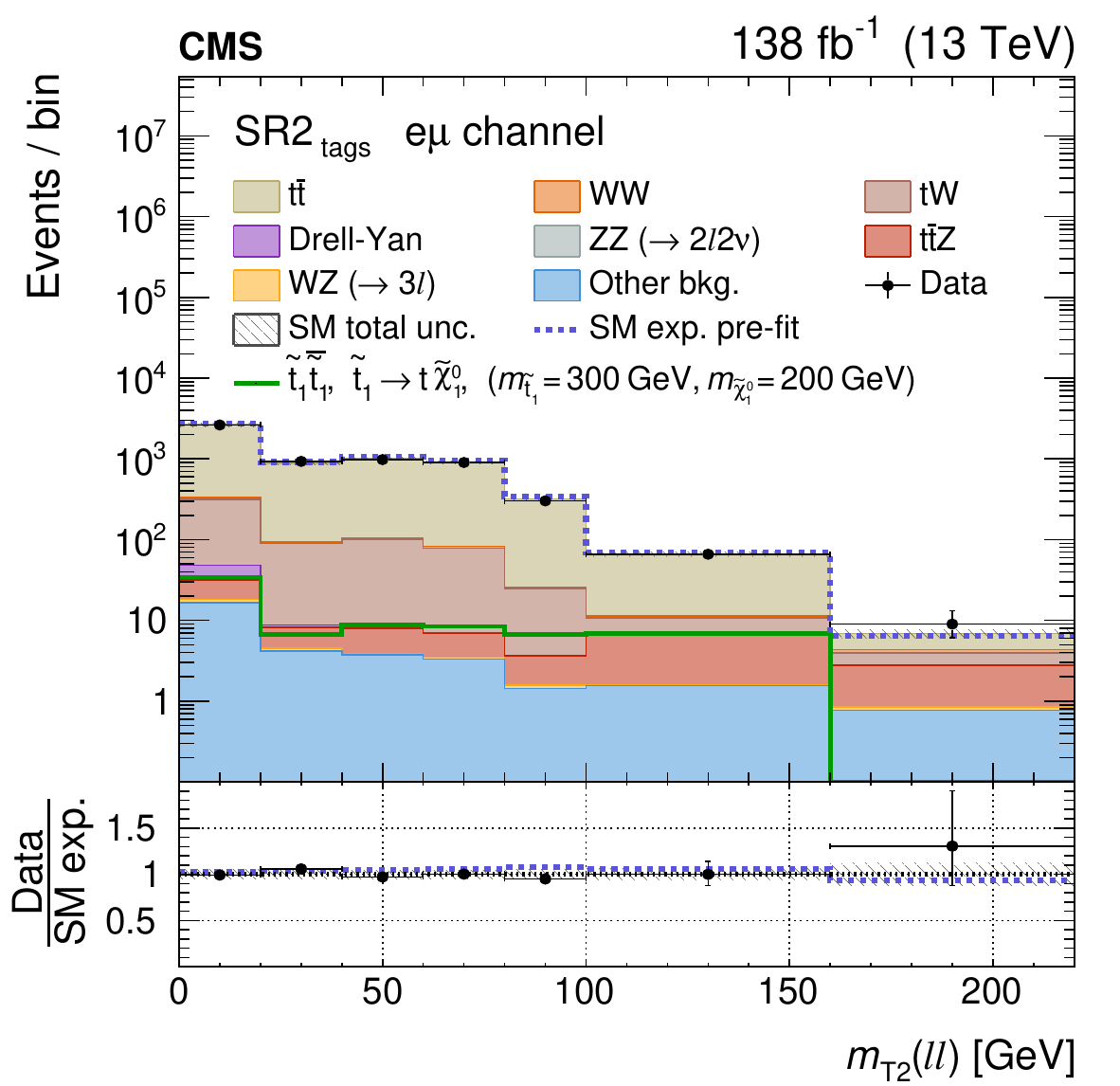}

\includegraphics[width=0.49\textwidth]{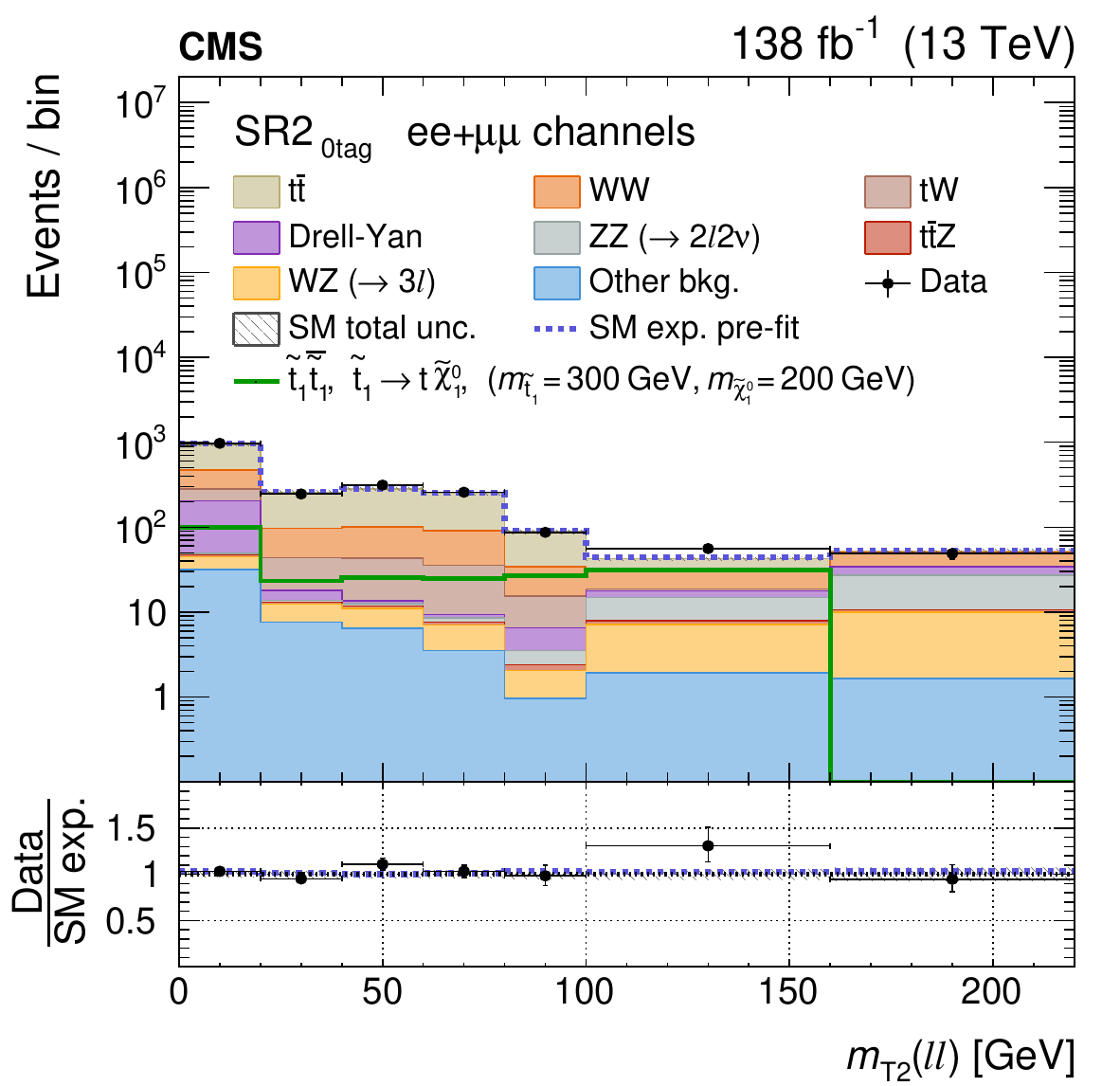}
\includegraphics[width=0.49\textwidth]{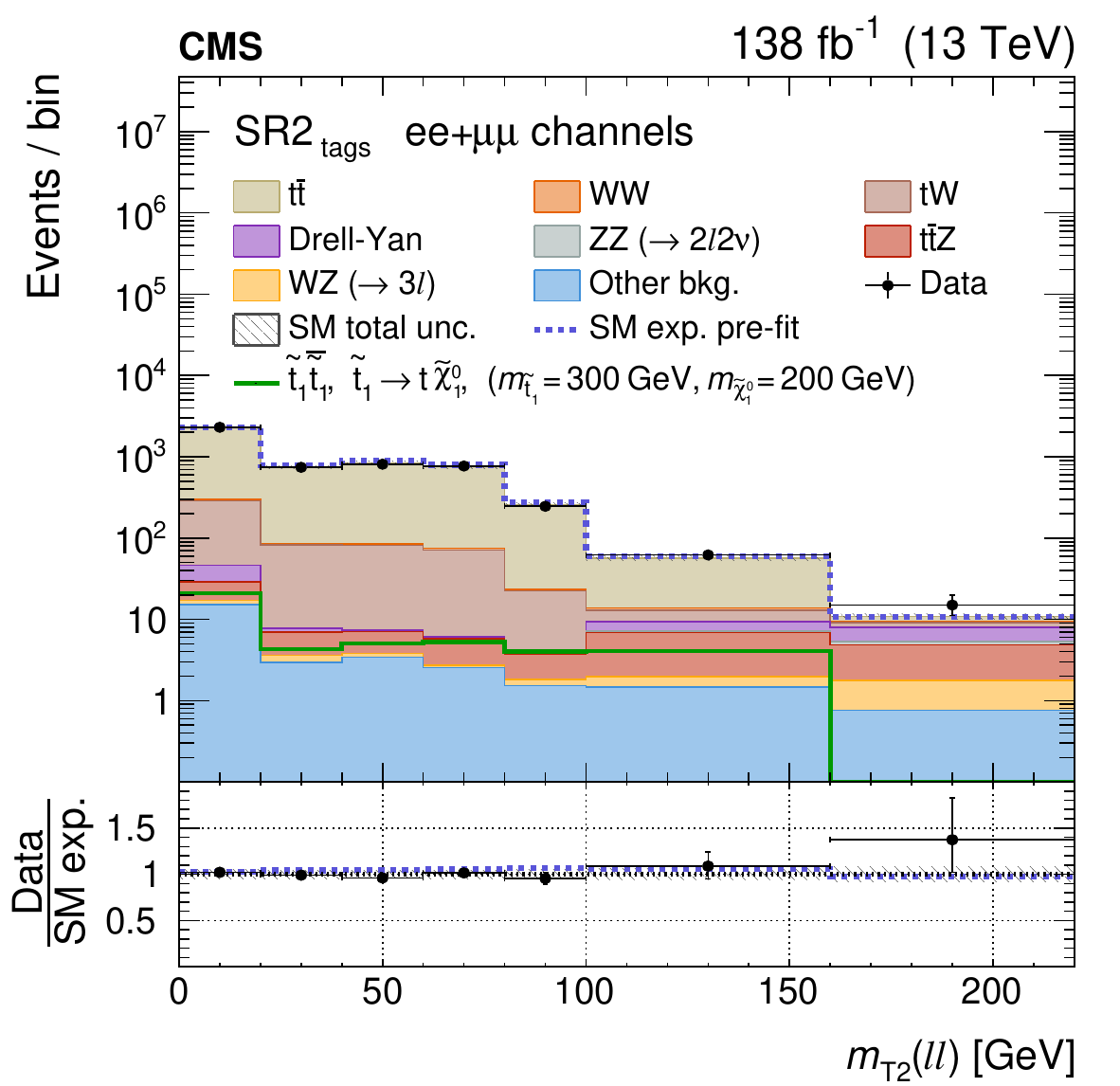}
	\caption{Distributions of \mtll after the fit to data in the top squark SRs. The first row shows DF events in SR2, divided into \Reg{SR2}{0tag}{} (left) and \Reg{SR2}{tags}{} (right). The second row shows the same for SF events. The magenta histogram shows the expected \mtll distribution for top squark pair production with $\invM{\stone}=300\GeV$ and $\invM{\PSGczDo}=200\GeV$. The expected total SM contributions before the fit (blue dashed line) is also shown. The last bin includes the overflow entries. In the bottom panels, the ratio of data and SM expectations is shown for the expected total SM contribution after the fit using the background-only hypothesis (black markers with error bars). Also shown is the ratio of the SM expectations before and after the fit (blue dashed line). The hatched band represents the total uncertainty after the fit..} 
\label{Fig:MT2ll_T2_SR2}
\end{figure}

\begin{figure}
  \centering
  \includegraphics[width=0.49\textwidth]{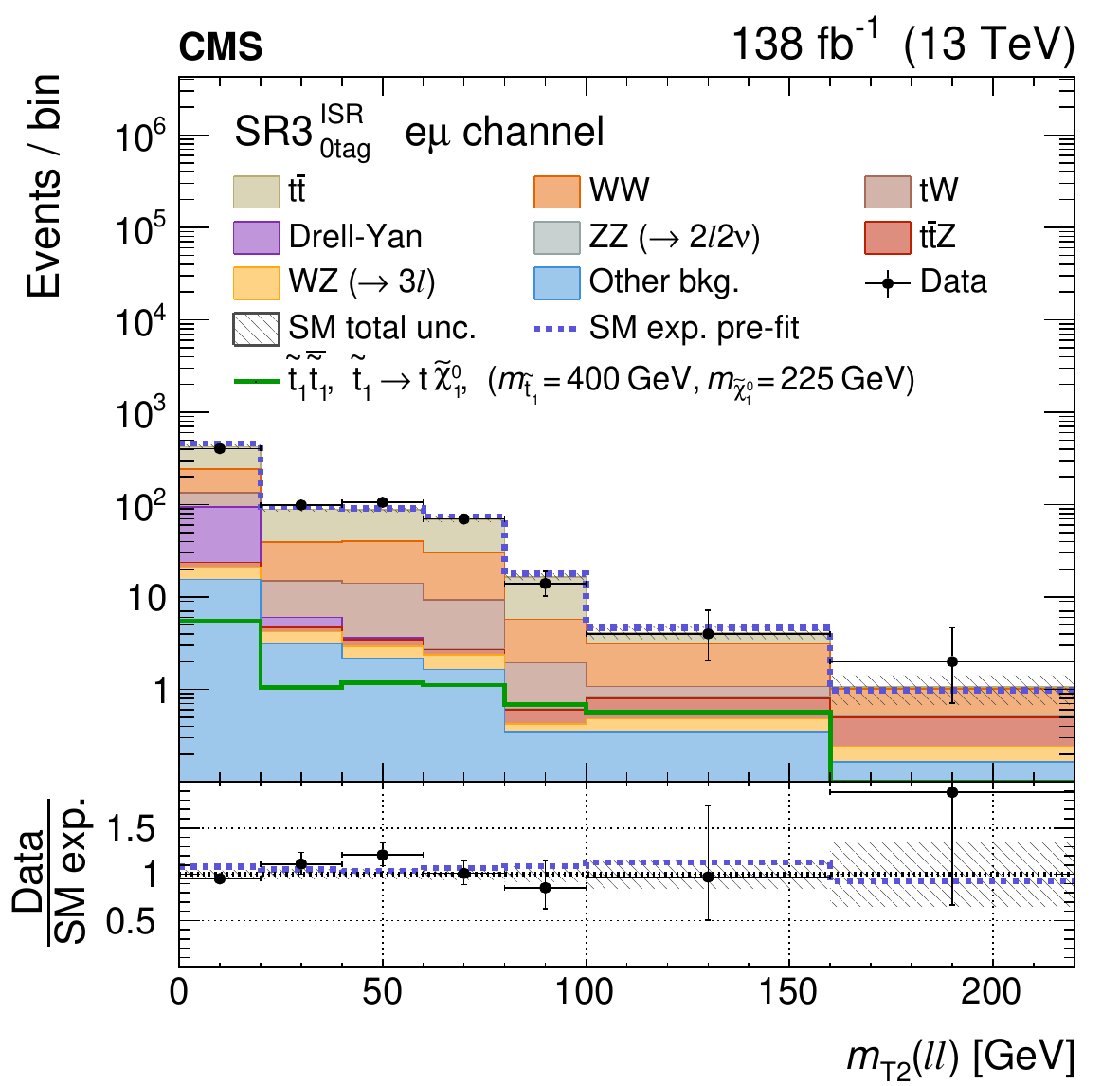}
  \includegraphics[width=0.49\textwidth]{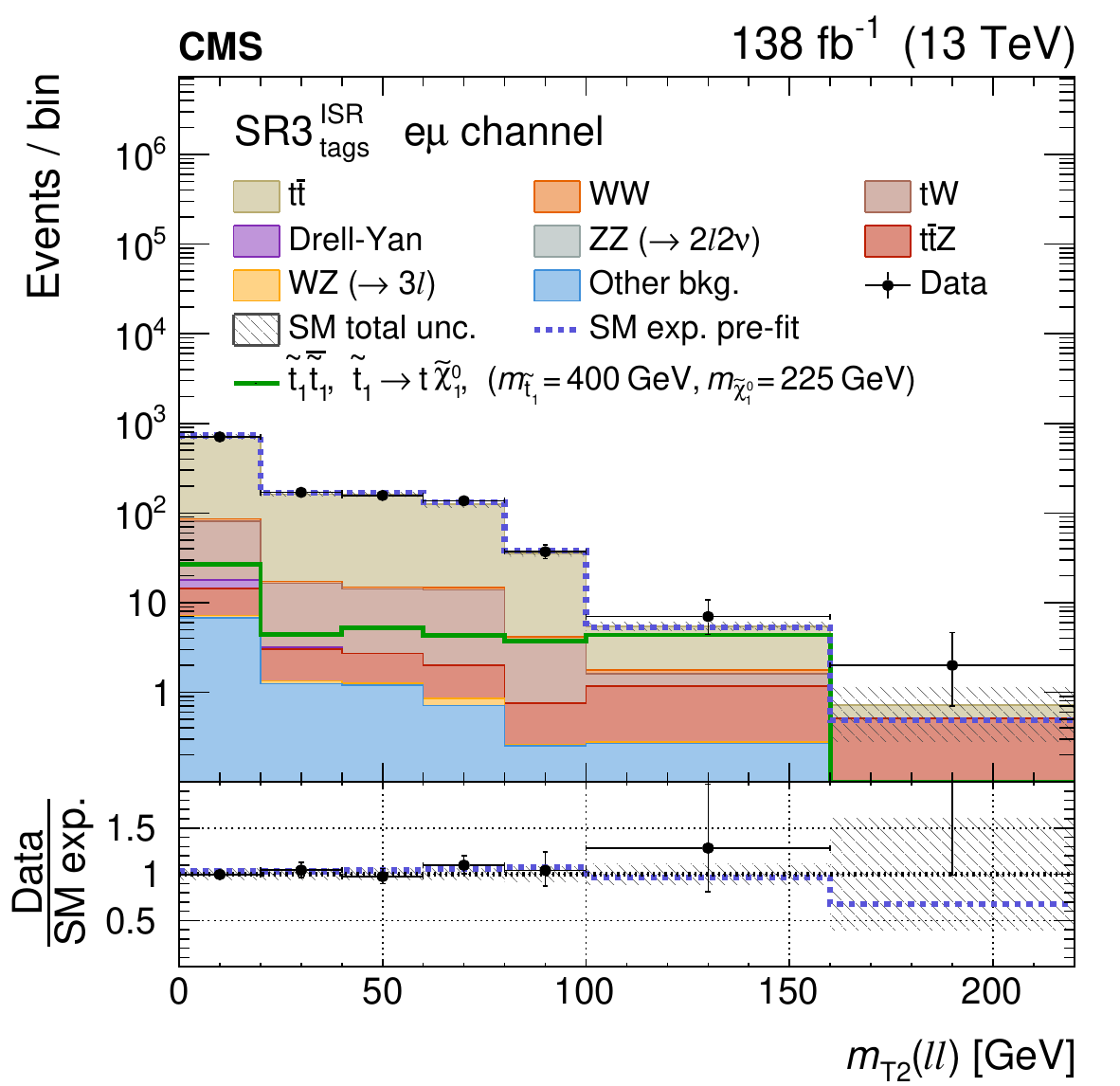}

  \includegraphics[width=0.49\textwidth]{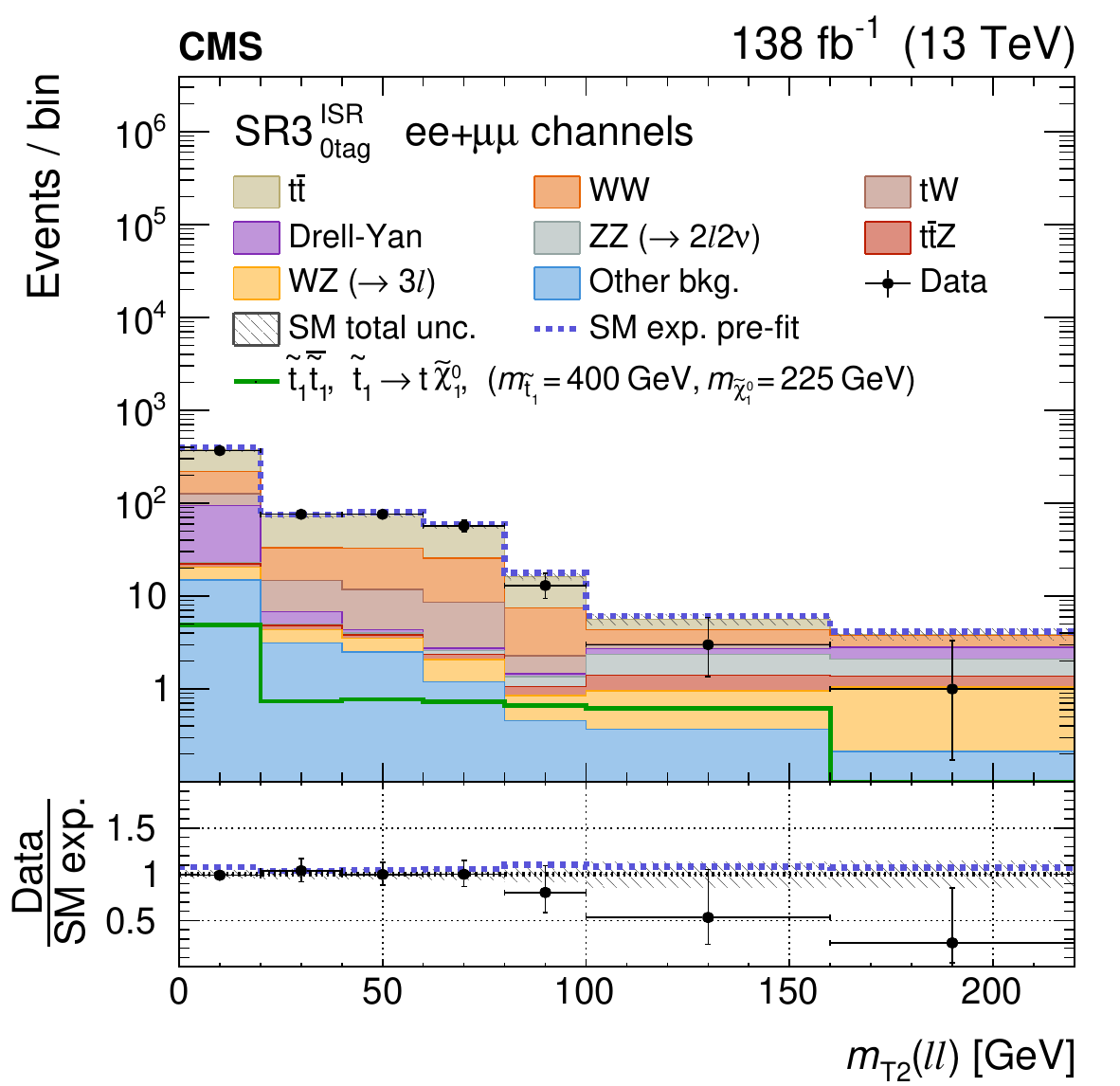}
  \includegraphics[width=0.49\textwidth]{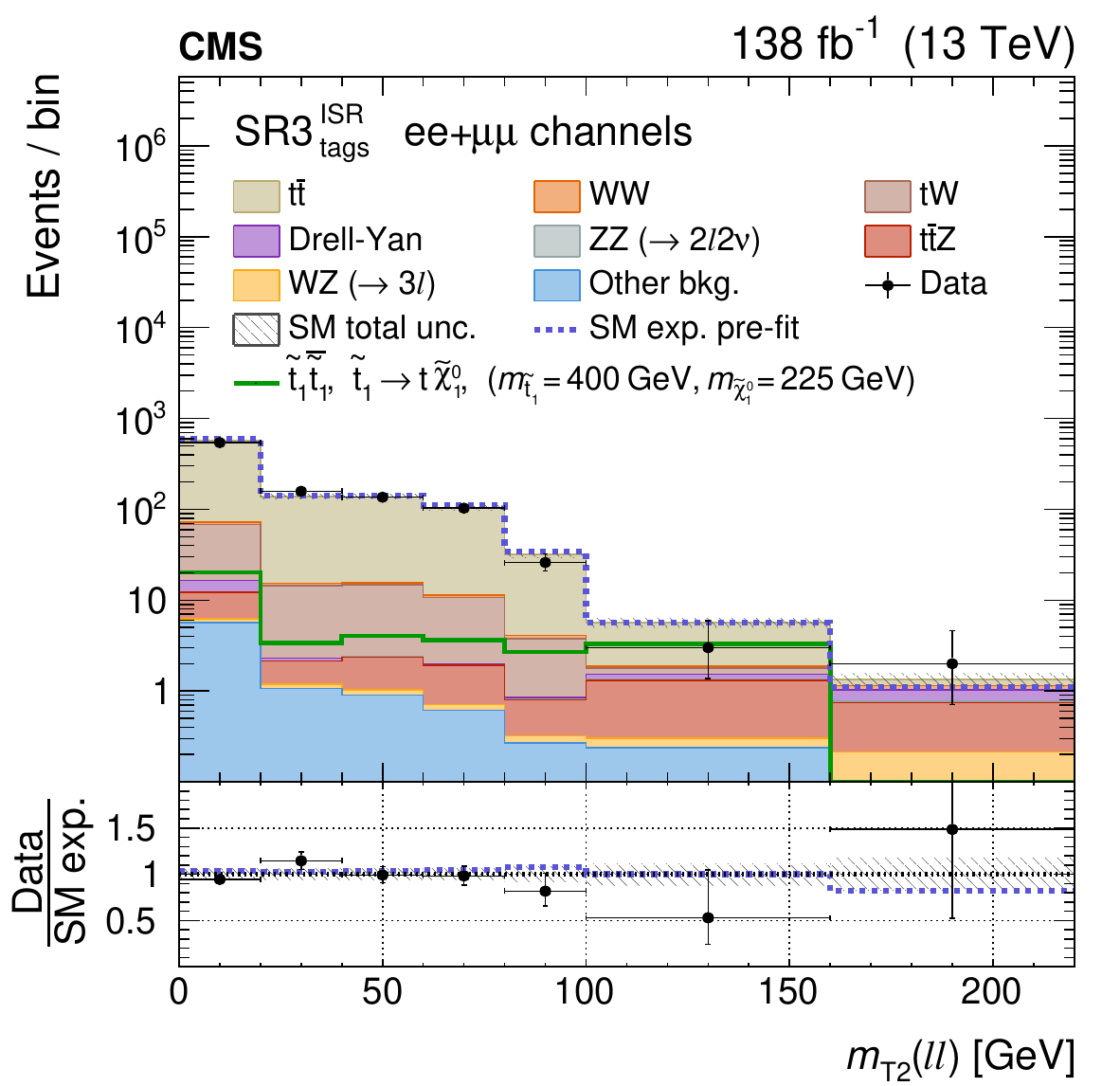}
    \caption{Distributions of \mtll after the fit to data in the top squark SRs. The first row shows DF events in SR3, divided into \Reg{SR3}{0tag}{ISR} (left) and \Reg{SR3}{tags}{ISR} (right). The second row shows the same for SF events. The magenta histogram shows the expected \mtll distribution for top squark pair production with $\invM{\stone}=400\GeV$ and $\invM{\PSGczDo}=225\GeV$. The expected total SM contributions before the fit (blue dashed line) is also shown. The last bin includes the overflow entries. In the bottom panels, the ratio of data and SM expectations is shown for the expected total SM contribution after the fit using the background-only hypothesis (black markers with error bars). Also shown is the ratio of the SM expectations before and after the fit (blue dashed line). The hatched band represents the total uncertainty after the fit.}
   \label{Fig:MT2ll_T2_SR3}
\end{figure}

\begin{figure}
  \centering

  \includegraphics[width=0.49\textwidth]{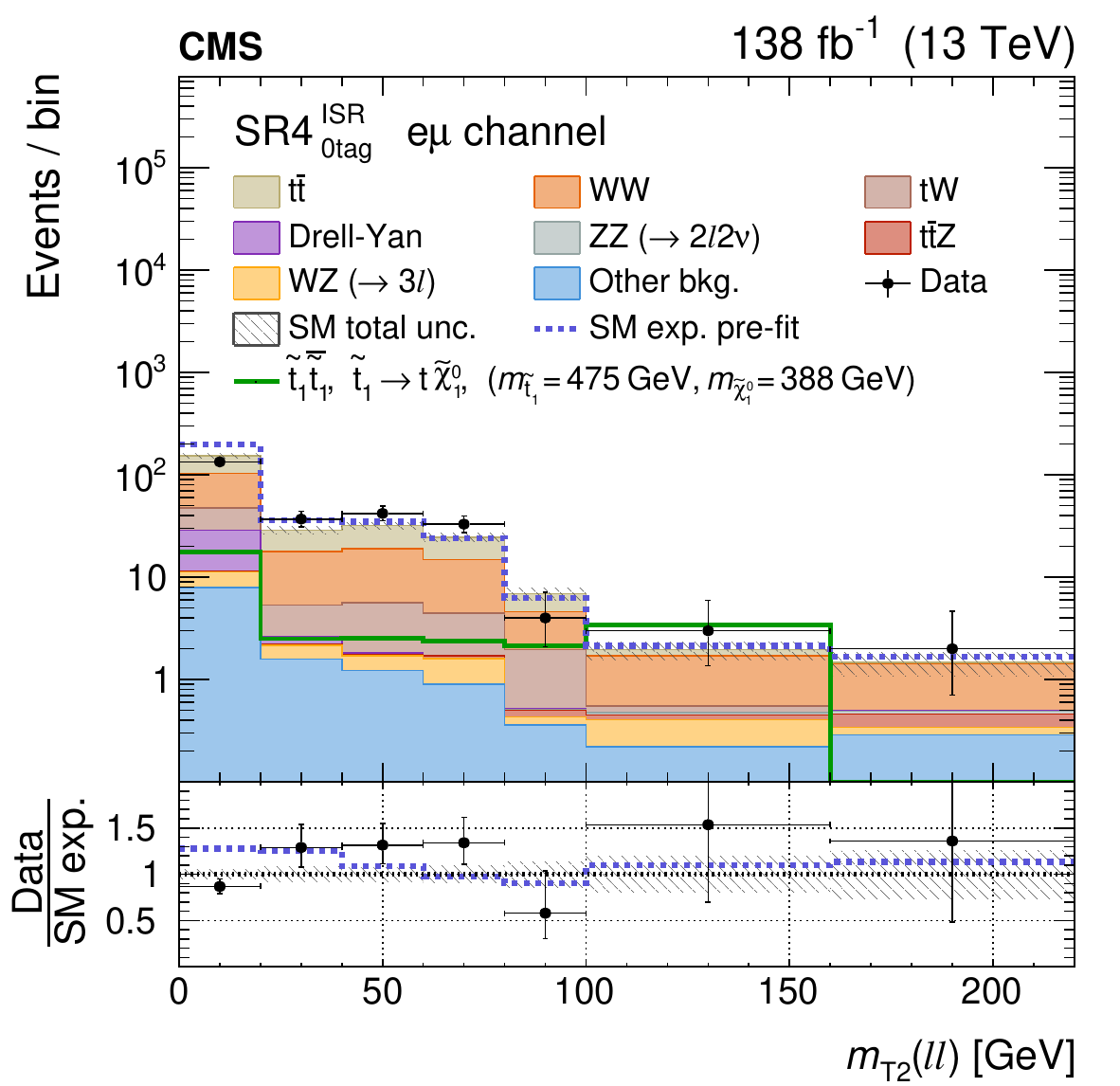}
  \includegraphics[width=0.49\textwidth]{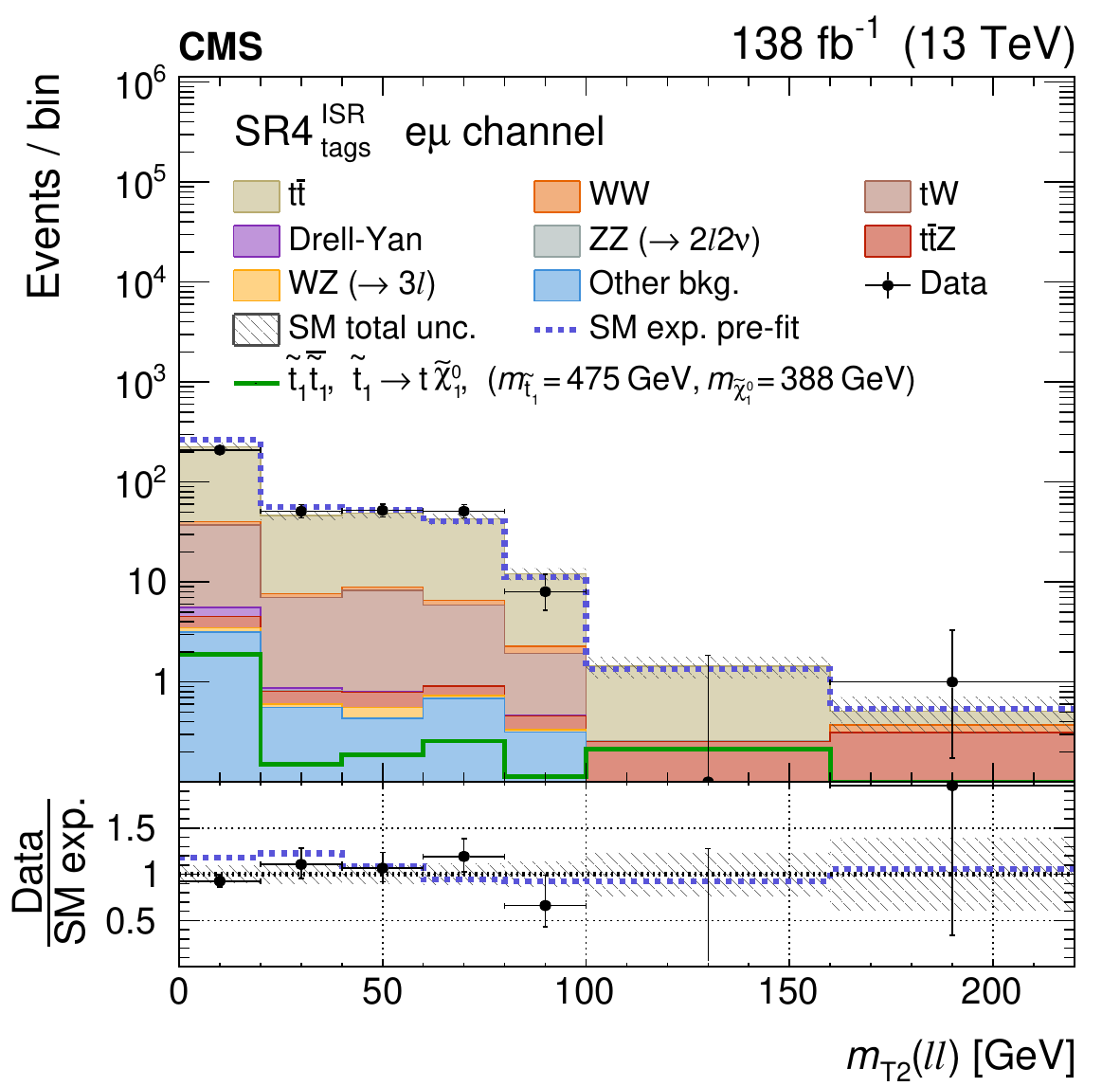}

  \includegraphics[width=0.49\textwidth]{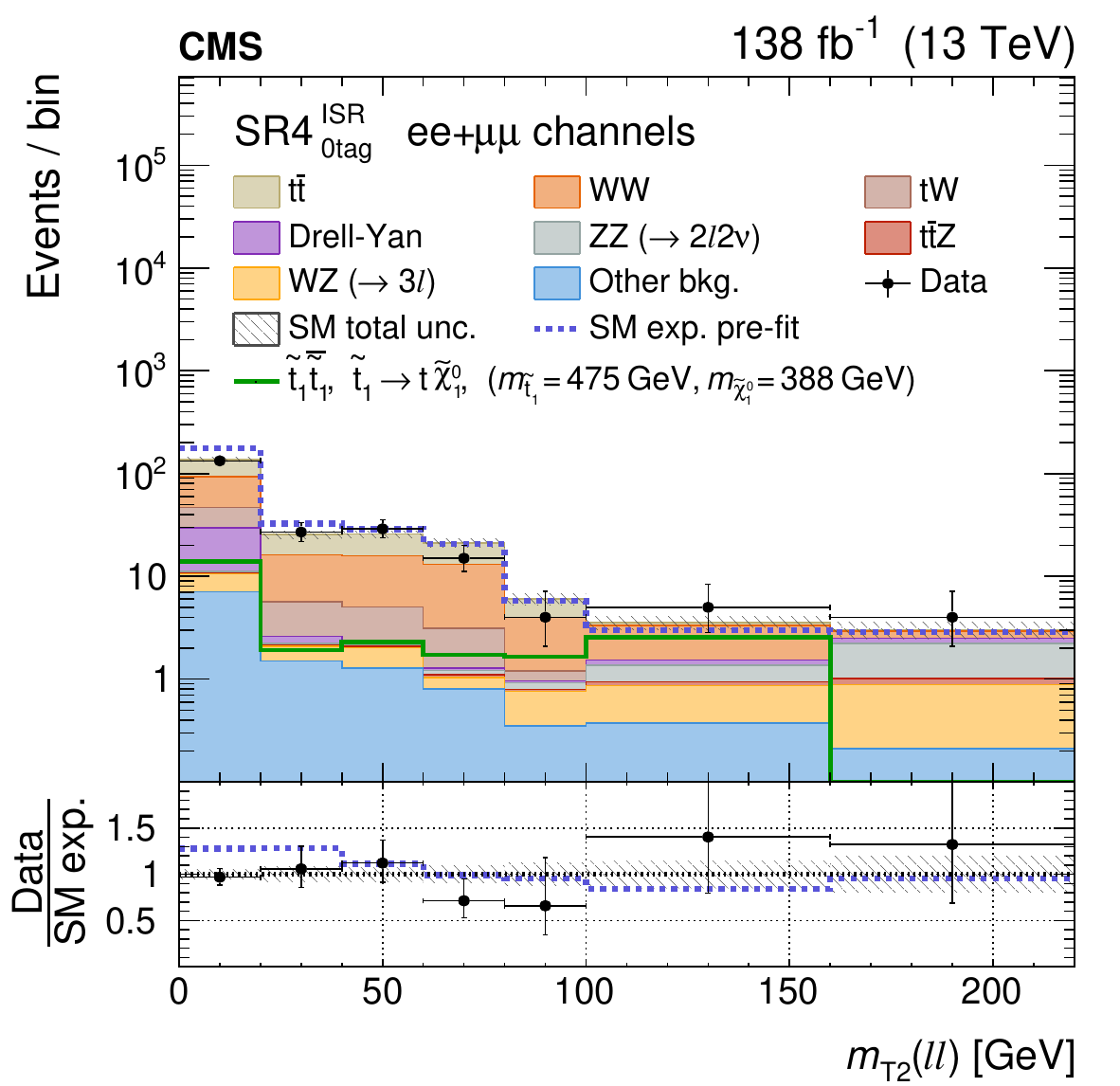}
  \includegraphics[width=0.49\textwidth]{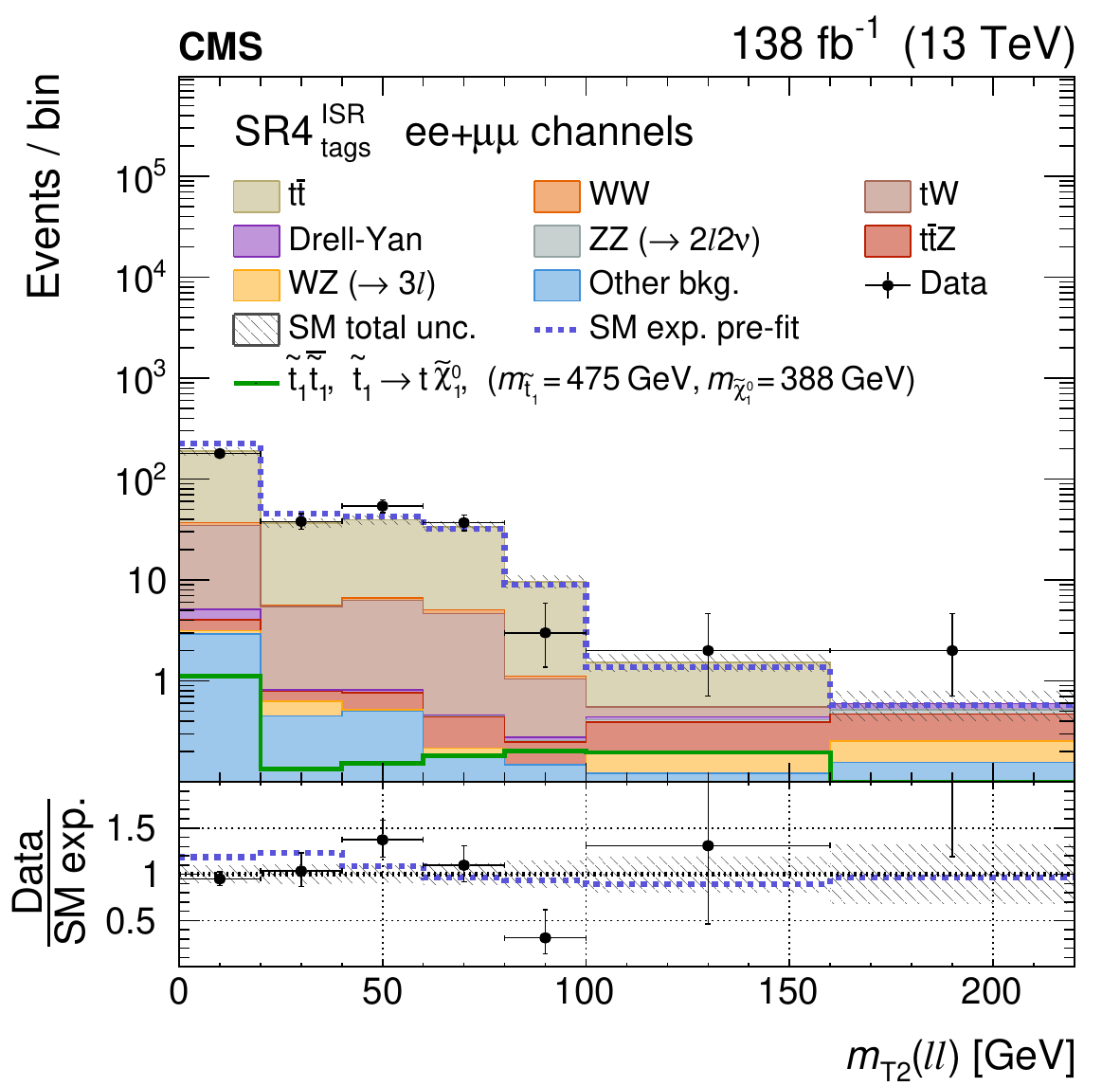}
    \caption{Distributions of \mtll after the fit to data in the top squark SRs. The first row shows DF events in SR4, divided into \Reg{SR4}{0tag}{ISR} (left) and \Reg{SR4}{tags}{ISR} (right). The second row shows the same for SF events. The magenta histogram shows the expected \mtll distribution for top squark pair production with $\invM{\stone}=475\GeV$ and $\invM{\PSGczDo}=338\GeV$. The expected total SM contributions before the fit (blue dashed line) is also shown. The last bin includes the overflow entries. In the bottom panels, the ratio of data and SM expectations is shown for the expected total SM contribution after the fit using the background-only hypothesis (black markers with error bars). Also shown is the ratio of the SM expectations before and after the fit (blue dashed line). The hatched band represents the total uncertainty after the fit.}
   \label{Fig:MT2ll_T2_SR4}
\end{figure}

The results of the fit in the SRs for the chargino search are shown in Figs.~\ref{Fig:MT2ll_TChi_SR1} to~\ref{Fig:MT2ll_TChi_SR34} for DF and SF events, respectively. The results for the top squark search are shown in Figs.~\ref{Fig:MT2ll_T2_SR1} to~\ref{Fig:MT2ll_T2_SR4}.
Each figure compares the number of observed events in the SRs with the expected yields from SM processes after a background-only fit. 
As a comparison, the expected yields for a representative signal point are given. The signal point is chosen among those providing the largest contribution to each SR and lying close to the expected sensitivity reach.
The total expected SM contributions before the fit are also shown.
Detailed information on the observed and expected yields after the background-only fit are given in Tables~\ref{tab:PostfitTChill_Short} and \ref{tab:PostfitT2ttll_Short} for all dilepton final states and all SRs. 

No excess over SM prediction is observed in data.
Upper limits at the 95\% confidence level (\CL) on the production cross sections for the different signal models considered are set using the \CLs criterion~\cite{bib:CLs,bib:CL2} with the profile likelihood ratio modified for upper limits~\cite{CMS-NOTE-2011-005} as the test statistic. The asymptotic approximation is used in the limit setting procedure~\cite{bib:AsymptCL}.
The following results have been determined using the CMS statistical analysis tool \textsc{Combine}~\cite{CMS:2024onh}, which is based on the \textsc{RooFit}~\cite{Verkerke:2003ir} and \textsc{RooStats}~\cite{Moneta:2010pm} frameworks.

\begin{table}[htb]
\centering
\topcaption{Observed and expected yields of DF (upper half) and SF (lower half) events after the background-only fit in the chargino and slepton SRs. The quoted uncertainties in the background predictions include statistical and systematic contributions.}
\cmsTable{
\begin{tabular}{llccccccccc}
\hline
\multicolumn{2}{c}{\mtll [{\GeVns}]} & 0--20 & 20--40 & 40--60 & 60--80 & 80--100 & 100--160 & 160--240 & 240--370 & ${\ge}370$ \\
\hline
DF events &  &  &  &  &  &  &  &  &  & \\
\multirow{2}{*}{\Reg{SR1}{0tag}{0jet}} & SM Processes & $101\pm 8$ & $67.7\pm 4.5$ & $80.1\pm 4.9$ & $90.9\pm 5.9$ & $52.4\pm 3.7$ & $52.5\pm 6.6$ & $12.2\pm 2.3$ \\
 & Data & $109$ & $59$ & $70$ & $97$ & $52$ & $47$ & $13$ \\
\multirow{2}{*}{\Reg{SR1}{0tag}{jets}} & SM Processes & $3470\pm 55$ & $1148\pm 20$ & $1388\pm 23$ & $1380\pm 20$ & $625\pm 13$ & $181\pm 7$ & $15.3\pm 2.1$ \\
 & Data & $3545$ & $1099$ & $1368$ & $1397$ & $653$ & $179$ & $9$ \\
\multirow{2}{*}{\Reg{SR2}{0tag}{0jet}} & SM Processes & $21.1\pm 2.5$ & $13.7\pm 1.8$ & $14.9\pm 2.0$ & $12.4\pm 1.8$ & $4.9\pm 0.9$ & $6.0\pm 1.5$ & $7.9\pm 1.9$ & $1.3\pm 0.7$ \\
 & Data & $21$ & $14$ & $12$ & $13$ & $5$ & $7$ & $8$ & $2$ \\
\multirow{2}{*}{\Reg{SR2}{0tag}{jets}} & SM Processes & $1037\pm 25$ & $288\pm 7$ & $311\pm 8$ & $267\pm 7$ & $87.8\pm 3.4$ & $28.5\pm 2.2$ & $9.4\pm 1.5$ & $1.4\pm 0.4$ \\
 & Data & $982$ & $286$ & $297$ & $278$ & $80$ & $33$ & $9$ & $0$ \\
\multirow{2}{*}{\Reg{SR3}{0tag}{}} & SM Processes & $525\pm 15$ & $124\pm 4$ & $126\pm 5$ & $105\pm 4$ & $27.2\pm 1.7$ & $10.3\pm 1.1$ & $7.7\pm 1.2$ & $6.5\pm 1.3$ \\
 & Data & $488$ & $134$ & $142$ & $109$ & $29$ & $8$ & $6$ & $7$ \\
\multirow{2}{*}{\Reg{SR4}{0tag}{}} & SM Processes & $173\pm 8$ & $33.5\pm 2.3$ & $38.8\pm 2.4$ & $30.6\pm 2.3$ & $8.9\pm 1.0$ & $3.7\pm 0.6$ & $3.6\pm 0.7$ & $3.4\pm 0.9$ & $2.3\pm 0.9$ \\
 & Data & $154$ & $43$ & $49$ & $35$ & $8$ & $4$ & $2$ & $4$ & $0$ \\[\cmsTabSkip]
SF events &  &  &  &  &  &  &  &  &  & \\
\multirow{2}{*}{\Reg{SR1}{0tag}{0jet}} & SM Processes & $122\pm 9$ & $62.9\pm 4.1$ & $75.9\pm 4.7$ & $81.8\pm 5.3$ & $43.8\pm 2.9$ & $65.2\pm 5.6$ & $27.2\pm 2.7$ \\
 & Data & $118$ & $73$ & $79$ & $82$ & $40$ & $70$ & $31$ \\
\multirow{2}{*}{\Reg{SR1}{0tag}{jets}} & SM Processes & $3053\pm 52$ & $999\pm 17$ & $1185\pm 20$ & $1153\pm 16$ & $546\pm 12$ & $215\pm 7$ & $64.3\pm 3.6$ \\
 & Data & $2992$ & $1002$ & $1142$ & $1202$ & $562$ & $231$ & $68$ \\
\multirow{2}{*}{\Reg{SR2}{0tag}{0jet}} & SM Processes & $20.6\pm 2.1$ & $13.9\pm 1.8$ & $14.3\pm 1.9$ & $14.0\pm 1.8$ & $4.4\pm 0.8$ & $7.2\pm 1.4$ & $12.6\pm 2.1$ & $2.8\pm 0.7$ \\
 & Data & $25$ & $16$ & $10$ & $15$ & $5$ & $10$ & $8$ & $2$ \\
\multirow{2}{*}{\Reg{SR2}{0tag}{jets}} & SM Processes & $918\pm 22$ & $248\pm 7$ & $269\pm 7$ & $236\pm 6$ & $83.5\pm 4.3$ & $35.4\pm 2.0$ & $25.8\pm 2.2$ & $9.9\pm 0.9$ \\
 & Data & $947$ & $230$ & $303$ & $243$ & $82$ & $46$ & $28$ & $11$ \\
\multirow{2}{*}{\Reg{SR3}{0tag}{}} & SM Processes & $461\pm 14$ & $108\pm 4$ & $111\pm 4$ & $86.6\pm 3.4$ & $26.6\pm 1.7$ & $15.2\pm 1.6$ & $16.5\pm 1.7$ & $15.9\pm 1.7$ \\
 & Data & $454$ & $119$ & $118$ & $82$ & $30$ & $15$ & $20$ & $13$ \\
\multirow{2}{*}{\Reg{SR4}{0tag}{}} & SM Processes & $158\pm 8$ & $30.4\pm 2.1$ & $32.0\pm 2.1$ & $28.5\pm 2.4$ & $7.4\pm 0.8$ & $6.2\pm 0.9$ & $3.8\pm 0.6$ & $7.5\pm 1.1$ & $5.6\pm 1.0$ \\
 & Data & $159$ & $34$ & $31$ & $25$ & $7$ & $6$ & $4$ & $10$ & $8$ \\
\hline
\end{tabular}
}
\label{tab:PostfitTChill_Short}
\end{table}

\begin{table}[htb]
\centering
\topcaption{Observed and expected yields of DF (upper half) and SF (lower half) events after the background-only fit in the top squark SRs. The quoted uncertainties in the background predictions include statistical and systematic contributions.}
\cmsTable{
\begin{tabular}{llccccccc}
\hline
\multicolumn{2}{c}{\mtll [{\GeVns}]} & 0--20 & 20--40 & 40--60 & 60--80 & 80--100 & 100--160 & ${>}160$ \\
\hline
DF events &  &  &  &  &  &  &  & \\
\multirow{2}{*}{\Reg{SR1}{tags}{}} & SM Processes & $10097\pm 302$ & $3987\pm 116$ & $5045\pm 142$ & $5275\pm 129$ & $2525\pm 59$ & $559\pm 18$ & $9.3\pm 1.0$ \\
 & Data & $10108$ & $4015$ & $5081$ & $5258$ & $2464$ & $534$ & $6$ \\
\multirow{2}{*}{\Reg{SR1}{0tag}{}} & SM Processes & $3587\pm 76$ & $1214\pm 30$ & $1468\pm 35$ & $1472\pm 30$ & $674\pm 16$ & $230\pm 9$ & $28.4\pm 3.1$ \\
 & Data & $3654$ & $1158$ & $1438$ & $1494$ & $705$ & $226$ & $22$ \\
\multirow{2}{*}{\Reg{SR2}{tags}{}} & SM Processes & $2649\pm 127$ & $880\pm 41$ & $1008\pm 46$ & $900\pm 41$ & $318\pm 15$ & $66.0\pm 4.0$ & $6.9\pm 0.8$ \\
 & Data & $2628$ & $931$ & $979$ & $902$ & $302$ & $66$ & $9$ \\
\multirow{2}{*}{\Reg{SR2}{0tag}{}} & SM Processes & $1062\pm 35$ & $300\pm 11$ & $324\pm 12$ & $280\pm 10$ & $92.8\pm 4.3$ & $35.1\pm 2.7$ & $20.5\pm 2.4$ \\
 & Data & $1003$ & $300$ & $309$ & $291$ & $85$ & $40$ & $19$ \\
\multirow{2}{*}{\Reg{SR3}{tags}{ISR}} & SM Processes & $709\pm 40$ & $163\pm 9$ & $161\pm 9$ & $125\pm 8$ & $35.5\pm 2.7$ & $5.5\pm 0.7$ & $0.72\pm 0.61$ \\
 & Data & $707$ & $170$ & $157$ & $137$ & $37$ & $7$ & $2$ \\
\multirow{2}{*}{\Reg{SR3}{0tag}{ISR}} & SM Processes & $425\pm 17$ & $89.1\pm 4.5$ & $87.7\pm 4.7$ & $69.4\pm 3.9$ & $16.4\pm 1.3$ & $4.1\pm 0.6$ & $1.1\pm 0.5$ \\
 & Data & $404$ & $99$ & $106$ & $70$ & $14$ & $4$ & $2$ \\
\multirow{2}{*}{\Reg{SR4}{tags}{ISR}} & SM Processes & $225\pm 19$ & $46.1\pm 4.3$ & $48.7\pm 4.4$ & $42.7\pm 4.5$ & $12.1\pm 1.6$ & $1.5\pm 0.3$ & $0.51\pm 0.20$ \\
 & Data & $209$ & $51$ & $52$ & $51$ & $8$ & $0$ & $1$ \\
\multirow{2}{*}{\Reg{SR4}{0tag}{ISR}} & SM Processes & $154\pm 9$ & $28.7\pm 2.3$ & $31.9\pm 2.3$ & $24.6\pm 2.2$ & $6.9\pm 0.9$ & $1.9\pm 0.4$ & $1.5\pm 0.4$ \\
 & Data & $134$ & $37$ & $42$ & $33$ & $4$ & $3$ & $2$ \\[\cmsTabSkip]
SF events &  &  &  &  &  &  &  & \\
\multirow{2}{*}{\Reg{SR1}{tags}{}} & SM Processes & $8416\pm 249$ & $3331\pm 97$ & $4194\pm 118$ & $4303\pm 106$ & $2101\pm 50$ & $474\pm 15$ & $20.9\pm 1.3$ \\
 & Data & $8405$ & $3351$ & $4211$ & $4232$ & $2173$ & $465$ & $25$ \\
\multirow{2}{*}{\Reg{SR1}{0tag}{}} & SM Processes & $3174\pm 68$ & $1062\pm 27$ & $1263\pm 30$ & $1238\pm 26$ & $588\pm 15$ & $277\pm 9$ & $90.3\pm 4.3$ \\
 & Data & $3110$ & $1075$ & $1221$ & $1284$ & $602$ & $301$ & $99$ \\
\multirow{2}{*}{\Reg{SR2}{tags}{}} & SM Processes & $2260\pm 107$ & $749\pm 34$ & $844\pm 38$ & $758\pm 34$ & $257\pm 13$ & $57.0\pm 3.5$ & $10.9\pm 0.9$ \\
 & Data & $2308$ & $742$ & $811$ & $770$ & $246$ & $62$ & $15$ \\
\multirow{2}{*}{\Reg{SR2}{0tag}{}} & SM Processes & $942\pm 29$ & $259\pm 9$ & $282\pm 10$ & $250\pm 9$ & $88.4\pm 4.6$ & $42.8\pm 2.3$ & $51.8\pm 3.5$ \\
 & Data & $972$ & $246$ & $313$ & $258$ & $87$ & $56$ & $49$ \\
\multirow{2}{*}{\Reg{SR3}{tags}{ISR}} & SM Processes & $573\pm 32$ & $138\pm 8$ & $137\pm 8$ & $105\pm 6$ & $31.8\pm 2.4$ & $5.7\pm 0.7$ & $1.3\pm 0.2$ \\
 & Data & $542$ & $158$ & $136$ & $103$ & $26$ & $3$ & $2$ \\
\multirow{2}{*}{\Reg{SR3}{0tag}{ISR}} & SM Processes & $372\pm 15$ & $73.2\pm 3.7$ & $76.1\pm 4.0$ & $57.0\pm 3.2$ & $16.2\pm 1.4$ & $5.6\pm 0.8$ & $3.9\pm 0.5$ \\
 & Data & $368$ & $76$ & $76$ & $57$ & $13$ & $3$ & $1$ \\
\multirow{2}{*}{\Reg{SR4}{tags}{ISR}} & SM Processes & $188\pm 16$ & $36.7\pm 3.5$ & $39.3\pm 3.6$ & $33.7\pm 3.6$ & $9.6\pm 1.3$ & $1.5\pm 0.3$ & $0.59\pm 0.19$ \\
 & Data & $179$ & $38$ & $54$ & $37$ & $3$ & $2$ & $2$ \\
\multirow{2}{*}{\Reg{SR4}{0tag}{ISR}} & SM Processes & $137\pm 8$ & $25.5\pm 2.1$ & $25.8\pm 1.9$ & $21.0\pm 2.1$ & $6.1\pm 0.8$ & $3.6\pm 0.5$ & $3.0\pm 0.6$ \\
 & Data & $133$ & $27$ & $29$ & $15$ & $4$ & $5$ & $4$ \\
\hline
\end{tabular}
}
\label{tab:PostfitT2ttll_Short}
\end{table}

\begin{figure}[hbtp]
  \centering
  \includegraphics[width=0.48\textwidth]{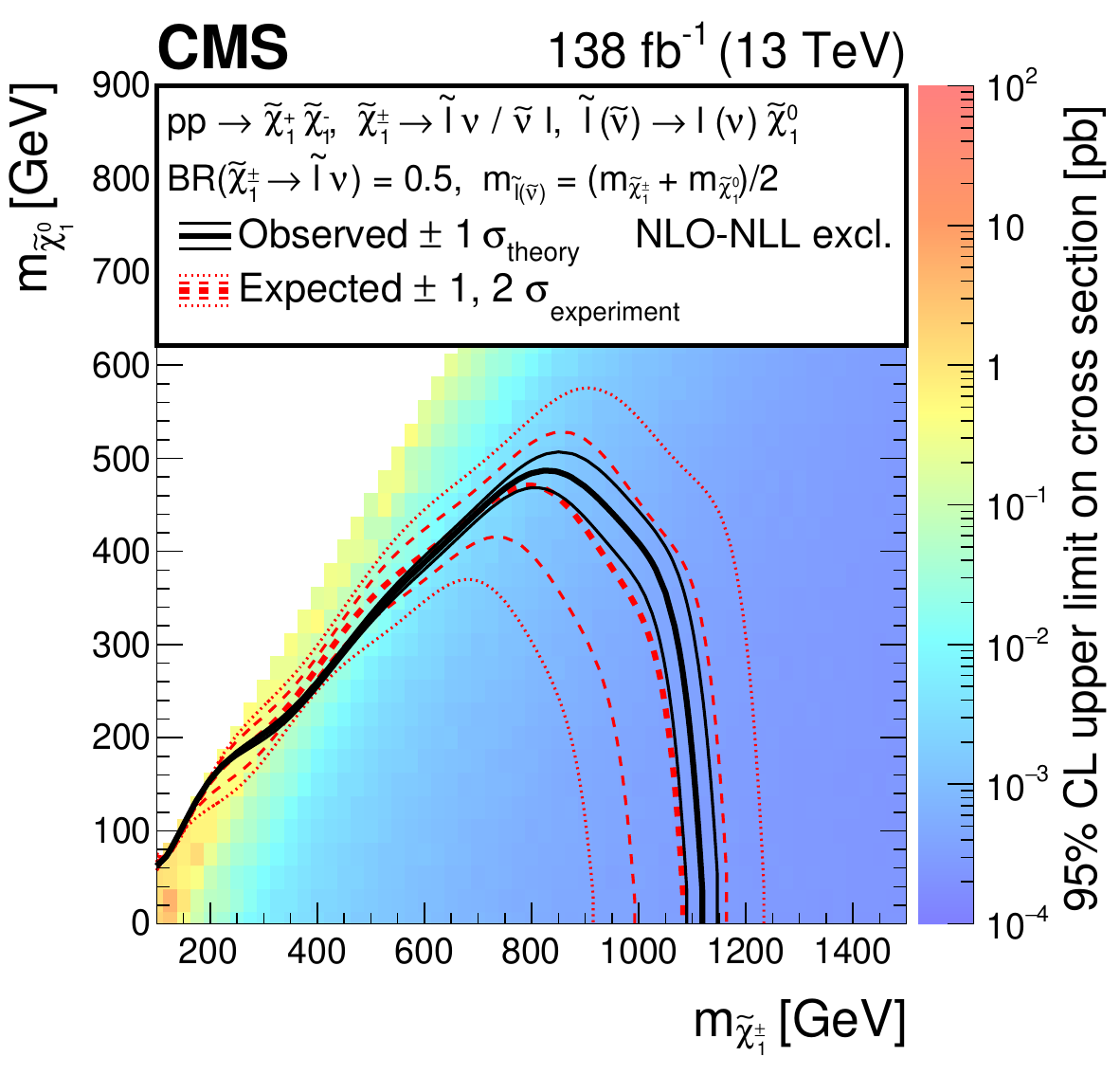}
  \includegraphics[width=0.48\textwidth]{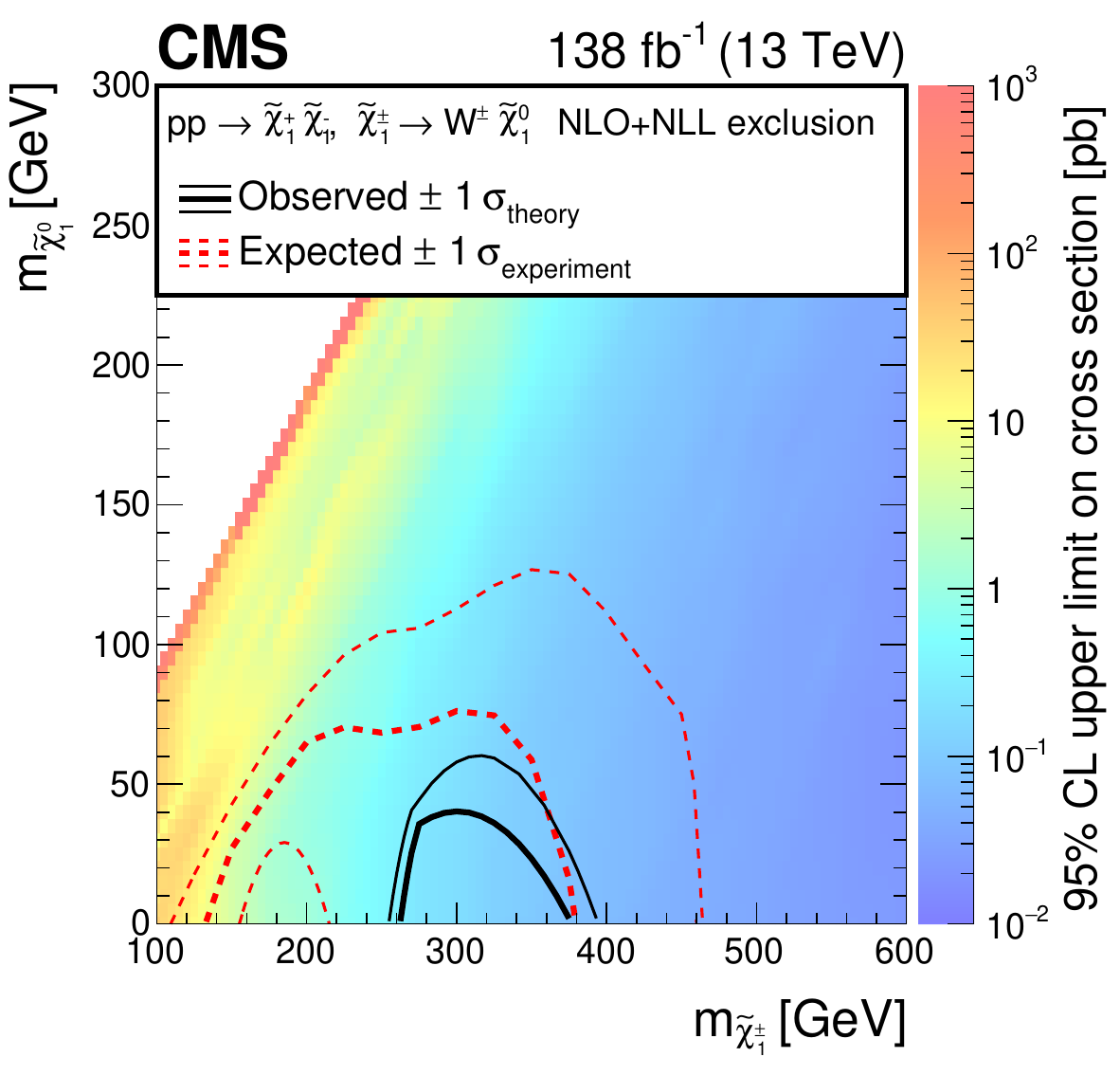}
	\caption{Upper limits at 95\% \CL on the chargino pair production cross section as a function of the chargino and neutralino masses. The plot on the left shows the results when the chargino undergoes a cascade decay \TChipmDecay. The plot on the right gives the limits assuming that the chargino decays into a neutralino and a \PW boson ($\PSGcpmDo\to\PWpm\PSGczDo$). Exclusion regions in the plane (\invM{\PSGcpmDo}, \invM{\PSGczDo}) are determined by comparing the upper limits with the NLO+NLL production cross sections. The thick dashed red line shows the expected exclusion region. The thin dashed (dotted) red lines show the $1\sigma$ ($2\sigma$) variation of the expected exclusion regions due to the experimental uncertainties. The thick black line shows the observed exclusion region, while the thin black lines show the variation of the observed exclusion regions due to the theoretical uncertainties in the production cross section.}
    \label{Fig:TChipmLimit}
\end{figure}

The 95\% \CL upper limits on the chargino pair production cross sections with the chargino decaying into sleptons are shown in Fig.~\ref{Fig:TChipmLimit} (left). The \TChipmSlep and \TChipmSneu decay chains are given a \BF of 50\% each, and the sleptons are assumed to be degenerate, with a mass equal to the average of the chargino and neutralino masses. 
Observed and expected exclusion regions in the (\invM{\PSGcpmDo}, \invM{\PSGczDo}) plane are also determined, from a comparison of the upper limits with the $\pp\to\PSGcpDo\PSGcmDo$ production cross sections.
Masses are excluded up to values of about 1120\GeV for the chargino and 490\GeV for the neutralino. 
These results complement the ones in the low mass region ($\invM{\PSGcpmDo}<480\GeV$) from the compressed mass spectra search~\cite{CMSCompressedSUS} and extend the previous CMS constraints~\cite{CMSTChipm13TeV} in the high mass region by 320 (170)\GeV for the chargino (neutralino).
Figure~\ref{Fig:TChipmLimit} (right) shows the results obtained when the chargino is assumed to decay into a \PW boson and the lightest neutralino. More limited sensitivity is found in this scenario, because of the relatively small branching fraction for the leptonic decay of the \PW boson. 
In general, the observed limits agree with the expected limits within the experimental uncertainties. 
A localized deviation from the expected sensitivity is observed for chargino masses of 200--400\GeV, driven primarily by SF events with $100<\mtll<160\GeV$ in SR2.
The corresponding exclusion regions partially cover the gap between those set by CMS in compressed mass spectra ($\invM{\PSGcpmDo}<200\GeV$~\cite{CMSCompressedSUS}) and in hadronic final states ($290<\invM{\PSGcpmDo}<670\GeV$~\cite{CMSTChiWWHadrWX}).                                                                                                                                                        

\begin{figure}[hbtp]
  \centering
  \includegraphics[width=0.48\textwidth]{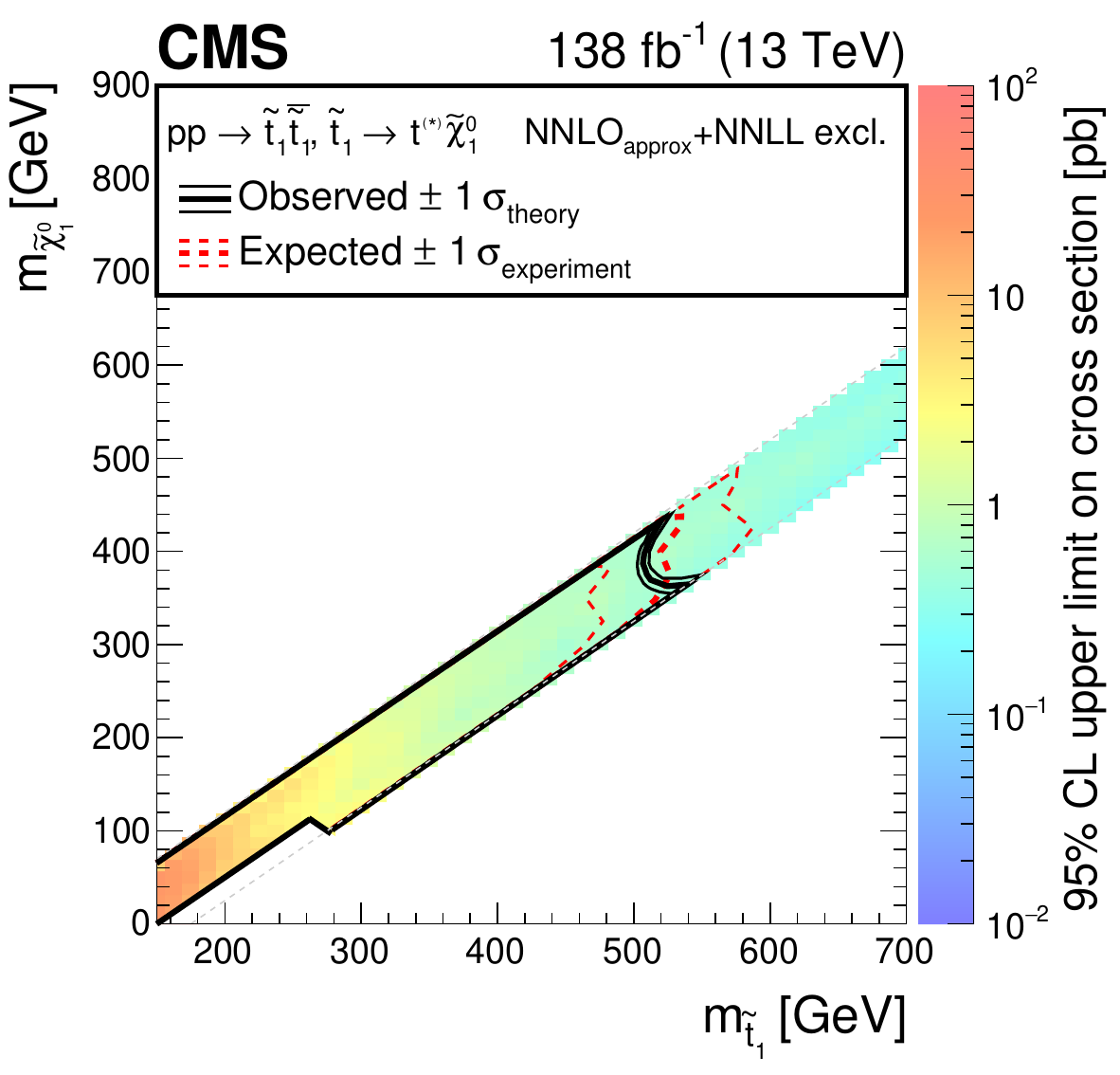}
  \includegraphics[width=0.48\textwidth]{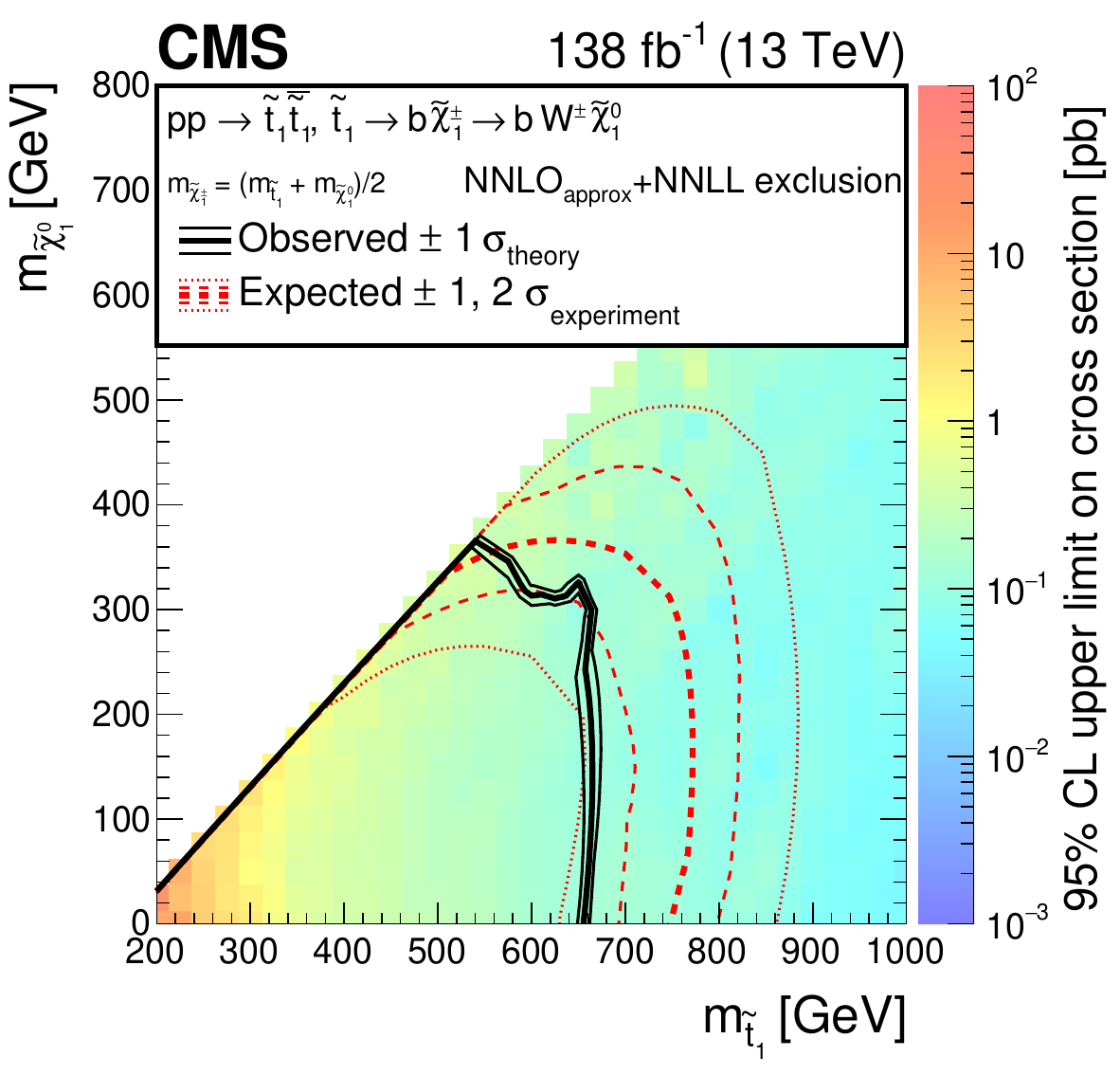}
	\caption{Upper limits at 95\% \CL on the top squark pair production cross section as a function of the top squark and neutralino masses. The plot on the left shows the results when the top squark decays into a top quark and a neutralino are assumed. The two diagonal gray dashed lines enclose the compressed region where $\invM\PW<\invM{\stone}-\invM{\PSGczDo}\lesssim \invM{\PQt}$. No result is given in the region with $\abs{\invM{\stone}-\invM{\PSGczDo}-\invM{\PQt}}<25\GeV$ and $\invM{\PSGczDo}<100\GeV$ as explained in the text. The plot on the right gives the limits for top squarks decaying into a bottom quark and a chargino, with the latter decaying into a \PW boson and a neutralino. The mass of the chargino is assumed to be equal to the average of the top squark and neutralino masses. Exclusion regions in the plane (\invM{\stone}, \invM{\PSGczDo}) are determined by comparing the upper limits with the NNLO$_{\mathrm{approx}}$+NNLL production cross sections. The thick dashed red line shows the expected exclusion region. The thin dashed (dotted) red lines show the $1\sigma$ ($2\sigma$) variation of the expected exclusion regions due to the experimental uncertainties. The thick black line shows the observed exclusion region, while the thin black lines show the variation of the observed exclusion regions due to the theoretical uncertainties in the production cross section.}
    \label{Fig:T2Limit}
\end{figure}

Figure~\ref{Fig:T2Limit} shows the observed and expected 95\% \CL upper limits on the top squark production cross section for the SMS where the top squark decays into top quark and neutralino (left) or via a chargino (right).
Also shown are the expected and observed exclusion regions when assuming top squark pair production cross sections.
In the compressed-mass region where the analysis has been optimized, top squark masses are excluded up to about 500--540 (540)\GeV for the $\stone\to\PGt^*\PSGczDo$ ($\,\stone\to \PQb\PSGcpmDo\to \PQb\PW\PSGczDo$) decay mode.
These results extend by 100\GeV the previous constraints set by CMS on the top squark mass for compressed mass scenarios with dilepton final states.
No result is presented for the scenario with the top squark decaying into top quark and neutralino when $\abs{\invM{\stone}-\invM{\PSGczDo}-\invM{\PQt}}<25\GeV$ and $\invM{\PSGczDo}<100\GeV$. In this region, signal events are very similar to \ttbar ones, with little contribution from the neutralinos to \ptvecmiss. A dedicated search has been developed by CMS to constrain the top squark pair production in this phase space~\cite{CMSStopComb}, based on a detailed modeling of the kinematic differences between top squark and top quark pair production.
In the region with large mass splitting between the top squark and the neutralino, a reduction in the exclusion reach relative to the expectation is observed, associated primarily with DF events with $\mtll>160\GeV$ in \Reg{SR3}{tags}{ISR} and SF events with $\mtll>160\GeV$ in \Reg{SR4}{tags}{ISR}.

\begin{figure}[hbtp]
  \centering
  \includegraphics[width=0.48\textwidth]{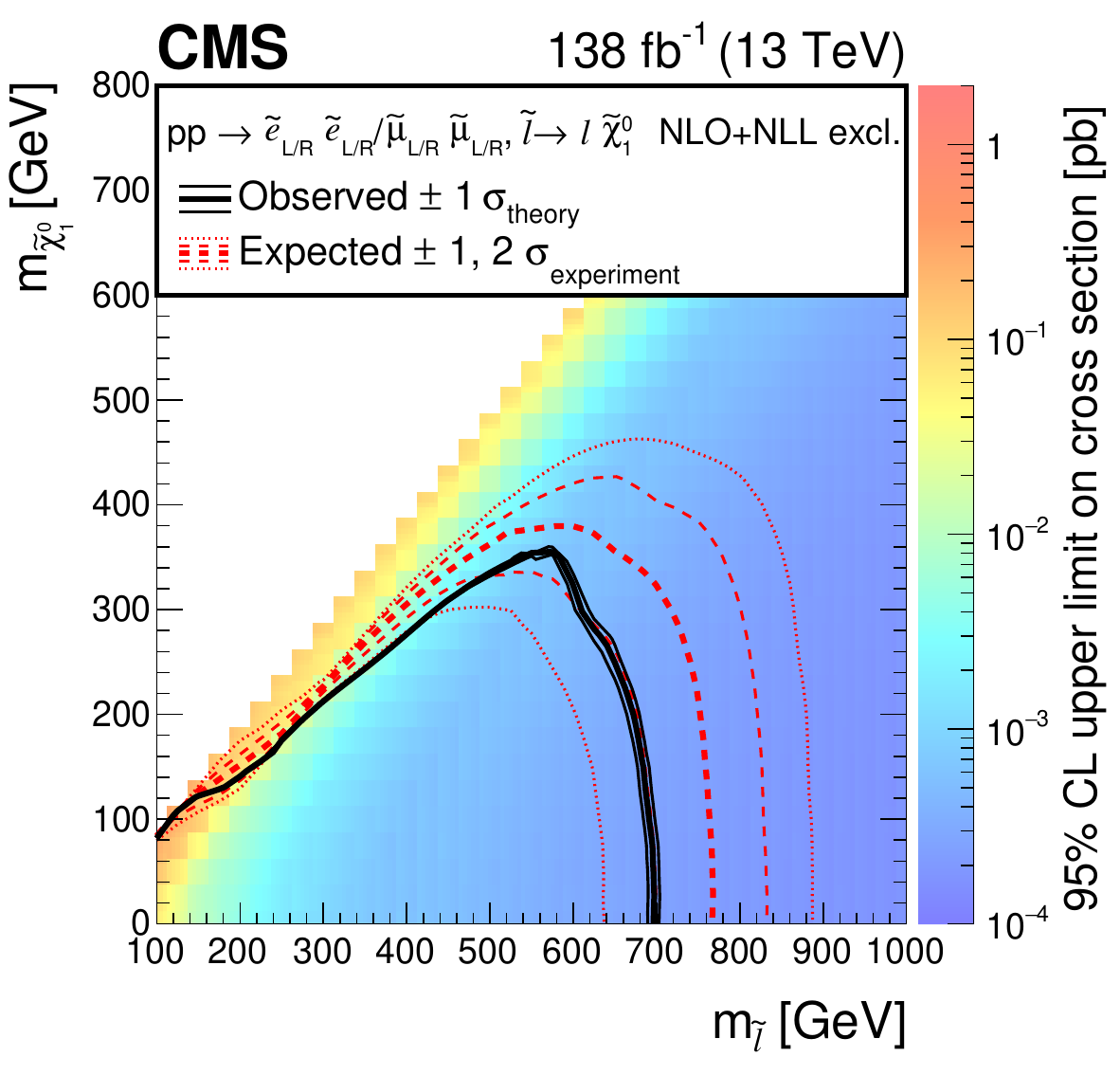}
	\caption{Upper limits at 95\% \CL on the slepton pair production cross section as a function of the slepton and neutralino masses. Exclusion regions in the plane (\invM{\PSell}, \invM{\PSGczDo}) are determined by comparing the upper limits with the NLO+NLL production cross sections. The thick dashed red line shows the expected exclusion region. The thin dashed (dotted) red lines show the $1\sigma$ ($2\sigma$) variation of the expected exclusion regions due to the experimental uncertainties. The thick black line shows the observed exclusion region, while the thin black lines show the variation of the observed exclusion regions due to the theoretical uncertainties in the production cross section.}
  \label{Fig:TSlepLimit}
\end{figure}

Finally, 95\% CL upper limits in the (\invM{\PSell}, \invM{\PSGczDo}) plane on slepton pair production cross sections are shown in Fig.~\ref{Fig:TSlepLimit}, along with the observed and expected exclusion regions. 
Masses are excluded up to values of about 700\GeV for the slepton and 360\GeV for the neutralino. 
In the region $\invM{\PSell}<300\GeV$, this search extends previous CMS boundaries set in a similar phase space~\cite{edgerun2} by up to 60\GeV, complementing the constraints set in the compressed mass spectra search~\cite{CMSCompressedSUS}.

\section{Summary}\label{sec:summary}

A search for beyond the standard model phenomena in final states with two isolated leptons of opposite charge and high missing transverse momentum has been performed using proton-proton collisions at a center-of-mass energy of 13\TeV. The data sample has been collected with the CMS detector at the LHC, and corresponds to an integrated luminosity of 138\fbinv.
The selected events have been categorized in several search regions according to the flavor of the leptons, the value of the missing transverse momentum, and the multiplicities of jets and \PQb-tagged jets.
The observed data were found to be consistent with the predictions of the standard model in all the search regions. 
Upper limits at 95\% confidence level (CL) have been set on the production cross sections for several supersymmetric models with simplified mass spectra, and have been interpreted as 95\% CL exclusion regions in the plane defined by the masses of the promptly produced supersymmetric particle and the lightest neutralino, taken to be the lightest supersymmetric particle. 

The exclusion regions depend on the specific supersymmetric mass spectra assumed in the interpretation.
When the three slepton generations are taken to be degenerate, with their masses being the average of those of the chargino and the neutralino, the chargino's decay mode through intermediate sleptons opens up. In this decay scenario, chargino masses up to 1120\GeV are excluded for a massless neutralino, while masses of the neutralino are excluded up to 490\GeV for a chargino mass around 820\GeV.
For a decay mode involving a neutralino and a \PW boson, chargino masses of 260--380\GeV are excluded when the neutralino mass is below 25\GeV.                              
Upper limits on the production of a pair of top squarks were derived considering the decay of the top squark to a top quark and a lightest neutralino. 
In the compressed region where the mass splitting between the top squark and the neutralino lies between the \PW boson and top quark masses, top squark masses up to 500--540\GeV are excluded. 
A model with a chargino that has a mass intermediate between the top squark and the lightest neutralino masses has also been investigated, where the top squark decays into a bottom quark and a chargino that subsequently decays into a \PW boson and a neutralino. In this scenario, top squark masses up to 540\GeV are excluded in the compressed region where the mass splitting between the chargino and the neutralino is just above the \PW boson mass.

Lastly, slepton pair production has been considered, excluding slepton masses up to 700\GeV for a massless neutralino.
For slepton masses below 150\GeV, slepton-neutralino mass splittings greater than 30\GeV are excluded.

These results constitute the tightest constraints currently available for chargino pair production with decay modes via intermediate sleptons.

\begin{acknowledgments}
We congratulate our colleagues in the CERN accelerator departments for the excellent performance of the LHC and thank the technical and administrative staffs at CERN and at other CMS institutes for their contributions to the success of the CMS effort. In addition, we gratefully acknowledge the computing centers and personnel of the Worldwide LHC Computing Grid and other centers for delivering so effectively the computing infrastructure essential to our analyses. Finally, we acknowledge the enduring support for the construction and operation of the LHC, the CMS detector, and the supporting computing infrastructure provided by the following funding agencies: SC (Armenia), BMFWF and FWF (Austria); FNRS and FWO (Belgium); CNPq, CAPES, FAPERJ, FAPERGS, and FAPESP (Brazil); MES and BNSF (Bulgaria); CERN; CAS, MoST, and NSFC (China); MINCIENCIAS (Colombia); MSES and CSF (Croatia); RIF (Cyprus); SENESCYT (Ecuador); ERC PRG and PSG, TARISTU24-TK10 and MoER TK202 (Estonia); Academy of Finland, MEC, and HIP (Finland); CEA and CNRS/IN2P3 (France); SRNSF (Georgia); BMFTR, DFG, and HGF (Germany); GSRI (Greece); MATE and NKFIH (Hungary); DAE and DST (India); IPM (Iran); SFI (Ireland); INFN (Italy); MSIT and NRF (Republic of Korea); MES (Latvia); LMTLT (Lithuania); MOE and UM (Malaysia); BUAP, CINVESTAV, CONACYT, LNS, SEP, and UASLP-FAI (Mexico); MOS (Montenegro); MBIE (New Zealand); PAEC (Pakistan); MSHE, NSC, and NAWA (Poland); FCT (Portugal); MESTD (Serbia); MICIU/AEI and PCTI (Spain); MOSTR (Sri Lanka); Swiss Funding Agencies (Switzerland); NSTC (Taiwan); MHESI (Thailand); TUBITAK and TENMAK (T\"{u}rkiye); NASU (Ukraine); STFC (United Kingdom); DOE and NSF (USA).

\begin{sloppypar}
\setlength\emergencystretch{\hsize}
\hyphenation{Rachada-pisek} Individuals have received support from the Marie-Curie program and the European Research Council and Horizon 2020 Grant, contract Nos.\ 675440, 724704, 752730, 758316, 765710, 824093, 101115353, 101002207, 101001205, and COST Action CA16108 (European Union); the Leventis Foundation; the Alfred P.\ Sloan Foundation; the Alexander von Humboldt Foundation; the Science Committee, project no. 22rl-037 (Armenia); the Fonds pour la Formation \`a la Recherche dans l'Industrie et dans l'Agriculture (FRIA) and Fonds voor Wetenschappelijk Onderzoek contract No. 1228724N (Belgium); the Beijing Municipal Science \& Technology Commission, No. Z191100007219010, the Fundamental Research Funds for the Central Universities, the Ministry of Science and Technology of China under Grant No. 2023YFA1605804, the Natural Science Foundation of China under Grant No. 12535004, and USTC Research Funds of the Double First-Class Initiative No.\ YD2030002017 (China); the Ministry of Education, Youth and Sports (MEYS) of the Czech Republic; the Shota Rustaveli National Science Foundation (Georgia); the Deutsche Forschungsgemeinschaft (DFG), among others, under Germany's Excellence Strategy -- EXC 2121 ``Quantum Universe" -- 390833306, and under project number 400140256 - GRK2497; the Hellenic Foundation for Research and Innovation (HFRI), Project Number 2288 (Greece); the Hungarian Academy of Sciences, the New National Excellence Program - \'UNKP, the NKFIH research grants K 131991, K 138136, K 143460, K 143477, K 147557, K 146913, K 146914, K 147048, TKP2021-NKTA-64, 2026-4.1.1-MISSZI\'{O}K-2026-00005 and 2025-1.1.5-NEMZ\_KI-2025-00004, and MATE KKP and KKPCs Research Excellence and Flagship Research Groups grants (Hungary); the Council of Science and Industrial Research, India; ICSC -- National Research Center for High Performance Computing, Big Data and Quantum Computing, FAIR -- Future Artificial Intelligence Research, and CUP I53D23001070006 (Mission 4 Component 1), funded by the NextGenerationEU program, the Italian Ministry of University and Research (MUR) under Bando PRIN 2022 -- CUP I53C24002390006, PRIN PRIMULA 2022RBYK7T (Italy); the Latvian Council of Science; the Ministry of Science and Higher Education, project no. 2022/WK/14, and the National Science Center, contracts Opus 2021/41/B/ST2/01369, 2021/43/B/ST2/01552, 2023/49/B/ST2/03273, and the NAWA contract BPN/PPO/2021/1/00011 (Poland); the Funda\c{c}\~ao para a Ci\^encia e a Tecnologia (Portugal); the National Priorities Research Program by Qatar National Research Fund; MICIU/AEI/10.13039/501100011033, ERDF/EU, ``European Union NextGenerationEU/PRTR", projects PID2022-142604OB-C21, PID2022-139519OB-C21, PID2023-147706NB-I00, PID2023-148896NB-I00, PID2023-146983NB-I00, PID2023-147115NB-I00, PID2023-148418NB-C41, PID2023-148418NB-C42, PID2023-148418NB-C43, PID2023-148418NB-C44, PID2024-158190NB-C22, RYC2021-033305-I, RYC2024-048719-I, CNS2023-144781, CNS2024-154769 and Plan de Ciencia, Tecnolog{\'i}a e Innovaci{\'o}n de Asturias, Spain; the Chulalongkorn Academic into Its 2nd Century Project Advancement Project, the National Science, Research and Innovation Fund program IND\_FF\_68\_369\_2300\_097, and the Program Management Unit for Human Resources \& Institutional Development, Research and Innovation, grant B39G680009 (Thailand); the Eric \& Wendy Schmidt Fund for Strategic Innovation through the CERN Next Generation Triggers project under grant agreement number SIF-2023-004; the Kavli Foundation; the Nvidia Corporation; the SuperMicro Corporation; the Welch Foundation, contract C-1845; and the Weston Havens Foundation (USA).
\end{sloppypar}
\end{acknowledgments}\section*{Data availability} Release and preservation of data used by the CMS Collaboration as the basis for publications is guided by the  \href{https://doi.org/10.7483/OPENDATA.CMS.1BNU.8V1W}{CMS data preservation, re-use and open access policy}.

\bibliography{auto_generated}

@ARTICLE{DarkMatter1,
	AUTHOR=	"Bertone, G. and Hooper, D. and Silk, J.",
	TITLE=	"Particle dark matter: evidence, candidates and constraints",
	JOURNAL=	"Phys. Rept.",
	VOLUME=	"405",
	PAGES=	"279",
	DOI=	"10.1016/j.physrep.2004.08.031",
	YEAR=	"2005",
	EPRINT=	"hep-ph/0404175",
	ARCHIVEPREFIX=	"arXiv",
	PRIMARYCLASS=	"hep-ph",
	SLACCITATION=	"%%CITATION = ARXIV:0404175;%%",
}

@ARTICLE{DarkMatter2,
	AUTHOR=	"Feng, J. L. ",
	TITLE=	"Dark matter candidates from particle physics and methods of detection",
	JOURNAL=	"Ann. Rev. Astron. Astrophys.",
	VOLUME=	"48",
	PAGES=	"495",
	DOI=	"10.1146/annurev-astro-082708-101659",
	YEAR=	"2010",
	EPRINT=	"1003.0904",
	ARCHIVEPREFIX=	"arXiv",
	PRIMARYCLASS=	"astro-ph",
	SLACCITATION=	"%%CITATION = ARXIV:1003.0904;%%",
}

@ARTICLE{DarkMatter3,
	AUTHOR=	"Porter, T. A. and Johnson, R. P. and Graham, P. W.",
	TITLE=	"Dark matter searches with astroparticle data",
	JOURNAL=	"Ann. Rev. Astron. Astrophys.",
	VOLUME=	"49",
	PAGES=	"155",
	DOI=	"10.1146/annurev-astro-081710-102528",
	YEAR=	"2011",
	EPRINT=	"1104.2836",
	ARCHIVEPREFIX=	"arXiv",
	PRIMARYCLASS=	"astro-ph",
	SLACCITATION=	"%%CITATION = ARXIV:1104.2836;%%",
}

@ARTICLE{PhysRevD.13.3333,
	TITLE=	"Symmetry breaking and scalar bosons",
	AUTHOR=	"Gildener, Eldad and Weinberg, Steven",
	JOURNAL=	"Phys. Rev. D",
	VOLUME=	"13",
	PAGES=	"3333",
	YEAR=	"1976",
	DOI=	"10.1103/PhysRevD.13.3333",
}

@ARTICLE{Hierarchy1,
	AUTHOR=	"'t Hooft, G.",
	TITLE=	"Naturalness, chiral symmetry, and spontaneous chiral symmetry breaking",
	JOURNAL=	"NATO Sci. Ser. B",
	VOLUME=	"59",
	PAGES=	"135",
	DOI=	"10.1007/978-1-4684-7571-5_9",
	YEAR=	"1980",
}

@ARTICLE{Ramond:1971gb,
	AUTHOR=	"Ramond, Pierre",
	TITLE=	"Dual theory for free fermions",
	JOURNAL=	"Phys. Rev. D",
	VOLUME=	"3",
	PAGES=	"2415",
	DOI=	"10.1103/PhysRevD.3.2415",
	YEAR=	"1971",
	REPORTNUMBER=	"FERMILAB-PUB-70-008-T, FERMILAB-PUB-70-008-THY, THY-8",
	SLACCITATION=	"%%CITATION = PHRVA,D3,2415;%%",
}

@ARTICLE{Golfand:1971iw,
	AUTHOR=	"Golfand, Yu. A. and Likhtman, E. P.",
	TITLE=	"Extension of the algebra of {P}oincar\'{e} group generators and violation of {P} invariance",
	DOI=	"10.1142/9789814542340_0001",
	JOURNAL=	"JETP Lett.",
	VOLUME=	"13",
	PAGES=	"323",
	YEAR=	"1971",
	SLACCITATION=	"%%CITATION = JTPLA,13,323;%%",
}

@ARTICLE{Neveu:1971rx,
	AUTHOR=	"Neveu, A. and Schwarz, J. H.",
	TITLE=	"Factorizable dual model of pions",
	JOURNAL=	"Nucl. Phys. B",
	VOLUME=	"31",
	PAGES=	"86",
	DOI=	"10.1016/0550-3213(71)90448-2",
	YEAR=	"1971",
	SLACCITATION=	"%%CITATION = NUPHA,B31,86;%%",
}

@ARTICLE{Volkov:1972jx,
	AUTHOR=	"Volkov, D. V. and Akulov, V. P.",
	TITLE=	"Possible universal neutrino interaction",
	JOURNAL=	"JETP Lett.",
	VOLUME=	"16",
	PAGES=	"438",
	YEAR=	"1972",
	URL=	"http://jetpletters.ru/ps/0/article_26864.shtml",
	SLACCITATION=	"%%CITATION = JTPLA,16,438;%%",
}

@ARTICLE{Wess:1973kz,
	AUTHOR=	"Wess, J. and Zumino, B.",
	TITLE=	"A {L}agrangian model invariant under supergauge transformations",
	JOURNAL=	"Phys. Lett. B",
	VOLUME=	"49",
	PAGES=	"52",
	DOI=	"10.1016/0370-2693(74)90578-4",
	YEAR=	"1974",
	REPORTNUMBER=	"CERN-TH-1794",
	SLACCITATION=	"%%CITATION = PHLTA,B49,52;%%",
}

@ARTICLE{Wess:1974tw,
	AUTHOR=	"Wess, J. and Zumino, B.",
	TITLE=	"Supergauge transformations in four dimensions",
	JOURNAL=	"Nucl. Phys. B",
	VOLUME=	"70",
	PAGES=	"39",
	DOI=	"10.1016/0550-3213(74)90355-1",
	YEAR=	"1974",
	SLACCITATION=	"%%CITATION = NUPHA,B70,39;%%",
}

@ARTICLE{Fayet:1974pd,
	AUTHOR=	"Fayet, Pierre",
	TITLE=	"Supergauge invariant extension of the {H}iggs mechanism and a model for the electron and its neutrino",
	JOURNAL=	"Nucl. Phys. B",
	VOLUME=	"90",
	PAGES=	"104",
	DOI=	"10.1016/0550-3213(75)90636-7",
	YEAR=	"1975",
	SLACCITATION=	"%%CITATION = NUPHA,B90,104;%%",
}

@ARTICLE{Nilles:1983ge,
	AUTHOR=	"Nilles, Hans Peter",
	TITLE=	"Supersymmetry, supergravity and particle physics",
	JOURNAL=	"Phys. Rep.",
	VOLUME=	"110",
	PAGES=	"1",
	DOI=	"10.1016/0370-1573(84)90008-5",
	YEAR=	"1984",
	SLACCITATION=	"%%CITATION = PRPLC,110,1;%%",
}

@ARTICLE{SUSYPrimer,
	AUTHOR=	"Martin, Stephen P.",
	TITLE=	"A supersymmetry primer",
	JOURNAL=	"Adv. Ser. Direct. High Energy Phys.",
	VOLUME=	"21",
	PAGES=	"1",
	DOI=	"10.1142/9789814307505_0001",
	YEAR=	"2010",
	EPRINT=	"hep-ph/9709356",
	ARCHIVEPREFIX=	"arXiv",
	PRIMARYCLASS=	"hep-ph",
	SLACCITATION=	"%%CITATION = ARXIV:hep-ph/9709356;%%",
}

@ARTICLE{Farrar:1978xj,
	AUTHOR=	"Farrar, Glennys R. and Fayet, Pierre",
	TITLE=	"Phenomenology of the production, decay, and detection of new hadronic states associated with supersymmetry",
	JOURNAL=	"Phys. Lett. B",
	VOLUME=	"76",
	PAGES=	"575",
	DOI=	"10.1016/0370-2693(78)90858-4",
	YEAR=	"1978",
	REPORTNUMBER=	"CALT-68-648",
	SLACCITATION=	"%%CITATION = PHLTA,B76,575;%%",
}

@ARTICLE{Hierarchy2,
	AUTHOR=	"Witten, E.",
	TITLE=	"Dynamical breaking of supersymmetry",
	JOURNAL=	"Nucl. Phys. B",
	VOLUME=	"188",
	PAGES=	"513",
	DOI=	"10.1016/0550-3213(81)90006-7",
	YEAR=	"1981",
}

@ARTICLE{Hierarchy3,
	AUTHOR=	"Dimopoulos, S. and Georgi, H.",
	TITLE=	"Softly broken supersymmetry and {SU}(5)",
	JOURNAL=	"Nucl. Phys. B",
	VOLUME=	"193",
	PAGES=	"150",
	DOI=	"10.1016/0550-3213(81)90522-8",
	YEAR=	"1981",
}

@ARTICLE{Hierarchy4,
	AUTHOR=	"{Kaul, R. K. and Majumdar, P.}",
	TITLE=	"Cancellation of quadratically divergent mass corrections in globally supersymmetric spontaneously broken gauge theories",
	JOURNAL=	"Nucl. Phys. B",
	VOLUME=	"199",
	PAGES=	"36",
	DOI=	"10.1016/0550-3213(82)90565-X",
	YEAR=	"1982",
}

@ARTICLE{Chatrchyan:2008zzk,
	AUTHOR=	"{CMS Collaboration}",
	TITLE=	"The {CMS} experiment at the {CERN} {LHC}",
	JOURNAL=	"JINST",
	VOLUME=	"3",
	YEAR=	"2008",
	PAGES=	"S08004",
	DOI=	"10.1088/1748-0221/3/08/S08004",
	SLACCITATION=	"%%CITATION = JINST,3,S08004;%%",
}

@ARTICLE{CMS-LUM-17-003,
	AUTHOR=	"{CMS Collaboration}",
	TITLE=	"Precision luminosity measurement in proton-proton collisions at $\sqrt{s} =$ 13 {TeV} in 2015 and 2016 at {CMS}",
	EPRINT=	"2104.01927",
	ARCHIVEPREFIX=	"arXiv",
	PRIMARYCLASS=	"hep-ex",
	REPORTNUMBER=	"CMS-LUM-17-003, CERN-EP-2021-033",
	YEAR=	"2021",
	DOI=	"10.1140/epjc/s10052-021-09538-2",
	JOURNAL=	"Eur. Phys. J. C",
	VOLUME=	"81",
	PAGES=	"800",
}

@UNPUBLISHED{CMS-LUM-20-001,
	AUTHOR=	"{CMS Collaboration}",
	TITLE=	"Precision luminosity measurement in proton-proton collisions at a center-of-mass energy of 13 {TeV} with the {CMS} detector at the {Large Hadron Collider}",
	EPRINT=	"2606.26832",
	ARCHIVEPREFIX=	"arXiv",
	PRIMARYCLASS=	"hep-ex",
	REPORTNUMBER=	"CMS-LUM-20-001, CERN-EP-2026-134",
	YEAR=	"2026",
}

@ARTICLE{CMSTChipm13TeV,
	AUTHOR=	"{CMS Collaboration}",
	TITLE=	"Searches for pair production of charginos and top squarks in final states with two oppositely charged leptons in proton-proton collisions at $\sqrt{s}=13$ {TeV}",
	JOURNAL=	"JHEP",
	VOLUME=	"11",
	YEAR=	"2018",
	PAGES=	"079",
	DOI=	"10.1007/JHEP11(2018)079",
	ARCHIVEPREFIX=	"arXiv",
	EPRINT=	"1807.07799",
	PRIMARYCLASS=	"hep-ex",
	SLACCITATION=	"%%CITATION = ARXIV:1807.07799;%%",
}

@ARTICLE{SMS1,
	AUTHOR=	"Alwall, J. and Schuster, P. and Toro, N.",
	TITLE=	"Simplified models for a first characterization of new physics at the {LHC}",
	JOURNAL=	"Phys. Rev. D",
	VOLUME=	"79",
	PAGES=	"075020",
	DOI=	"10.1103/PhysRevD.79.075020",
	YEAR=	"2009",
	EPRINT=	"0810.3921",
	ARCHIVEPREFIX=	"arXiv",
	PRIMARYCLASS=	"hep-ph",
	SLACCITATION=	"%%CITATION = ARXIV:0810.3921;%%",
}

@ARTICLE{SMS2,
	AUTHOR=	"Alwall, J. and Le, M.-P. and Lisanti, M. and Wacker, J. G.",
	TITLE=	"Model-independent jets plus missing energy searches",
	JOURNAL=	"Phys. Rev. D",
	VOLUME=	"79",
	PAGES=	"015005",
	DOI=	"10.1103/PhysRevD.79.015005",
	YEAR=	"2009",
	EPRINT=	"0809.3264",
	ARCHIVEPREFIX=	"arXiv",
	PRIMARYCLASS=	"hep-ph",
	SLACCITATION=	"%%CITATION = ARXIV:0809.3264;%%",
}

@ARTICLE{SMS3,
	AUTHOR=	"{LHC New Physics Working Group}",
	EDITOR=	"Arkani-Hamed, Nima and others",
	TITLE=	"{Simplified models for LHC new physics searches}",
	EPRINT=	"1105.2838",
	ARCHIVEPREFIX=	"arXiv",
	PRIMARYCLASS=	"hep-ph",
	REPORTNUMBER=	"SLAC-PUB-15045, FERMILAB-PUB-11-842-A-PPD",
	DOI=	"10.1088/0954-3899/39/10/105005",
	JOURNAL=	"J. Phys. G",
	VOLUME=	"39",
	PAGES=	"105005",
	YEAR=	"2012",
}

@ARTICLE{CMSTChiWWHadrWX,
	AUTHOR=	"{CMS Collaboration}",
	TITLE=	"{Search for electroweak production of charginos and neutralinos at $\sqrt{s}=13$ TeV in final states containing hadronic decays of WW, WZ, or WH and missing transverse momentum}",
	EPRINT=	"2205.09597",
	ARCHIVEPREFIX=	"arXiv",
	PRIMARYCLASS=	"hep-ex",
	REPORTNUMBER=	"CMS-SUS-21-002, CERN-EP-2022-031",
	DOI=	"10.1016/j.physletb.2022.137460",
	JOURNAL=	"Phys. Lett. B",
	VOLUME=	"842",
	PAGES=	"137460",
	YEAR=	"2023",
}

@ARTICLE{CMSCompressedSUS,
	TITLE=	"{A general search for supersymmetric particles in scenarios with compressed mass spectra using proton-proton collisions at $\sqrt{s}=13~\mathrm{TeV}$}",
	AUTHOR=	"{CMS Collaboration}",
	EPRINT=	"2508.13900",
	ARCHIVEPREFIX=	"arXiv",
	PRIMARYCLASS=	"hep-ex",
	JOURNAL=	"Phys. Rev. D",
	VOLUME=	"112",
	PAGES=	"112023",
	YEAR=	"2025",
	DOI=	"10.1103/b26z-zmpy",
	SLACCITATION=	"%%CITATION = ARXIV:2508.13900;%%",
}

@ARTICLE{ref:ATLASTChiWWSlep2lepCompressedRun2,
	AUTHOR=	"{ATLAS Collaboration}",
	TITLE=	"{Search for direct pair production of sleptons and charginos decaying to two leptons and neutralinos with mass splittings near the W-boson mass in $ \sqrt{s} $ = 13 TeV pp collisions with the ATLAS detector}",
	EPRINT=	"2209.13935",
	ARCHIVEPREFIX=	"arXiv",
	PRIMARYCLASS=	"hep-ex",
	REPORTNUMBER=	"CERN-EP-2022-132",
	DOI=	"10.1007/JHEP06(2023)031",
	JOURNAL=	"JHEP",
	VOLUME=	"06",
	PAGES=	"031",
	YEAR=	"2023",
}

@ARTICLE{ref:ATLASTChiSlep2lepRun2,
	AUTHOR=	"{ATLAS Collaboration}",
	TITLE=	"{Search for electroweak production of charginos and sleptons decaying into final states with two leptons and missing transverse momentum in $\sqrt{s}=13$ TeV $pp$ collisions using the ATLAS detector}",
	EPRINT=	"1908.08215",
	ARCHIVEPREFIX=	"arXiv",
	PRIMARYCLASS=	"hep-ex",
	REPORTNUMBER=	"CERN-EP-2019-106",
	DOI=	"10.1140/epjc/s10052-019-7594-6",
	JOURNAL=	"Eur. Phys. J. C",
	VOLUME=	"80",
	PAGES=	"123",
	YEAR=	"2020",
}

@ARTICLE{ref:ATLASTChiSlepTauhRun2,
	AUTHOR=	"{ATLAS Collaboration}",
	TITLE=	"{Search for electroweak production of supersymmetric particles in final states with two \ensuremath{\tau}-leptons in $ \sqrt{s} $ = 13 TeV pp collisions with the ATLAS detector}",
	EPRINT=	"2402.00603",
	ARCHIVEPREFIX=	"arXiv",
	PRIMARYCLASS=	"hep-ex",
	REPORTNUMBER=	"CERN-EP-2023-295",
	DOI=	"10.1007/JHEP05(2024)150",
	JOURNAL=	"JHEP",
	VOLUME=	"05",
	PAGES=	"150",
	INSTITUTION=	"CERN",
	YEAR=	"2024",
}

@ARTICLE{ATLASTChiWW1lep,
	AUTHOR=	"{ATLAS Collaboration}",
	TITLE=	"{Search for direct production of electroweakinos in final states with one lepton, jets and missing transverse momentum in pp collisions at $ \sqrt{s} $ = 13 TeV with the ATLAS detector}",
	EPRINT=	"2310.08171",
	ARCHIVEPREFIX=	"arXiv",
	PRIMARYCLASS=	"hep-ex",
	REPORTNUMBER=	"CERN-EP-2023-174",
	DOI=	"10.1007/JHEP12(2023)167",
	JOURNAL=	"JHEP",
	VOLUME=	"12",
	PAGES=	"167",
	YEAR=	"2023",
}

@ARTICLE{ATLASTChipmWWCombi,
	AUTHOR=	"{ATLAS Collaboration}",
	TITLE=	"{A statistical combination of ATLAS Run 2 searches for charginos and neutralinos at the LHC}",
	EPRINT=	"2402.08347",
	ARCHIVEPREFIX=	"arXiv",
	PRIMARYCLASS=	"hep-ex",
	REPORTNUMBER=	"CERN-EP-2024-018",
	JOURNAL=	"Phys. Rev. Lett.",
	VOLUME=	"133",
	YEAR=	"2024",
	PAGES=	"031802",
	DOI=	"10.1103/PhysRevLett.133.031802",
}

@ARTICLE{ref:t2tt2lep,
	AUTHOR=	"{CMS Collaboration}",
	TITLE=	"Search for top squarks and dark matter particles in opposite-charge dilepton final states at $\sqrt{s} = 13$ {TeV}",
	JOURNAL=	"Phys. Rev. D",
	VOLUME=	"97",
	YEAR=	"2018",
	PAGES=	"032009",
	EPRINT=	"1711.00752",
	ARCHIVEPREFIX=	"arXiv",
	PRIMARYCLASS=	"hep-ex",
	DOI=	"10.1103/PhysRevD.97.032009",
	SLACCITATION=	"%%CITATION = ARXIV:1711.00752;%%",
}

@ARTICLE{ref:t2tt2lepRun2,
	AUTHOR=	"{CMS Collaboration}",
	TITLE=	"{Search for top squark pair production using dilepton final states in ${\text {p}}{\text {p}}$ collision data collected at $\sqrt{s}=13\,\text {TeV} $}",
	EPRINT=	"2008.05936",
	ARCHIVEPREFIX=	"arXiv",
	PRIMARYCLASS=	"hep-ex",
	REPORTNUMBER=	"CMS-SUS-19-011, CERN-EP-2020-131",
	DOI=	"10.1140/epjc/s10052-020-08701-5",
	JOURNAL=	"Eur. Phys. J. C",
	VOLUME=	"81",
	PAGES=	"3",
	YEAR=	"2021",
}

@ARTICLE{ref:t2tt0lep,
	AUTHOR=	"{CMS Collaboration}",
	TITLE=	"Search for direct production of supersymmetric partners of the top quark in the all-jets final state in proton-proton collisions at $\sqrt{s} = 13$ {TeV}",
	JOURNAL=	"JHEP",
	VOLUME=	"10",
	YEAR=	"2017",
	PAGES=	"005",
	EPRINT=	"1707.03316",
	ARCHIVEPREFIX=	"arXiv",
	PRIMARYCLASS=	"hep-ex",
	DOI=	"10.1007/JHEP10(2017)005",
	SLACCITATION=	"%%CITATION = ARXIV:1707.03316;%%",
}

@ARTICLE{ref:t2tt1lep,
	AUTHOR=	"{CMS Collaboration}",
	TITLE=	"Search for top squark pair production in pp collisions at $\sqrt{s} = 13$ {TeV} using single lepton events",
	JOURNAL=	"JHEP",
	VOLUME=	"10",
	YEAR=	"2017",
	PAGES=	"019",
	EPRINT=	"1706.04402",
	ARCHIVEPREFIX=	"arXiv",
	PRIMARYCLASS=	"hep-ex",
	DOI=	"10.1007/JHEP10(2017)019",
	SLACCITATION=	"%%CITATION = ARXIV:1706.04402;%%",
}

@ARTICLE{ref:t2tt1lepRun2,
	AUTHOR=	"{CMS Collaboration}",
	TITLE=	"{Search for direct top squark pair production in events with one lepton, jets, and missing transverse momentum at 13 TeV with the CMS experiment}",
	EPRINT=	"1912.08887",
	ARCHIVEPREFIX=	"arXiv",
	PRIMARYCLASS=	"hep-ex",
	REPORTNUMBER=	"CMS-SUS-19-009, CERN-EP-2019-233",
	DOI=	"10.1007/JHEP05(2020)032",
	JOURNAL=	"JHEP",
	VOLUME=	"05",
	PAGES=	"032",
	YEAR=	"2020",
}

@ARTICLE{CMSStopComb,
	AUTHOR=	"{CMS Collaboration}",
	TITLE=	"Combined searches for the production of supersymmetric top quark partners in protonproton collisions at $\sqrt{s}=13$ {TeV}",
	JOURNAL=	"Eur. Phys. J. C",
	VOLUME=	"81",
	YEAR=	"2021",
	PAGES=	"970",
	EPRINT=	"2107.10892",
	ARCHIVEPREFIX=	"arXiv",
	PRIMARYCLASS=	"hep-ex",
	DOI=	"10.1140/epjc/s10052-021-09721-5",
	SLACCITATION=	"%%CITATION = ARXIV:2107.10892;%%",
}

@ARTICLE{ref:ATLAS0lep,
	AUTHOR=	"{ATLAS Collaboration}",
	TITLE=	"Search for a scalar partner of the top quark in the jets plus missing transverse momentum final state at $\sqrt{s}=13$ {TeV} with the {ATLAS} detector",
	JOURNAL=	"JHEP",
	VOLUME=	"12",
	YEAR=	"2017",
	PAGES=	"085",
	EPRINT=	"1709.04183",
	ARCHIVEPREFIX=	"arXiv",
	PRIMARYCLASS=	"hep-ex",
	DOI=	"10.1007/JHEP12(2017)085",
	SLACCITATION=	"%%CITATION = ARXIV:1709.04183;%%",
}

@ARTICLE{ref:ATLAS1lep,
	AUTHOR=	"{ATLAS Collaboration}",
	TITLE=	"Search for top-squark pair production in final states with one lepton, jets, and missing transverse momentum using $36\mathrm{fb}^{-1}$ of $\sqrt{s}=13$ {TeV} pp collision data with the {ATLAS} detector",
	EPRINT=	"1711.11520",
	ARCHIVEPREFIX=	"arXiv",
	PRIMARYCLASS=	"hep-ex",
	REPORTNUMBER=	"CERN-EP-2017-246",
	DOI=	"10.1007/JHEP06(2018)108",
	JOURNAL=	"JHEP",
	VOLUME=	"06",
	PAGES=	"108",
	YEAR=	"2018",
}

@ARTICLE{ATLASStop,
	AUTHOR=	"{ATLAS Collaboration}",
	TITLE=	"Search for direct top squark pair production in final states with two leptons in $\sqrt{s}=13$ {TeV} pp collisions with the {ATLAS} detector",
	JOURNAL=	"Eur. Phys. J. C",
	VOLUME=	"77",
	YEAR=	"2017",
	PAGES=	"898",
	EPRINT=	"1708.03247",
	ARCHIVEPREFIX=	"arXiv",
	PRIMARYCLASS=	"hep-ex",
	DOI=	"10.1140/epjc/s10052-017-5445-x",
	SLACCITATION=	"%%CITATION = ARXIV:1708.03247;%%",
}

@ARTICLE{ref:ATLAS0lepRun2,
	AUTHOR=	"{ATLAS Collaboration}",
	TITLE=	"{Search for a scalar partner of the top quark in the all-hadronic $t{\bar{t}}$ plus missing transverse momentum final state at $\sqrt{s}=13$ TeV with the ATLAS detector}",
	EPRINT=	"2004.14060",
	ARCHIVEPREFIX=	"arXiv",
	PRIMARYCLASS=	"hep-ex",
	REPORTNUMBER=	"CERN-EP-2020-044",
	DOI=	"10.1140/epjc/s10052-020-8102-8",
	JOURNAL=	"Eur. Phys. J. C",
	VOLUME=	"80",
	PAGES=	"737",
	YEAR=	"2020",
}

@ARTICLE{ref:ATLAS1lepRun2,
	AUTHOR=	"{ATLAS Collaboration}",
	TITLE=	"{Search for new phenomena with top quark pairs in final states with one lepton, jets, and missing transverse momentum in $pp$ collisions at $ \sqrt{s} $ = 13 TeV with the ATLAS detector}",
	EPRINT=	"2012.03799",
	ARCHIVEPREFIX=	"arXiv",
	PRIMARYCLASS=	"hep-ex",
	REPORTNUMBER=	"CERN-EP-2020-177",
	DOI=	"10.1007/JHEP04(2021)174",
	JOURNAL=	"JHEP",
	VOLUME=	"04",
	PAGES=	"174",
	YEAR=	"2021",
}

@ARTICLE{ATLASStopRun2,
	AUTHOR=	"{ATLAS Collaboration}",
	TITLE=	"{Search for new phenomena in events with two opposite-charge leptons, jets and missing transverse momentum in pp collisions at $ \sqrt{\mathrm{s}} $ = 13 TeV with the ATLAS detector}",
	EPRINT=	"2102.01444",
	ARCHIVEPREFIX=	"arXiv",
	PRIMARYCLASS=	"hep-ex",
	REPORTNUMBER=	"CERN-EP-2020-243",
	DOI=	"10.1007/JHEP04(2021)165",
	JOURNAL=	"JHEP",
	VOLUME=	"04",
	PAGES=	"165",
	YEAR=	"2021",
}

@ARTICLE{ATLASStop02LepMET,
	AUTHOR=	"{ATLAS Collaboration}",
	TITLE=	"{Search for new phenomena with top-quark pairs and large missing transverse momentum using 140 $\mathrm{fb}^{-1}$ of $pp$ collision data at $\sqrt{s}=13$ TeV with the ATLAS detector}",
	ARCHIVEPREFIX=	"arXiv",
	EPRINT=	"2401.13430",
	REPORTNUMBER=	"CERN-EP-2024-003",
	JOURNAL=	"JHEP",
	VOLUME=	"03",
	PAGES=	"139",
	YEAR=	"2024",
	DOI=	"10.1007/JHEP03(2024)139",
}

@ARTICLE{edgerun2,
	AUTHOR=	"{CMS Collaboration}",
	TITLE=	"Search for supersymmetry in final states with two oppositely charged same-flavor leptons and missing transverse momentum in proton-proton collisions at $\sqrt{s}=13$ {TeV}",
	JOURNAL=	"JHEP",
	VOLUME=	"04",
	YEAR=	"2021",
	PAGES=	"123",
	EPRINT=	"2012.08600",
	ARCHIVEPREFIX=	"arXiv",
	PRIMARYCLASS=	"hep-ex",
	DOI=	"10.1007/JHEP04(2021)123",
	SLACCITATION=	"%%CITATION = ARXIV:2012.08600;%%",
}

@ARTICLE{ref:TwoThreeLeptoSoft,
	AUTHOR=	"{CMS Collaboration}",
	TITLE=	"{Search for supersymmetry in final states with two or three soft leptons and missing transverse momentum in proton-proton collisions at $ \sqrt{s} $ = 13 TeV}",
	EPRINT=	"2111.06296",
	ARCHIVEPREFIX=	"arXiv",
	PRIMARYCLASS=	"hep-ex",
	REPORTNUMBER=	"CMS-SUS-18-004, CERN-EP-2021-168",
	DOI=	"10.1007/JHEP04(2022)091",
	JOURNAL=	"JHEP",
	VOLUME=	"04",
	PAGES=	"091",
	YEAR=	"2022",
}

@ARTICLE{CMSEWKCombination,
	AUTHOR=	"{CMS Collaboration}",
	TITLE=	"{Combined search for electroweak production of winos, binos, higgsinos, and sleptons in proton-proton collisions at s=13\,\,TeV}",
	EPRINT=	"2402.01888",
	ARCHIVEPREFIX=	"arXiv",
	PRIMARYCLASS=	"hep-ex",
	REPORTNUMBER=	"CMS-SUS-21-008, CERN-EP-2023-238",
	DOI=	"10.1103/PhysRevD.109.112001",
	JOURNAL=	"Phys. Rev. D",
	VOLUME=	"109",
	PAGES=	"112001",
	YEAR=	"2024",
}

@ARTICLE{ref:ATLASCompSlep,
	AUTHOR=	"{ATLAS Collaboration}",
	TITLE=	"{Searches for direct slepton production in the compressed-mass corridor in $\sqrt{s}=13$ TeV $pp$ collisions with the ATLAS detector}",
	EPRINT=	"2503.17186",
	ARCHIVEPREFIX=	"arXiv",
	PRIMARYCLASS=	"hep-ex",
	REPORTNUMBER=	"CERN-EP-2025-039",
	DOI=	"10.1007/JHEP08(2025)053",
	JOURNAL=	"JHEP",
	VOLUME=	"08",
	PAGES=	"053",
	YEAR=	"2025",
}

@MISC{hepdata,
	HOWPUBLISHED=	"{HEPD}ata record for this analysis",
	DOI=	"10.17182/hepdata.181478",
	YEAR=	"2026",
}

@ARTICLE{CMS:2020cmk,
	AUTHOR=	"{CMS Collaboration}",
	TITLE=	"{Performance of the CMS Level-1 trigger in proton-proton collisions at $\sqrt{s} = 13$\,TeV}",
	JOURNAL=	"JINST",
	VOLUME=	"15",
	PAGES=	"P10017",
	YEAR=	"2020",
	DOI=	"10.1088/1748-0221/15/10/P10017",
	EPRINT=	"2006.10165",
	ARCHIVEPREFIX=	"arXiv",
	PRIMARYCLASS=	"hep-ex",
	REPORTNUMBER=	"CMS-TRG-17-001, CERN-EP-2020-065",
}

@ARTICLE{CMS:2016ngn,
	AUTHOR=	"{CMS Collaboration}",
	TITLE=	"{The CMS trigger system}",
	JOURNAL=	"JINST",
	VOLUME=	"12",
	PAGES=	"P01020",
	DOI=	"10.1088/1748-0221/12/01/P01020",
	YEAR=	"2017",
	EPRINT=	"1609.02366",
	ARCHIVEPREFIX=	"arXiv",
	PRIMARYCLASS=	"physics.ins-det",
	REPORTNUMBER=	"CMS-TRG-12-001, CERN-EP-2016-160",
	SLACCITATION=	"%%CITATION = ARXIV:1609.02366;%%",
}

@ARTICLE{CMS:2024aqx,
	AUTHOR=	"{CMS Collaboration}",
	TITLE=	"Performance of the {CMS} high-level trigger during {LHC} Run 2",
	EPRINT=	"2410.17038",
	ARCHIVEPREFIX=	"arXiv",
	PRIMARYCLASS=	"physics.ins-det",
	REPORTNUMBER=	"CMS-TRG-19-001, CERN-EP-2024-259",
	DOI=	"10.1088/1748-0221/19/11/P11021",
	JOURNAL=	"JINST",
	VOLUME=	"19",
	PAGES=	"P11021",
	YEAR=	"2024",
}

@ARTICLE{bib:powhegv2a,
	AUTHOR=	"Nason, Paolo",
	TITLE=	"{A new method for combining NLO QCD with shower Monte Carlo algorithms}",
	EPRINT=	"hep-ph/0409146",
	ARCHIVEPREFIX=	"arXiv",
	REPORTNUMBER=	"BICOCCA-FT-04-11",
	DOI=	"10.1088/1126-6708/2004/11/040",
	JOURNAL=	"JHEP",
	VOLUME=	"11",
	PAGES=	"040",
	YEAR=	"2004",
}

@ARTICLE{bib:powhegv2b,
	AUTHOR=	"Frixione, Stefano and Nason, Paolo and Oleari, Carlo",
	TITLE=	"{Matching NLO QCD computations with Parton Shower simulations: the POWHEG method}",
	EPRINT=	"0709.2092",
	ARCHIVEPREFIX=	"arXiv",
	PRIMARYCLASS=	"hep-ph",
	REPORTNUMBER=	"BICOCCA-FT-07-9, GEF-TH-21-2007",
	DOI=	"10.1088/1126-6708/2007/11/070",
	JOURNAL=	"JHEP",
	VOLUME=	"11",
	PAGES=	"070",
	YEAR=	"2007",
}

@ARTICLE{bib:powhegv2c,
	AUTHOR=	"Alioli, Simone and Nason, Paolo and Oleari, Carlo and Re, Emanuele",
	TITLE=	"{A general framework for implementing NLO calculations in shower Monte Carlo programs: the POWHEG BOX}",
	EPRINT=	"1002.2581",
	ARCHIVEPREFIX=	"arXiv",
	PRIMARYCLASS=	"hep-ph",
	REPORTNUMBER=	"DESY-10-018, SFB-CPP-10-22, IPPP-10-11, DCPT-10-22",
	DOI=	"10.1007/JHEP06(2010)043",
	JOURNAL=	"JHEP",
	VOLUME=	"06",
	PAGES=	"043",
	YEAR=	"2010",
}

@ARTICLE{Czakon:2011xx,
	AUTHOR=	"Czakon, Michal and Mitov, Alexander",
	TITLE=	"Top++: {A} program for the calculation of the top-pair cross-section at hadron colliders",
	JOURNAL=	"Comput. Phys. Commun.",
	VOLUME=	"185",
	PAGES=	"2930",
	DOI=	"10.1016/j.cpc.2014.06.021",
	YEAR=	"2014",
	EPRINT=	"1112.5675",
	ARCHIVEPREFIX=	"arXiv",
	PRIMARYCLASS=	"hep-ph",
	REPORTNUMBER=	"CERN-PH-TH-2011-303, TTK-11-58",
	SLACCITATION=	"%%CITATION = ARXIV:1112.5675;%%",
}

@ARTICLE{bib:powheg3,
	AUTHOR=	"Re, E.",
	TITLE=	"Single-top {W}t-channel production matched with parton showers using the {POWHEG} method",
	JOURNAL=	"Eur. Phys. J. C",
	VOLUME=	"71",
	YEAR=	"2011",
	PAGES=	"1547",
	EPRINT=	"1009.2450",
	ARCHIVEPREFIX=	"arXiv",
	PRIMARYCLASS=	"hep-ph",
	DOI=	"10.1140/epjc/s10052-011-1547-z",
}

@ARTICLE{bib:tWxsec,
	AUTHOR=	"Kidonakis, N.",
	TITLE=	"{NNLL} threshold resummation for top-pair and single-top production",
	JOURNAL=	"Phys. Part. Nucl.",
	VOLUME=	"45",
	YEAR=	"2014",
	PAGES=	"714",
	DOI=	"10.1134/S1063779614040091",
	EPRINT=	"1210.7813",
	ARCHIVEPREFIX=	"arXiv",
	PRIMARYCLASS=	"hep-ph",
	SLACCITATION=	"%%CITATION = ARXIV:1210.7813;%%",
}

@ARTICLE{bib:powhegVVa,
	AUTHOR=	"Melia, T. and Nason, P. and Rontsch, R. and Zanderighi, G.",
	TITLE=	"{W}$^+${W}$^-$, {WZ} and {ZZ} production in the {POWHEG} {BOX}",
	JOURNAL=	"JHEP",
	VOLUME=	"11",
	YEAR=	"2011",
	PAGES=	"078",
	EPRINT=	"1107.5051",
	ARCHIVEPREFIX=	"arXiv",
	PRIMARYCLASS=	"hep-ph",
	DOI=	"10.1007/JHEP11(2011)078",
}

@ARTICLE{bib:powhegVVb,
	AUTHOR=	"Nason, P. and Zanderighi, G.",
	TITLE=	"{W}$^+${W}$^-$, {WZ} and {ZZ} production in the {POWHEG}-{BOX}-{V}2",
	JOURNAL=	"Eur. Phys. J. C",
	VOLUME=	"74",
	YEAR=	"2014",
	PAGES=	"2702",
	EPRINT=	"1311.1365",
	ARCHIVEPREFIX=	"arXiv",
	PRIMARYCLASS=	"hep-ph",
	DOI=	"10.1140/epjc/s10052-013-2702-5",
}

@ARTICLE{bib:WWxsec,
	AUTHOR=	"Gehrmann, T. and Grazzini, M. and Kallweit, S. and Maierh{\"o}fer, P. and von Manteuffel, A. and Pozzorini, S. and Rathlev, D. and Tancredi, L.",
	TITLE=	"{W}$^{+}${W}$^{-}$ production at hadron colliders in {NNLO} {QCD}",
	JOURNAL=	"Phys. Rev. Lett.",
	VOLUME=	"113",
	YEAR=	"2014",
	PAGES=	"212001",
	DOI=	"10.1103/PhysRevLett.113.212001",
	EPRINT=	"1408.5243",
	ARCHIVEPREFIX=	"arXiv",
	PRIMARYCLASS=	"hep-ph",
	SLACCITATION=	"%%CITATION = ARXIV:1408.5243;%%",
}

@ARTICLE{HIG16041,
	AUTHOR=	"{CMS Collaboration}",
	TITLE=	"Measurements of properties of the {H}iggs boson decaying into the four-lepton final state in pp collisions at $\sqrt{s}=13$ {TeV}",
	JOURNAL=	"JHEP",
	VOLUME=	"11",
	PAGES=	"047",
	DOI=	"10.1007/JHEP11(2017)047",
	YEAR=	"2017",
	EPRINT=	"1706.09936",
	ARCHIVEPREFIX=	"arXiv",
	PRIMARYCLASS=	"hep-ex",
	REPORTNUMBER=	"CERN-EP-2017-123 ",
	SLACCITATION=	"%%CITATION = ARXIV:1706.09936;%%",
}

@ARTICLE{bib:MCFM,
	AUTHOR=	"Campbell, J. M. and Ellis, R. K.",
	TITLE=	"{MCFM} for the {T}evatron and the {LHC}",
	JOURNAL=	"Nucl. Phys. Proc. Suppl.",
	VOLUME=	"205",
	YEAR=	"2010",
	PAGES=	"10",
	EPRINT=	"1007.3492",
	ARCHIVEPREFIX=	"arXiv",
	PRIMARYCLASS=	"hep-ph",
	DOI=	"10.1016/j.nuclphysbps.2010.08.011",
}

@ARTICLE{bib:ggWWkfactors,
	AUTHOR=	"Caola, F. and Melnikov, K. and Rtsch, R. and Tancredi, L.",
	TITLE=	"{QCD} corrections to {W}$^+${W}$^{-}$ production through gluon fusion",
	JOURNAL=	"Phys. Lett. B",
	VOLUME=	"754",
	YEAR=	"2016",
	PAGES=	"275",
	DOI=	"10.1016/j.physletb.2016.01.046",
	EPRINT=	"1511.08617",
	ARCHIVEPREFIX=	"arXiv",
	PRIMARYCLASS=	"hep-ph",
	SLACCITATION=	"%%CITATION = ARXIV:1511.08617;%%",
}

@ARTICLE{Artoisenet:2012st,
	AUTHOR=	"Artoisenet, Pierre and Frederix, Rikkert and Mattelaer, Olivier and Rietkerk, Robbert",
	TITLE=	"Automatic spin-entangled decays of heavy resonances in {Monte Carlo} simulations",
	JOURNAL=	"JHEP",
	VOLUME=	"03",
	YEAR=	"2013",
	PAGES=	"015",
	DOI=	"10.1007/JHEP03(2013)015",
	EPRINT=	"1212.3460",
	ARCHIVEPREFIX=	"arXiv",
	PRIMARYCLASS=	"hep-ph",
}

@ARTICLE{bib:DYxsec,
	AUTHOR=	"Gavin, R. and Li, Y. and Petriello, F. and Quackenbush, S.",
	TITLE=	"{FEWZ} 2.0: {A} code for hadronic {Z} production at next-to-next-to-leading order",
	JOURNAL=	"Comput. Phys. Commun.",
	VOLUME=	"182",
	YEAR=	"2011",
	PAGES=	"2388",
	EPRINT=	"1011.3540",
	ARCHIVEPREFIX=	"arXiv",
	PRIMARYCLASS=	"hep-ph",
	DOI=	"10.1016/j.cpc.2011.06.008",
}

@ARTICLE{bib:ttVxsec,
	AUTHOR=	"Garzelli, M. V. and Kardos, A. and Papadopoulos, C. G. and Trocsanyi, Z.",
	TITLE=	"$\ttbar${W}$^{\pm}$ and $\ttbar${Z} hadroproduction at {NLO} accuracy in {QCD} with parton shower and hadronization effects",
	JOURNAL=	"JHEP",
	VOLUME=	"11",
	YEAR=	"2012",
	PAGES=	"056",
	EPRINT=	"1208.2665",
	ARCHIVEPREFIX=	"arXiv",
	PRIMARYCLASS=	"hep-ph",
	DOI=	"10.1007/JHEP11(2012)056",
}

@ARTICLE{bib:HWWxsec,
	AUTHOR=	"{LHC Higgs Cross Section Working Group}",
	TITLE=	"Handbook of {LHC Higgs} cross sections: 3. {Higgs properties}",
	JOURNAL=	"CERN",
	DOI=	"10.5170/CERN-2013-004",
	EPRINT=	"1307.1347",
	ARCHIVEPREFIX=	"arXiv",
	PRIMARYCLASS=	"hep-ph",
	REPORTNUMBER=	"FERMILAB-CONF-13-667-T, CERN-2013-004",
	YEAR=	"2013",
	SLACCITATION=	"%%CITATION = ARXIV:1307.1347;%%",
}

@ARTICLE{Beenakker:1999xh,
	AUTHOR=	"Beenakker, W. and Klasen, M. and Kramer, M. and Plehn, T. and Spira, M. and Zerwas, P. M.",
	TITLE=	"Production of charginos, neutralinos, and sleptons at hadron colliders",
	JOURNAL=	"Phys. Rev. Lett.",
	VOLUME=	"83",
	YEAR=	"1999",
	PAGES=	"3780",
	DOI=	"10.1103/PhysRevLett.83.3780",
	NOTE=	"[Erratum: \DOI{10.1103/PhysRevLett.100.029901}]",
	EPRINT=	"hep-ph/9906298",
	ARCHIVEPREFIX=	"arXiv",
	PRIMARYCLASS=	"hep-ph",
	REPORTNUMBER=	"CERN-TH-99-159, DESY-99-055, DTP-99-44, MADPH-99-1114, ANL-HEP-PR-99-71",
	SLACCITATION=	"%%CITATION = HEP-PH/9906298;%%",
}

@ARTICLE{Fuks:2012qx,
	AUTHOR=	"Fuks, Benjamin and Klasen, Michael and Lamprea, David R. and Rothering, Marcel",
	TITLE=	"Gaugino production in proton-proton collisions at a center-of-mass energy of 8 {TeV}",
	JOURNAL=	"JHEP",
	VOLUME=	"10",
	PAGES=	"081",
	DOI=	"10.1007/JHEP10(2012)081",
	YEAR=	"2012",
	EPRINT=	"1207.2159",
	ARCHIVEPREFIX=	"arXiv",
	PRIMARYCLASS=	"hep-ph",
	REPORTNUMBER=	"IPHC-PHENO-12-07, MS-TP-12-05",
	SLACCITATION=	"%%CITATION = ARXIV:1207.2159;%%",
}

@ARTICLE{Fuks:2013vua,
	AUTHOR=	"Fuks, Benjamin and Klasen, Michael and Lamprea, David R. and Rothering, Marcel",
	TITLE=	"Precision predictions for electroweak superpartner production at hadron colliders with {\sc {R}esummino}",
	JOURNAL=	"Eur. Phys. J. C",
	VOLUME=	"73",
	PAGES=	"2480",
	DOI=	"10.1140/epjc/s10052-013-2480-0",
	YEAR=	"2013",
	EPRINT=	"1304.0790",
	ARCHIVEPREFIX=	"arXiv",
	PRIMARYCLASS=	"hep-ph",
	REPORTNUMBER=	"CERN-PH-TH-2013-064, IPHC-PHENO-13-02, MS-TP-13-06",
	SLACCITATION=	"%%CITATION = ARXIV:1304.0790;%%",
}

@ARTICLE{Fuks:2013lya,
	AUTHOR=	"Fuks, Benjamin and Klasen, Michael and Lamprea, David R. and Rothering, Marcel",
	TITLE=	"{Revisiting slepton pair production at the Large Hadron Collider}",
	JOURNAL=	"JHEP",
	VOLUME=	"01",
	PAGES=	"168",
	YEAR=	"2014",
	DOI=	"10.1007/JHEP01(2014)168",
	EPRINT=	"1310.2621",
	ARCHIVEPREFIX=	"arXiv",
	PRIMARYCLASS=	"hep-ph",
	REPORTNUMBER=	"CERN-PH-TH-2013-240, MS-TP-13-35",
	SLACCITATION=	"%%CITATION = ARXIV:1310.2621;%%",
}

@ARTICLE{Beenakker:1996ch,
	AUTHOR=	"Beenakker, W. and Hopker, R. and Spira, M. and Zerwas, P. M.",
	TITLE=	"{Squark and gluino production at hadron colliders}",
	JOURNAL=	"Nucl. Phys. B",
	VOLUME=	"492",
	YEAR=	"1997",
	PAGES=	"51",
	DOI=	"10.1016/S0550-3213(97)80027-2",
	EPRINT=	"hep-ph/9610490",
	ARCHIVEPREFIX=	"arXiv",
	PRIMARYCLASS=	"hep-ph",
	REPORTNUMBER=	"DESY-96-150, CERN-TH-96-215",
	SLACCITATION=	"%%CITATION = HEP-PH/9610490;%%",
}

@ARTICLE{Kulesza:2008jb,
	AUTHOR=	"Kulesza, A. and Motyka, L.",
	TITLE=	"{Threshold resummation for squark-antisquark and gluino-pair production at the LHC}",
	JOURNAL=	"Phys. Rev. Lett.",
	VOLUME=	"102",
	YEAR=	"2009",
	PAGES=	"111802",
	DOI=	"10.1103/PhysRevLett.102.111802",
	EPRINT=	"0807.2405",
	ARCHIVEPREFIX=	"arXiv",
	PRIMARYCLASS=	"hep-ph",
	REPORTNUMBER=	"DESY-08-096",
	SLACCITATION=	"%%CITATION = ARXIV:0807.2405;%%",
}

@ARTICLE{Kulesza:2009kq,
	AUTHOR=	"Kulesza, A. and Motyka, L.",
	TITLE=	"{Soft gluon resummation for the production of gluino-gluino and squark-antisquark pairs at the LHC}",
	JOURNAL=	"Phys. Rev. D",
	VOLUME=	"80",
	YEAR=	"2009",
	PAGES=	"095004",
	DOI=	"10.1103/PhysRevD.80.095004",
	EPRINT=	"0905.4749",
	ARCHIVEPREFIX=	"arXiv",
	PRIMARYCLASS=	"hep-ph",
	REPORTNUMBER=	"DESY-09-078, PITHA-09-11",
	SLACCITATION=	"%%CITATION = ARXIV:0905.4749;%%",
}

@ARTICLE{Beenakker:2009ha,
	AUTHOR=	"Beenakker, Wim and Brensing, Silja and Kramer, Michael and Kulesza, Anna and Laenen, Eric and Niessen, Irene",
	TITLE=	"{Soft-gluon resummation for squark and gluino hadroproduction}",
	JOURNAL=	"JHEP",
	VOLUME=	"12",
	YEAR=	"2009",
	PAGES=	"041",
	DOI=	"10.1088/1126-6708/2009/12/041",
	EPRINT=	"0909.4418",
	ARCHIVEPREFIX=	"arXiv",
	PRIMARYCLASS=	"hep-ph",
	REPORTNUMBER=	"ITP-UU-09-33, NIKHEF-2009-015",
	SLACCITATION=	"%%CITATION = ARXIV:0909.4418;%%",
}

@ARTICLE{Beenakker:2011fu,
	AUTHOR=	"Beenakker, W. and Brensing, S. and Kramer, M. n and Kulesza, A. and Laenen, E. and Motyka, L. and Niessen, I.",
	TITLE=	"{Squark and gluino hadroproduction}",
	JOURNAL=	"Int. J. Mod. Phys. A",
	VOLUME=	"26",
	YEAR=	"2011",
	PAGES=	"2637",
	DOI=	"10.1142/S0217751X11053560",
	EPRINT=	"1105.1110",
	ARCHIVEPREFIX=	"arXiv",
	PRIMARYCLASS=	"hep-ph",
	REPORTNUMBER=	"DESY-11-060, ITFA-11-09, ITP-UU-11-13, NIKHEF-2011-011, TTK-11-12",
	SLACCITATION=	"%%CITATION = ARXIV:1105.1110;%%",
}

@ARTICLE{Borschensky:2014cia,
	AUTHOR=	"Borschensky, Christoph and Kramer, Michael and Kulesza, Anna and Mangano, Michelangelo and Padhi, Sanjay and Plehn, Tilman and Portell, Xavier",
	TITLE=	"{Squark and gluino production cross sections in pp collisions at $\sqrt{s}$ = 13, 14, 33 and 100 TeV}",
	JOURNAL=	"Eur. Phys. J. C",
	VOLUME=	"74",
	YEAR=	"2014",
	PAGES=	"3174",
	DOI=	"10.1140/epjc/s10052-014-3174-y",
	EPRINT=	"1407.5066",
	ARCHIVEPREFIX=	"arXiv",
	PRIMARYCLASS=	"hep-ph",
	REPORTNUMBER=	"MS-TP-14-25, CERN-PH-TH-2014-137, TTK-14-13",
	SLACCITATION=	"%%CITATION = ARXIV:1407.5066;%%",
}

@ARTICLE{Beenakker:2011sf,
	AUTHOR=	"Beenakker, Wim and Brensing, Silja and Kramer, Michael and Kulesza, Anna and Laenen, Eric and Niessen, Irene",
	TITLE=	"{NNLL resummation for squark-antisquark pair production at the LHC}",
	JOURNAL=	"JHEP",
	VOLUME=	"01",
	YEAR=	"2012",
	PAGES=	"076",
	DOI=	"10.1007/JHEP01(2012)076",
	EPRINT=	"1110.2446",
	ARCHIVEPREFIX=	"arXiv",
	PRIMARYCLASS=	"hep-ph",
	REPORTNUMBER=	"DESY-11-175, ITP-UU-11-36, ITFA-11-14, NIKHEF-2011-027, TTK-11-47",
	SLACCITATION=	"%%CITATION = ARXIV:1110.2446;%%",
}

@ARTICLE{Beenakker:2013mva,
	AUTHOR=	"Beenakker, Wim and Janssen, Tim and Lepoeter, Susanne and Kr{\"a}mer, Michael and Kulesza, Anna and Laenen, Eric and Niessen, Irene and Thewes, Silja and Van Daal, Tom",
	TITLE=	"{Towards NNLL resummation: hard matching coefficients for squark and gluino hadroproduction}",
	JOURNAL=	"JHEP",
	VOLUME=	"10",
	YEAR=	"2013",
	PAGES=	"120",
	DOI=	"10.1007/JHEP10(2013)120",
	EPRINT=	"1304.6354",
	ARCHIVEPREFIX=	"arXiv",
	PRIMARYCLASS=	"hep-ph",
	REPORTNUMBER=	"NIKHEF-2012-016, MS-TP-13-10, TTK-13-10, DESY-13-072",
	SLACCITATION=	"%%CITATION = ARXIV:1304.6354;%%",
}

@ARTICLE{Beenakker:2014sma,
	AUTHOR=	"Beenakker, Wim and Borschensky, Christoph and Kr{\"a}mer, Michael and Kulesza, Anna and Laenen, Eric and Theeuwes, Vincent and Thewes, Silja",
	TITLE=	"{NNLL resummation for squark and gluino production at the LHC}",
	JOURNAL=	"JHEP",
	VOLUME=	"12",
	YEAR=	"2014",
	PAGES=	"023",
	DOI=	"10.1007/JHEP12(2014)023",
	EPRINT=	"1404.3134",
	ARCHIVEPREFIX=	"arXiv",
	PRIMARYCLASS=	"hep-ph",
	REPORTNUMBER=	"MS-TP-14-16, TTK-14-08, NIKHEF-2014-009, ITP-UU-14-11",
	SLACCITATION=	"%%CITATION = ARXIV:1404.3134;%%",
}

@ARTICLE{Beenakker:2016lwe,
	AUTHOR=	"Beenakker, Wim and Borschensky, Christoph and Kr{\"a}mer, Michael and Kulesza, Anna and Laenen, Eric",
	TITLE=	"{NNLL-fast: predictions for coloured supersymmetric particle production at the LHC with threshold and Coulomb resummation}",
	JOURNAL=	"JHEP",
	VOLUME=	"12",
	YEAR=	"2016",
	PAGES=	"133",
	DOI=	"10.1007/JHEP12(2016)133",
	EPRINT=	"1607.07741",
	ARCHIVEPREFIX=	"arXiv",
	PRIMARYCLASS=	"hep-ph",
	REPORTNUMBER=	"MS-TP-16-17, NIKHEF-2016-037, TTK-16-28",
	SLACCITATION=	"%%CITATION = ARXIV:1607.07741;%%",
}

@ARTICLE{Beenakker:1997ut,
	AUTHOR=	"Beenakker, W. and Kramer, M. and Plehn, T. and Spira, M. and Zerwas, P. M.",
	TITLE=	"{Stop production at hadron colliders}",
	JOURNAL=	"Nucl. Phys. B",
	VOLUME=	"515",
	YEAR=	"1998",
	PAGES=	"3",
	DOI=	"10.1016/S0550-3213(98)00014-5",
	EPRINT=	"hep-ph/9710451",
	ARCHIVEPREFIX=	"arXiv",
	PRIMARYCLASS=	"hep-ph",
	REPORTNUMBER=	"DESY-97-214, CERN-TH-97-177, RAL-TR-97-056",
	SLACCITATION=	"%%CITATION = HEP-PH/9710451;%%",
}

@ARTICLE{Beenakker:2010nq,
	AUTHOR=	"Beenakker, Wim and Brensing, Silja and Kramer, Michael and Kulesza, Anna and Laenen, Eric and Niessen, Irene",
	TITLE=	"{Supersymmetric top and bottom squark production at hadron colliders}",
	JOURNAL=	"JHEP",
	VOLUME=	"08",
	YEAR=	"2010",
	PAGES=	"098",
	DOI=	"10.1007/JHEP08(2010)098",
	EPRINT=	"1006.4771",
	ARCHIVEPREFIX=	"arXiv",
	PRIMARYCLASS=	"hep-ph",
	REPORTNUMBER=	"CERN-PH-TH-2010-142, ITFA-2010-015, ITP-UU-10-17, NIKHEF-2010-016, TTK-10-33",
	SLACCITATION=	"%%CITATION = ARXIV:1006.4771;%%",
}

@ARTICLE{Beenakker:2016gmf,
	AUTHOR=	"Beenakker, Wim and Borschensky, Christoph and Heger, Raphael and Kr{\"a}mer, Michael and Kulesza, Anna and Laenen, Eric",
	TITLE=	"{NNLL resummation for stop pair-production at the LHC}",
	JOURNAL=	"JHEP",
	VOLUME=	"05",
	YEAR=	"2016",
	PAGES=	"153",
	DOI=	"10.1007/JHEP05(2016)153",
	EPRINT=	"1601.02954",
	ARCHIVEPREFIX=	"arXiv",
	PRIMARYCLASS=	"hep-ph",
	REPORTNUMBER=	"MS-TP-16-01, TTK-16-01, NIKHEF-2016-002",
	SLACCITATION=	"%%CITATION = ARXIV:1601.02954;%%",
}

@ARTICLE{Beenakker:2024,
	AUTHOR=	"Beenakker, Wim and Borschensky, Christoph and Kr{\"a}mer, Michael and Kulesza, Anna and Laenen, Eric and Mamu{\check{z}}i{\check{c}}, Judita and Valero, Laura Moreno",
	TITLE=	"{NNLL-fast 2.0: Coloured sparticle production at the LHC run 3 with $\sqrt{s}=$ 13.6 TeV}",
	JOURNAL=	"SciPost Phys. Core",
	VOLUME=	"7",
	PAGES=	"072",
	YEAR=	"2024",
	PUBLISHER=	"SciPost",
	DOI=	"10.21468/SciPostPhysCore.7.4.072",
	EPRINT=	"2404.18837",
	ARCHIVEPREFIX=	"arXiv",
	PRIMARYCLASS=	"hep-ph",
}

@ARTICLE{NNPDF31,
	AUTHOR=	"Ball, Richard D. and others",
	COLLABORATION=	"NNPDF",
	TITLE=	"Parton distributions from high-precision collider data",
	JOURNAL=	"Eur. Phys. J. C",
	VOLUME=	"77",
	YEAR=	"2017",
	PAGES=	"663",
	DOI=	"10.1140/epjc/s10052-017-5199-5",
	EPRINT=	"1706.00428",
	ARCHIVEPREFIX=	"arXiv",
	PRIMARYCLASS=	"hep-ph",
	REPORTNUMBER=	"CAVENDISH-HEP-17-06, CERN-TH-2017-077, Edinburgh 2017/08, Nikhef/2017-006, OUTP-17-04P,TIF-UNIMI-2017-3",
	SLACCITATION=	"%%CITATION = ARXIV:1706.00428;%%",
}

@ARTICLE{Sjostrand:2014zea,
	AUTHOR=	"Sj{\"o}strand, Torbj{\"o}rn and Ask, Stefan and Christiansen, Jesper R. and Corke, Richard and Desai, Nishita and Ilten, Philip and Mrenna, Stephen and Prestel, Stefan and Rasmussen, Christine O. and Skands, Peter Z.",
	TITLE=	"An {I}ntroduction to {PYTHIA} 8.2",
	JOURNAL=	"Comput. Phys. Commun.",
	VOLUME=	"191",
	YEAR=	"2015",
	PAGES=	"159",
	DOI=	"10.1016/j.cpc.2015.01.024",
	EPRINT=	"1410.3012",
	ARCHIVEPREFIX=	"arXiv",
	PRIMARYCLASS=	"hep-ph",
	REPORTNUMBER=	"LU-TP-14-36, MCNET-14-22, CERN-PH-TH-2014-190, FERMILAB-PUB-14-316-CD, DESY-14-178, SLAC-PUB-16122",
	SLACCITATION=	"%%CITATION = ARXIV:1410.3012;%%",
}

@ARTICLE{CMS:2019csb,
	AUTHOR=	"{CMS Collaboration}",
	TITLE=	"Extraction and validation of a new set of {CMS} {PYTHIA8} tunes from underlying-event measurements",
	EPRINT=	"1903.12179",
	ARCHIVEPREFIX=	"arXiv",
	PRIMARYCLASS=	"hep-ex",
	REPORTNUMBER=	"CMS-GEN-17-001, CERN-EP-2019-007",
	DOI=	"10.1140/epjc/s10052-019-7499-4",
	JOURNAL=	"Eur. Phys. J. C",
	VOLUME=	"80",
	PAGES=	"4",
	YEAR=	"2020",
}

@ARTICLE{Agostinelli:2002hh,
	AUTHOR=	"Agostinelli, S. and others",
	COLLABORATION=	"GEANT4",
	TITLE=	"{\GEANTfour}---a simulation toolkit",
	JOURNAL=	"Nucl. Instrum. Meth. A",
	VOLUME=	"506",
	YEAR=	"2003",
	PAGES=	"250",
	DOI=	"10.1016/S0168-9002(03)01368-8",
	SLACCITATION=	"%%CITATION = NUIMA,A506,250;%%",
}

@ARTICLE{Abdullin:2011zz,
	AUTHOR=	"Abdullin, S. and Azzi, P. and Beaudette, F. and Janot, P. and Perrotta, A.",
	TITLE=	"The {Fast Simulation} of the {CMS} {Detector} at {LHC}",
	JOURNAL=	"J. Phys.: Conf. Ser.",
	VOLUME=	"331",
	YEAR=	"2011",
	PAGES=	"032049",
	DOI=	"10.1088/1742-6596/331/3/032049",
}

@ARTICLE{CMS:2017yfk,
	AUTHOR=	"{CMS Collaboration}",
	TITLE=	"Particle-flow reconstruction and global event description with the {CMS} detector",
	JOURNAL=	"JINST",
	VOLUME=	"12",
	YEAR=	"2017",
	PAGES=	"P10003",
	DOI=	"10.1088/1748-0221/12/10/P10003",
	EPRINT=	"1706.04965",
	ARCHIVEPREFIX=	"arXiv",
	PRIMARYCLASS=	"physics.ins-det",
	REPORTNUMBER=	"CMS-PRF-14-001, CERN-EP-2017-110",
	SLACCITATION=	"%%CITATION = ARXIV:1706.04965;%%",
}

@TECHREPORT{CMS-TDR-15-02,
	AUTHOR=	"{CMS Collaboration}",
	TITLE=	"Technical proposal for the {Phase-II} upgrade of the {Compact Muon Solenoid}",
	DOI=	"10.17181/CERN.VU8I.D59J",
	TYPE=	"CMS Technical Proposal",
	NUMBER=	"CERN-LHCC-2015-010, CMS-TDR-15-02",
	YEAR=	"2015",
}

@ARTICLE{CMS:2020uim,
	AUTHOR=	"{CMS Collaboration}",
	TITLE=	"Electron and photon reconstruction and identification with the {CMS} experiment at the {CERN} {LHC}",
	EPRINT=	"2012.06888",
	JOURNAL=	"JINST",
	VOLUME=	"16",
	PAGES=	"P05014",
	YEAR=	"2021",
	ARCHIVEPREFIX=	"arXiv",
	PRIMARYCLASS=	"hep-ex",
	REPORTNUMBER=	"CMS-EGM-17-001, CERN-EP-2020-219",
	DOI=	"10.1088/1748-0221/16/05/P05014",
}

@TECHREPORT{CMS-DP-2020-021,
	TITLE=	"{ECAL} 2016 refined calibration and {Run2} summary plots",
	AUTHOR=	"{CMS Collaboration}",
	URL=	"https://cds.cern.ch/record/2717925",
	TYPE=	"CMS Detector Performance Note",
	NUMBER=	"CMS-DP-2020-021",
	YEAR=	"2020",
}

@ARTICLE{CMS:2018rym,
	AUTHOR=	"{CMS Collaboration}",
	TITLE=	"{Performance of the CMS muon detector and muon reconstruction with proton-proton collisions at $\sqrt{s}=$ 13 TeV}",
	EPRINT=	"1804.04528",
	ARCHIVEPREFIX=	"arXiv",
	PRIMARYCLASS=	"physics.ins-det",
	REPORTNUMBER=	"CMS-MUO-16-001, CERN-EP-2018-058",
	DOI=	"10.1088/1748-0221/13/06/P06015",
	JOURNAL=	"JINST",
	VOLUME=	"13",
	PAGES=	"P06015",
	YEAR=	"2018",
}

@ARTICLE{Cacciari:2008gp,
	AUTHOR=	"Cacciari, Matteo and Salam, Gavin P. and Soyez, Gregory",
	TITLE=	"{The anti-\kt jet clustering algorithm}",
	EPRINT=	"0802.1189",
	ARCHIVEPREFIX=	"arXiv",
	PRIMARYCLASS=	"hep-ph",
	REPORTNUMBER=	"LPTHE-07-03",
	DOI=	"10.1088/1126-6708/2008/04/063",
	JOURNAL=	"JHEP",
	VOLUME=	"04",
	PAGES=	"063",
	YEAR=	"2008",
}

@ARTICLE{Cacciari:2011ma,
	AUTHOR=	"Cacciari, Matteo and Salam, Gavin P. and Soyez, Gregory",
	TITLE=	"{FastJet} user manual",
	JOURNAL=	"Eur. Phys. J. C",
	VOLUME=	"72",
	PAGES=	"1896",
	DOI=	"10.1140/epjc/s10052-012-1896-2",
	YEAR=	"2012",
	EPRINT=	"1111.6097",
	ARCHIVEPREFIX=	"arXiv",
	PRIMARYCLASS=	"hep-ph",
	REPORTNUMBER=	"CERN-PH-TH-2011-297",
	SLACCITATION=	"%%CITATION = ARXIV:1111.6097;%%",
}

@ARTICLE{CMS:2020ebo,
	AUTHOR=	"{CMS Collaboration}",
	TITLE=	"{Pileup mitigation at CMS in 13\TeV data}",
	EPRINT=	"2003.00503",
	ARCHIVEPREFIX=	"arXiv",
	PRIMARYCLASS=	"hep-ex",
	REPORTNUMBER=	"CMS-JME-18-001, CERN-EP-2020-017",
	DOI=	"10.1088/1748-0221/15/09/P09018",
	JOURNAL=	"JINST",
	VOLUME=	"15",
	PAGES=	"P09018",
	YEAR=	"2020",
}

@ARTICLE{CMS:2016lmd,
	AUTHOR=	"{CMS Collaboration}",
	TITLE=	"{Jet energy scale and resolution in the CMS experiment in pp collisions at 8 TeV}",
	JOURNAL=	"JINST",
	VOLUME=	"12",
	YEAR=	"2017",
	PAGES=	"P02014",
	DOI=	"10.1088/1748-0221/12/02/P02014",
	EPRINT=	"1607.03663",
	ARCHIVEPREFIX=	"arXiv",
	PRIMARYCLASS=	"hep-ex",
	REPORTNUMBER=	"CMS-JME-13-004, CERN-PH-EP-2015-305",
	SLACCITATION=	"%%CITATION = ARXIV:1607.03663;%%",
}

@ARTICLE{BTV-16-002,
	AUTHOR=	"{CMS Collaboration}",
	TITLE=	"Identification of heavy-flavour jets with the {CMS} detector in pp collisions at 13 {TeV}",
	JOURNAL=	"JINST",
	VOLUME=	"13",
	PAGES=	"P05011",
	DOI=	"10.1088/1748-0221/13/05/P05011",
	YEAR=	"2018",
	EPRINT=	"1712.07158",
	ARCHIVEPREFIX=	"arXiv",
	PRIMARYCLASS=	"physics.ins-det",
	REPORTNUMBER=	"CMS-BTV-16-002, CERN-EP-2017-326",
	SLACCITATION=	"%%CITATION = ARXIV:1712.07158;%%",
}

@ARTICLE{CMS:2019ctu,
	AUTHOR=	"{CMS Collaboration}",
	TITLE=	"Performance of missing transverse momentum reconstruction in proton-proton collisions at $\sqrt{s} = 13$\,{TeV} using the {CMS} detector",
	JOURNAL=	"JINST",
	VOLUME=	"14",
	YEAR=	"2019",
	PAGES=	"P07004",
	DOI=	"10.1088/1748-0221/14/07/P07004",
	EPRINT=	"1903.06078",
	ARCHIVEPREFIX=	"arXiv",
	PRIMARYCLASS=	"hep-ex",
	REPORTNUMBER=	"CMS-JME-17-001, CERN-EP-2018-335",
	SLACCITATION=	"%%CITATION = ARXIV:1903.06078;%%",
}

@ARTICLE{PDG2026,
	AUTHOR=	"{Particle Data Group}",
	TITLE=	"Review of particle physics",
	DOI=	"10.1142/S0217751X26300115",
	JOURNAL=	"Int. J. Mod. Phys. A",
	VOLUME=	"41",
	PAGES=	"2630011",
	YEAR=	"2026",
}

@ARTICLE{Lester:1999tx,
	AUTHOR=	"Lester, C. G. and Summers, D. J.",
	TITLE=	"Measuring masses of semi-invisibly decaying particles pair produced at hadron colliders",
	JOURNAL=	"Phys. Lett. B",
	VOLUME=	"463",
	YEAR=	"1999",
	PAGES=	"99",
	EPRINT=	"hep-ph/9906349",
	ARCHIVEPREFIX=	"arXiv",
	DOI=	"10.1016/S0370-2693(99)00945-4",
	SLACCITATION=	"%%CITATION = HEP-PH/9906349;%%",
}

@ARTICLE{inppxsec13TeV,
	AUTHOR=	"{CMS Collaboration}",
	TITLE=	"Measurement of the inelastic proton-proton cross section at $\sqrt{s}=13$ {TeV}",
	JOURNAL=	"JHEP",
	VOLUME=	"07",
	YEAR=	"2018",
	PAGES=	"161",
	EPRINT=	"1802.02613",
	ARCHIVEPREFIX=	"arXiv",
	PRIMARYCLASS=	"hep-ex",
	DOI=	"10.1007/JHEP07(2018)161",
	SLACCITATION=	"%%CITATION = ARXIV:1802.02613;%%",
}

@ARTICLE{RFScale1,
	AUTHOR=	"Stefano Catani and Daniel de Florian and Massimiliano Grazzini and Paolo Nason",
	TITLE=	"{Soft-gluon resummation for Higgs boson production at hadron colliders}",
	JOURNAL=	"JHEP",
	VOLUME=	"07",
	PAGES=	"028",
	YEAR=	"2003",
	DOI=	"10.1088/1126-6708/2003/07/028",
	EPRINT=	"hep-ph/0306211",
	ARCHIVEPREFIX=	"arXiv",
}

@ARTICLE{RFScale2,
	AUTHOR=	"M Cacciari and S Frixione and M. L Mangano and P Nason and G Ridolfi",
	TITLE=	"{The $\ttbar$ cross-section at 1.8 and 1.96\TeV: a study of the systematics due to parton densities and scale dependence}",
	JOURNAL=	"JHEP",
	VOLUME=	"04",
	PAGES=	"068",
	YEAR=	"2004",
	DOI=	"10.1088/1126-6708/2004/04/068",
	EPRINT=	"hep-ph/0303085",
	ARCHIVEPREFIX=	"arXiv",
}

@ARTICLE{PDF4LHC,
	AUTHOR=	"Butterworth, Jon and Carrazza, Stefano and Cooper-Sarkar, Amanda and DeRoeck, Albert and Feltesse, Jo{\"e}l and Forte, Stefano and Gao, Jun and Glazov, Sasha and Huston, Joey and Kassabov, Zahari and McNulty, Ronan and Morsch, Andreas and Nadolsky, Voica and Rojo, Juan and Thorne, Robert",
	TITLE=	"{PDF4LHC recommendations for LHC Run II}",
	EPRINT=	"1510.03865",
	ARCHIVEPREFIX=	"arXiv",
	PRIMARYCLASS=	"hep-ph",
	REPORTNUMBER=	"OUTP-15-17P, SMU-HEP-15-12, TIF-UNIMI-2015-14, LCTS-2015-27, CERN-PH-TH-2015-249",
	DOI=	"10.1088/0954-3899/43/2/023001",
	JOURNAL=	"J. Phys. G",
	VOLUME=	"43",
	PAGES=	"023001",
	YEAR=	"2016",
}

@ARTICLE{bib:toppt1,
	AUTHOR=	"{CMS Collaboration}",
	TITLE=	"Measurement of differential cross sections for top quark pair production using the lepton+jets final state in proton-proton collisions at 13 {TeV}",
	JOURNAL=	"Phys. Rev. D",
	VOLUME=	"95",
	YEAR=	"2017",
	PAGES=	"092001",
	DOI=	"10.1103/PhysRevD.95.092001",
	EPRINT=	"1610.04191",
	ARCHIVEPREFIX=	"arXiv",
	PRIMARYCLASS=	"hep-ex",
	SLACCITATION=	"%%CITATION = ARXIV:1610.04191;%%",
}

@ARTICLE{bib:toppt3,
	AUTHOR=	"{CMS Collaboration}",
	TITLE=	"Measurement of the differential cross section for top quark pair production in pp collisions at $\sqrt{s}=8$ {TeV}",
	JOURNAL=	"Eur. Phys. J. C",
	VOLUME=	"75",
	YEAR=	"2015",
	PAGES=	"542",
	DOI=	"10.1140/epjc/s10052-015-3709-x",
	EPRINT=	"1505.04480",
	ARCHIVEPREFIX=	"arXiv",
	PRIMARYCLASS=	"hep-ex",
	SLACCITATION=	"%%CITATION = ARXIV:1505.04480;%%",
}

@ARTICLE{bib:toppt4,
	AUTHOR=	"{CMS Collaboration}",
	TITLE=	"Measurement of the $\ttbar$ production cross section in the all-jets final state in pp collisions at $\sqrt{s}=8$ {TeV}",
	JOURNAL=	"Eur. Phys. J. C",
	VOLUME=	"76",
	YEAR=	"2016",
	PAGES=	"128",
	DOI=	"10.1140/epjc/s10052-016-3956-5",
	EPRINT=	"1509.06076",
	ARCHIVEPREFIX=	"arXiv",
	PRIMARYCLASS=	"hep-ex",
	SLACCITATION=	"%%CITATION = ARXIV:1509.06076;%%",
}

@INPROCEEDINGS{ConwayNuis,
	AUTHOR=	"Conway, J. S.",
	TITLE=	"{Incorporating nuisance parameters in likelihoods for multisource spectra}",
	BOOKTITLE=	"{PHYSTAT 2011}",
	PAGES=	"115",
	YEAR=	"2011",
	DOI=	"10.5170/CERN-2011-006.115",
	EPRINT=	"1103.0354",
	ARCHIVEPREFIX=	"arXiv",
	PRIMARYCLASS=	"physics.data-an",
}

@ARTICLE{BarlowBeeston,
	AUTHOR=	"Barlow, Roger and Beeston, Christine",
	TITLE=	"{Fitting using finite Monte Carlo samples}",
	JOURNAL=	"Comput. Phys. Commun.",
	VOLUME=	"77",
	YEAR=	"1993",
	PAGES=	"219",
	DOI=	"10.1016/0010-4655(93)90005-W",
}

@ARTICLE{bib:CLs,
	AUTHOR=	"Read, Alexander L.",
	TITLE=	"Presentation of search results: The {CL$_{\textit{s}}$} technique",
	JOURNAL=	"J. Phys. G",
	VOLUME=	"28",
	YEAR=	"2002",
	PAGES=	"2693",
	DOI=	"10.1088/0954-3899/28/10/313",
	SLACCITATION=	"%%CITATION = JPAGA,G28,2693;%%",
}

@ARTICLE{bib:CL2,
	AUTHOR=	"Junk, Thomas",
	TITLE=	"Confidence level computation for combining searches with small statistics",
	JOURNAL=	"Nucl. Instrum. Meth. A",
	VOLUME=	"434",
	YEAR=	"1999",
	PAGES=	"435",
	DOI=	"10.1016/S0168-9002(99)00498-2",
	EPRINT=	"hep-ex/9902006",
	ARCHIVEPREFIX=	"arXiv",
	PRIMARYCLASS=	"hep-ex",
	REPORTNUMBER=	"CARLETON-OPAL-PHYS-99-01, CERN-EP-99-041",
	SLACCITATION=	"%%CITATION = HEP-EX/9902006;%%",
}

@TECHREPORT{CMS-NOTE-2011-005,
	TITLE=	"Procedure for the {LHC} {Higgs} boson search combination in {Summer} 2011",
	AUTHOR=	"{ATLAS and CMS Collaborations, and LHC Higgs Combination Group}",
	NUMBER=	"CMS-NOTE-2011-005, ATL-PHYS-PUB-2011-11",
	YEAR=	"2011",
	REPORTNUMBER=	"CMS-NOTE-2011-005",
	URL=	"https://cds.cern.ch/record/1379837",
}

@ARTICLE{bib:AsymptCL,
	AUTHOR=	"Cowan, Glen and Cranmer, Kyle and Gross, Eilam and Vitells, Ofer",
	TITLE=	"Asymptotic formulae for likelihood-based tests of new physics",
	JOURNAL=	"Eur. Phys. J. C",
	VOLUME=	"71",
	YEAR=	"2011",
	PAGES=	"1554",
	DOI=	"10.1140/epjc/s10052-011-1554-0",
	NOTE=	"[Erratum: \DOI{10.1140/epjc/s10052-013-2501-z}]",
	EPRINT=	"1007.1727",
	ARCHIVEPREFIX=	"arXiv",
	PRIMARYCLASS=	"physics.data-an",
	SLACCITATION=	"%%CITATION = ARXIV:1007.1727;%%",
}

@ARTICLE{CMS:2024onh,
	AUTHOR=	"{CMS Collaboration}",
	TITLE=	"{The CMS statistical analysis and combination tool: COMBINE}",
	EPRINT=	"2404.06614",
	ARCHIVEPREFIX=	"arXiv",
	PRIMARYCLASS=	"physics.data-an",
	REPORTNUMBER=	"CMS-CAT-23-001, CERN-EP-2024-078",
	DOI=	"10.1007/s41781-024-00121-4",
	JOURNAL=	"Comput. Softw. Big Sci.",
	VOLUME=	"8",
	PAGES=	"19",
	YEAR=	"2024",
}

@INPROCEEDINGS{Verkerke:2003ir,
	AUTHOR=	"Verkerke, W. and Kirkby, D.",
	TITLE=	"The \textsc{RooFit} toolkit for data modeling",
	BOOKTITLE=	"Proc. 13th International Conference on Computing in High Energy and Nuclear Physics (CHEP 2003)",
	ADDRESS=	"La Jolla, California, United States",
	EPRINT=	"physics/0306116",
	ARCHIVEPREFIX=	"arXiv",
	PRIMARYCLASS=	"physics.data-an",
	YEAR=	"2003",
	NOTE=	"[eConf C0303241 (2003) MOLT007]",
	URL=	"https://www.slac.stanford.edu/econf/C0303241/proc/papers/MOLT007.PDF",
}

@ARTICLE{Moneta:2010pm,
	AUTHOR=	"Moneta, Lorenzo and Belasco, Kevin and Cranmer, Kyle S. and Kreiss, S. and Lazzaro, Alfio and Piparo, Danilo and Schott, Gregory and Verkerke, Wouter and Wolf, Matthias",
	EDITOR=	"Speer, T. and Boudjema, F. and Lauret, Jerome and Naumann, Axel and Teodorescu, L. and Uwer, P.",
	TITLE=	"{The RooStats Project}",
	EPRINT=	"1009.1003",
	ARCHIVEPREFIX=	"arXiv",
	PRIMARYCLASS=	"physics.data-an",
	DOI=	"10.22323/1.093.0057",
	JOURNAL=	"PoS",
	VOLUME=	"ACAT2010",
	PAGES=	"057",
	YEAR=	"2010",
}

\cleardoublepage \appendix\section{The CMS Collaboration \label{app:collab}}\begin{sloppypar}\hyphenpenalty=5000\widowpenalty=500\clubpenalty=5000\cmsinstitute{Yerevan Physics Institute, Yerevan, Armenia}
{\tolerance=6000
V.~Chekhovsky, A.~Hayrapetyan, V.~Makarenko\cmsorcid{0000-0002-8406-8605}, A.~Tumasyan\cmsAuthorMark{1}\cmsorcid{0009-0000-0684-6742}
\par}
\cmsinstitute{Institut f\"{u}r Hochenergiephysik, Vienna, Austria}
{\tolerance=6000
W.~Adam\cmsorcid{0000-0001-9099-4341}, J.W.~Andrejkovic, L.~Benato\cmsorcid{0000-0001-5135-7489}, T.~Bergauer\cmsorcid{0000-0002-5786-0293}, K.~Damanakis\cmsorcid{0000-0001-5389-2872}, M.~Dragicevic\cmsorcid{0000-0003-1967-6783}, C.~Giordano\cmsorcid{0000-0001-6317-2481}, P.S.~Hussain\cmsorcid{0000-0002-4825-5278}, M.~Jeitler\cmsAuthorMark{2}\cmsorcid{0000-0002-5141-9560}, N.~Krammer\cmsorcid{0000-0002-0548-0985}, A.~Li\cmsorcid{0000-0002-4547-116X}, D.~Liko\cmsorcid{0000-0002-3380-473X}, I.~Mikulec\cmsorcid{0000-0003-0385-2746}, J.~Schieck\cmsAuthorMark{2}\cmsorcid{0000-0002-1058-8093}, D.~Schwarz\cmsorcid{0000-0002-3821-7331}, R.~Sch\"{o}fbeck\cmsAuthorMark{2}\cmsorcid{0000-0002-2332-8784}, M.~Sonawane\cmsorcid{0000-0003-0510-7010}, W.~Waltenberger\cmsorcid{0000-0002-6215-7228}, C.E.~Wulz\cmsAuthorMark{2}\cmsorcid{0000-0001-9226-5812}
% ...
\par}
$^{1}$Also at Yerevan State University, Yerevan, Armenia\\
$^{2}$Also at Technische Universit\"{a}t Wien, Vienna, Austria\\
% ...
\end{sloppypar}
\end{document}